\documentclass[showpacs,amsmath,amssymb,superscriptaddress,floatfix]{revtex4-2}

\makeatletter
\renewcommand*{\@fnsymbol}[1]{\ensuremath{\ifcase#1\or \dagger\or *\or    \ddagger\or \else\@ctrerr\fi}}
\makeatother

\usepackage{lineno}

\usepackage{setspace} 
\usepackage{color}         
\usepackage[pdftex]{graphicx}      
\usepackage{bm}
\usepackage{natbib}            
\usepackage[colorlinks,plainpages=false,linkcolor=black,urlcolor=black,citecolor=black,pdfpagemode=UseNone,pdfstartview=FitBH]{hyperref}
\usepackage{float}
\usepackage{placeins}
\usepackage{soul}

\usepackage[up,bf,raggedright]{titlesec}

\begin{document}

\title{Orbital-selective time-domain signature of nematicity dynamics in the charge-density-wave phase of La$_{1.65}$Eu$_{0.2}$Sr$_{0.15}$CuO$_4$}

\author{Martin~Bluschke}
\thanks{These authors contributed equally to this work.}
\affiliation{Quantum Matter Institute, University of British Columbia; Vancouver, British Columbia V6T 1Z4, Canada.}
\affiliation{Department of Physics \& Astronomy, University of British Columbia; Vancouver, British Columbia V6T 1Z1, Canada.}

\author{$\hspace{-0.5em}^{, *}$ Naman~K.~Gupta}
\thanks{These authors contributed equally to this work.}
\affiliation{Department of Physics \& Astronomy, University of Waterloo; Waterloo, Ontario N2L 3G1, Canada.}

\author{Hoyoung~Jang}
\affiliation{PAL-XFEL, Pohang Accelerator Laboratory; Pohang, Gyeongbuk 37673, Republic of Korea.}
\affiliation{Photon Science Center, Pohang University of Science and Technology; Pohang, Gyeongbuk 37673, Republic of Korea.}

\author{Ali.~A.~Husain}
\affiliation{Quantum Matter Institute, University of British Columbia; Vancouver, British Columbia V6T 1Z4, Canada.}
\affiliation{Department of Physics \& Astronomy, University of British Columbia; Vancouver, British Columbia V6T 1Z1, Canada.}

\author{Byungjune~Lee}
\affiliation{Max Planck POSTECH/Korea Research Initiative, Center for Complex Phase Materials; Pohang 37673, Republic of Korea.}
\affiliation{Department of Physics, Pohang University of Science and Technology; Pohang 37673, Republic of Korea.}

\author{MengXing~Na}
\affiliation{Quantum Matter Institute, University of British Columbia; Vancouver, British Columbia V6T 1Z4, Canada.}
\affiliation{Department of Physics \& Astronomy, University of British Columbia; Vancouver, British Columbia V6T 1Z1, Canada.}

\author{Brandon~Dos~Remedios}
\affiliation{Quantum Matter Institute, University of British Columbia; Vancouver, British Columbia V6T 1Z4, Canada.}
\affiliation{Department of Physics \& Astronomy, University of British Columbia; Vancouver, British Columbia V6T 1Z1, Canada.}

\author{Steef~Smit}
\affiliation{Quantum Matter Institute, University of British Columbia; Vancouver, British Columbia V6T 1Z4, Canada.}
\affiliation{Department of Physics \& Astronomy, University of British Columbia; Vancouver, British Columbia V6T 1Z1, Canada.}

\author{Peter~Moen}
\affiliation{Quantum Matter Institute, University of British Columbia; Vancouver, British Columbia V6T 1Z4, Canada.}
\affiliation{Department of Physics \& Astronomy, University of British Columbia; Vancouver, British Columbia V6T 1Z1, Canada.}

\author{Sang-Youn~Park}
\affiliation{PAL-XFEL, Pohang Accelerator Laboratory; Pohang, Gyeongbuk 37673, Republic of Korea.}

\author{Minseok~Kim}
\affiliation{PAL-XFEL, Pohang Accelerator Laboratory; Pohang, Gyeongbuk 37673, Republic of Korea.}

\author{Dogeun~Jang}
\affiliation{PAL-XFEL, Pohang Accelerator Laboratory; Pohang, Gyeongbuk 37673, Republic of Korea.}

\author{Hyeongi~Choi}
\affiliation{PAL-XFEL, Pohang Accelerator Laboratory; Pohang, Gyeongbuk 37673, Republic of Korea.}

\author{Ronny~Sutarto}
\affiliation{Canadian Light Source; Saskatoon, Saskatchewan S7N 2V3, Canada.}

\author{Alexander~H.~Reid}
\affiliation{Linac Coherent Light Source, SLAC National Accelerator Laboratory; Menlo Park, California 94025, USA.}

\author{Georgi~L.~Dakovski}
\affiliation{Linac Coherent Light Source, SLAC National Accelerator Laboratory; Menlo Park, California 94025, USA.}

\author{Giacomo~Coslovich}
\affiliation{Linac Coherent Light Source, SLAC National Accelerator Laboratory; Menlo Park, California 94025, USA.}

\author{Quynh~L.~Nguyen}
\affiliation{Linac Coherent Light Source, SLAC National Accelerator Laboratory; Menlo Park, California 94025, USA.}
\affiliation{Stanford PULSE Institute, Stanford University and SLAC National Accelerator Laboratory; Menlo Park, California 94025, USA.}

\author{Nicolas~G.~Burdet}
\affiliation{Linac Coherent Light Source, SLAC National Accelerator Laboratory; Menlo Park, California 94025, USA.}
\affiliation{Stanford Institute for Materials and Energy Sciences, SLAC National Accelerator Laboratory and Stanford University; Menlo Park, California 94025, USA.}

\author{Ming-Fu~Lin}
\affiliation{Linac Coherent Light Source, SLAC National Accelerator Laboratory; Menlo Park, California 94025, USA.}

\author{Alexandre~Revcolevschi}
\affiliation{Institut de Chimie Mol\'eculaire et des Mat\'eriaux d'Orsay, Universit\'e Paris-Saclay; CNRS, UMR 8182, 91405 Orsay, France.}

\author{Jae-Hoon~Park}
\affiliation{Max Planck POSTECH/Korea Research Initiative, Center for Complex Phase Materials; Pohang 37673, Republic of Korea.}
\affiliation{Department of Physics, Pohang University of Science and Technology; Pohang 37673, Republic of Korea.}

\author{Jochen~Geck}
\affiliation{Institute of Solid State and Materials Physics, TU Dresden; 01069 Dresden, Germany.}
\affiliation{Würzburg-Dresden Cluster of Excellence ct.qmat, Technische Universit\"at Dresden; 01062 Dresden, Germany.}

\author{Joshua~J.~Turner}
\affiliation{Linac Coherent Light Source, SLAC National Accelerator Laboratory; Menlo Park, California 94025, USA.}
\affiliation{Stanford Institute for Materials and Energy Sciences, SLAC National Accelerator Laboratory and Stanford University; Menlo Park, California 94025, USA.}

\author{Andrea~Damascelli}
\thanks{Corresponding authors: \newline martin.bluschke@ubc.ca~(M.B.); \newline damascelli@physics.ubc.ca~(A.D.); \newline  david.hawthorn@uwaterloo.ca~(D.G.H.)}
\affiliation{Quantum Matter Institute, University of British Columbia; Vancouver, British Columbia V6T 1Z4, Canada.}
\affiliation{Department of Physics \& Astronomy, University of British Columbia; Vancouver, British Columbia V6T 1Z1, Canada.}

\author{David~G.~Hawthorn}
\thanks{Corresponding authors: \newline martin.bluschke@ubc.ca~(M.B.); \newline damascelli@physics.ubc.ca~(A.D.); \newline  david.hawthorn@uwaterloo.ca~(D.G.H.)}
\affiliation{Department of Physics \& Astronomy, University of Waterloo; Waterloo, Ontario N2L 3G1, Canada.}

\maketitle

\newpage
\noindent\textbf{Understanding the interplay between charge, nematic, and structural ordering tendencies in cuprate superconductors is critical to unraveling their complex phase diagram. Using pump-probe time-resolved resonant x-ray scattering on the (0~0~1) Bragg peak at the Cu $L_3$ and O $K$ resonances, we investigate non-equilibrium dynamics of $Q_a$~=~$Q_b$~=~0 nematic order and its association with both charge density wave (CDW) order and lattice dynamics in La$_{1.65}$Eu$_{0.2}$Sr$_{0.15}$CuO$_4$. The orbital selectivity of the resonant x-ray scattering cross-section allows nematicity dynamics associated with the planar O 2$p$ and Cu 3$d$ states to be distinguished from the response of anisotropic lattice distortions. A direct time-domain comparison of CDW translational-symmetry breaking and nematic rotational-symmetry breaking reveals that these broken symmetries remain closely linked in the photoexcited state, consistent with the stability of CDW topological defects in the investigated pump fluence regime.}

\vskip 0.5cm
\vskip 0.5cm
\noindent Quantum materials with strong electronic correlations typically exhibit a variety of intertwined electronic ordering tendencies that have very similar, or even identical, energy and temperature scales~\cite{Keimer2017}. These include, for example, antiferromagnetism, charge- and spin-density waves, orbital order, superconductivity and nematicity. A famous case highlighting the importance of intertwined order is the cuprate high-temperature superconductors~\cite{Fradkin2015, Keimer2015}. Quite recently, charge density wave (CDW) order, a translational-symmetry-breaking modulation of low-energy charge degrees of freedom, was identified as a generic phase of the cuprates that co-exists and competes with superconductivity~\cite{Comin2016_CO_review}. This competition has been observed as a function of temperature, hole-doping, applied magnetic fields, uniaxial strain and optical pumping~\cite{Huecker2011, Ghiringhelli2012, Chang2012, Chang2016, Jang2016, Kim2018, Tabis2017, Bluschke2019, Wandel2022, Jang2022}. In addition to CDW order, cuprates also exhibit nematic order~\cite{Ando1999, Noda1999, Mook2000, Ando2002, Stock2004, Hinkov2007, Hinkov2008, Daou2010, Lawler2010, Mesaros2011, Wu2019, Auvray2019, Wang2023}, a breaking of the rotational symmetry of the electronic structure within the CuO$_2$ planes from four-fold to two-fold symmetric ($C_4$ rotational symmetry breaking). 

Electronic liquid crystal phases with varying combinations of broken rotational and translational symmetries have been theorized~\cite{Kivelson1998, Fradkin2010}, thus making it desirable to probe each broken symmetry independently. When present, electronic nematic order is able to couple to CDW order favoring a unidirectional character of the CDW. While spontaneous rotational symmetry breaking has been predicted in many early theoretical studies of doped antiferromagnets~\cite{Zaanen1989, Machida1989, Kato1990, Emery1993, White1998, Halboth2000, Lorenzana2002}, the effect of quenched disorder in real materials can be significant resulting in locally preferred stripe orientations and pinning of stripe fluctuations, which can obscure both macroscopic anisotropies as well as intrinsic energy and temperature scales~\cite{Kivelson2003}. Alternately, when the crystal structure explicitly breaks the $C_4$ rotational symmetry of the CuO$_2$ planes, the lattice itself provides an orienting potential for intrinsic nematic correlations. For example, this is the case in the 3-dimensional CDW state of orthorhombic YBa$_2$Cu$_3$O$_{6+\delta}$~(YBCO)~\cite{Gerber2015, Chang2016, Jang2016, Kim2021} as well as in the low-temperature tetragonal (LTT) phase of `214' cuprates such as La$_{2-x}$Ba$_{x}$CuO$_4$, La$_{1.6-x}$Nd$_{0.4}$Sr$_{x}$CuO$_4$ and La$_{1.8-x}$Eu$_{0.2}$Sr$_{x}$CuO$_4$. Although these orthorhombic crystal structures naturally induce rotational asymmetry in their electronic structure, nematicity is identified with an additional temperature dependent enhancement of this rotational asymmetry, driven by undirectional CDW formation or strong electronic correlations, such as nearest neighbour Coulomb repulsion of planar O states or exchange interactions~\cite{Kivelson2004, Fischer2014, Wang2023,Yamase2000,Yamase2021}.

\begin{figure*}[ht]
\includegraphics[trim={10cm 40cm 10cm 25cm},clip,width=\textwidth]{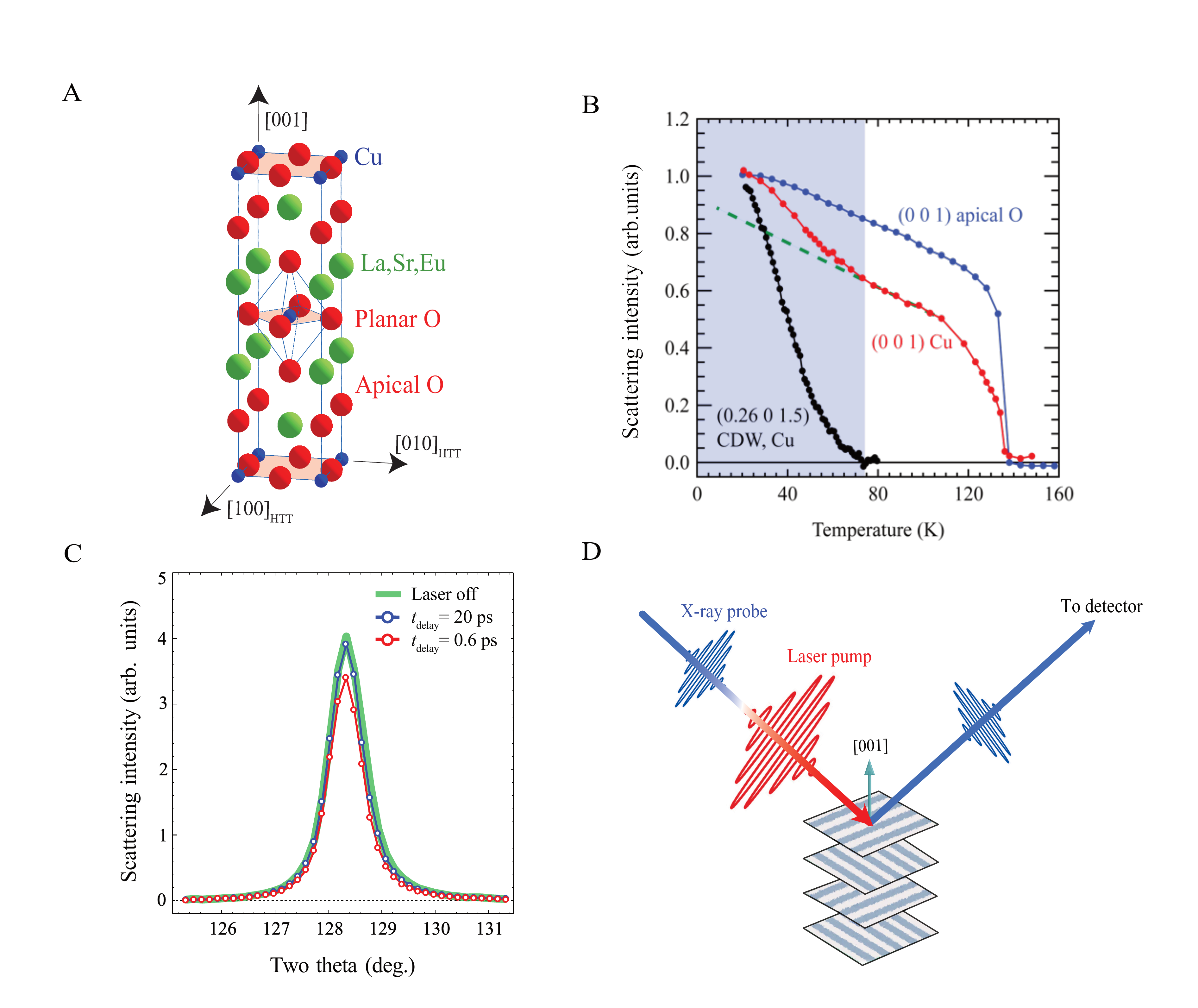}
\caption{\textbf{La$_{1.65}$Eu$_{0.2}$Sr$_{0.15}$CuO$_4$ structure, equilibrium RXS results and tr-RXS experiment.} (A) Crystal structure of LESCO. The crystal axes are labelled according to the high-temperature tetragonal (HTT) setting, and the CuO$_2$ planes are highlighted in pale red. (B) Equilibrium temperature dependence of the apical O $K$ RXS intensity at $Q=(001)$ compared with the Cu $L_3$ RXS intensity at $Q=(001)$ and $Q_{\text{CDW}}=(0.26~0~1.5)$. (C) Planar O $K$ tr-RXS intensity at $Q=(001)$ probed along the (00$L$) direction in a $\theta-2\theta$ scan. The equilibrium peak profile is compared with scans taken 0.6~ps and 20~ps after excitation with 1.55 eV pump photons at a fluence of 50 $\mu$J/cm$^{2}$. (D) Schematic depiction of the pump-probe geometry used to access the $Q=(001)$ peak originating from the layer alternating $ab$-plane anisotropy. 
\label{fig:fig1}}
\end{figure*}

A recent breakthrough involved the identification of an equilibrium resonant x-ray scattering (RXS) signature of nematic order in various 214 cuprate superconductors and its relationship to both the crystal structure and CDW order~\cite{Achkar2016, Gupta2021}. In these measurements, nematicity was probed by measuring the (001) Bragg peak using photons tuned to the Cu~$L_3$, the O~$K$ and the La~$M_5$ resonances. The (001) reflection is forbidden in the high temperature structural phases of these materials but is detectable on resonance in the LTT phase ($T~<~T_{\text{LTT}}~\approx$~135~K). Fig.~\ref{fig:fig1}(a) schematically depicts the layered 214 cuprate structure. The intensity of the (001) reflection is a measure of the contrast between the resonant scattering form factors of corresponding ions in neighbouring layers along the $c$ direction. From symmetry arguments, this contrast results from a local $ab$-plane anisotropy which rotates by $\pi/2$ around the $c$-axis with each consecutive layer~\cite{Achkar2016}. By choosing incident photon energies associated with the various accessible core-valence resonances it is possible to probe the local $ab$-plane anisotropy associated with the valence electronic states of each ion independently. When the incident photon energy matches the La~$M_5$ (835~eV) or the apical O~$K$ (532.4~eV) resonances the $Q=(001)$ RXS intensity is a measure of the anisotropic structural distortion associated with the LTT phase and a single order parameter-like onset of intensity is observed at $T_{\text{LTT}}$. In contrast, the (001) reflection measured at the Cu~$L_3$ resonance (931.7~eV) or the planar O~$K$ resonance (528.7~eV) is additionally sensitive to nematicity in the CuO$_2$ planes. Fig.~\ref{fig:fig1}(b) shows the equilibrium temperature-dependence of the $Q=(001)$ RXS intensity at the apical O $K$ and Cu~$L_3$ resonances in La$_{1.65}$Eu$_{0.2}$Sr$_{0.15}$CuO$_4$ (LESCO). At the Cu~$L_3$ resonance, the initial onset of intensity near 135~K corresponds to $T_{\text{LTT}}$, whereas the additional upturn near 75~K indicates the onset of nematic order, whose temperature dependence is correlated with the breaking of translational symmetry measured at $Q_{\text{CDW}}=(0.264~0~1.5)$.

\begin{figure}[ht]
\includegraphics[width=.5\textwidth]{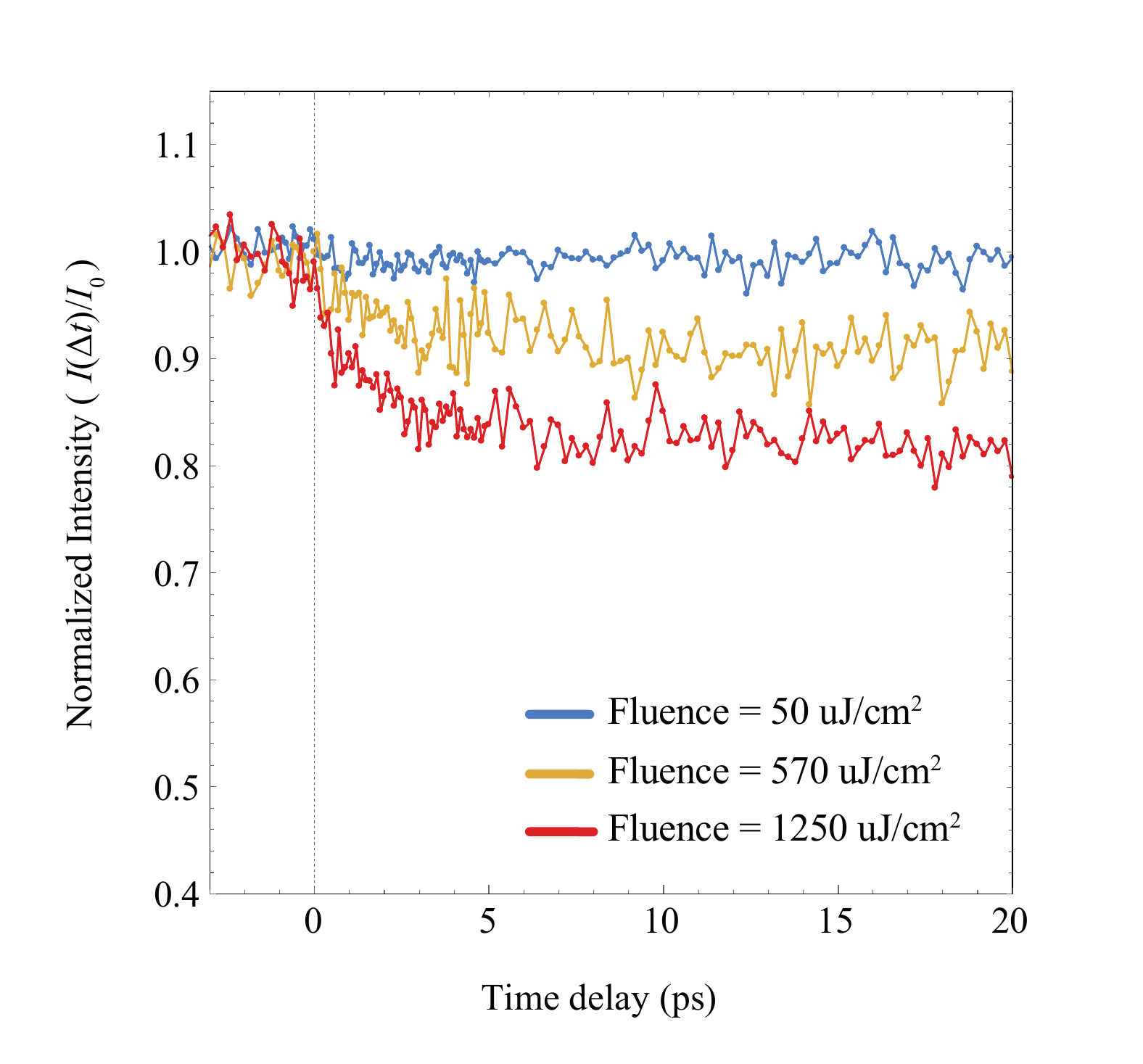}
\caption{\textbf{Fluence series at the apical O $K$ resonance.} LESCO $x=0.15$ at 20 K. Time-resolved resonant x-ray scattering (tr-RXS) intensity at the apical O $K$ resonance and $Q=(001)$ as a function of pump-probe delay. This pump-probe response is sensitive only to the anisotropy induced by the LTT structural distortion and becomes unobservable for low fluences.
\label{fig:fig2}}
\end{figure}

Equilibrium studies of the (001) Bragg peak require detailed temperature dependent measurements in order to isolate the electronic nematic component. Here we present a non-equilibrium measurement approach based on time-resolved resonant x-ray scattering (tr-RXS), in which we are able to disentangle electronic and lattice contributions to the (001) Bragg peak by virtue of their distinct responses to optical pumping combined with the orbital selectivity of the RXS cross section. Pump-probe experiments on cuprates make use of optical laser pulses to excite non-equilibrium populations of hot electrons or drive vibrational modes of the lattice. At low pump fluences these experiments have been successful at revealing important information about the low-energy excitations of equilibrium ordered states~\cite{Gedik2005, Torchinsky2013, Dakovski2015a, Boschini2018, Mitrano2019a, Mitrano2019b, Jang2022, Wandel2022}, whereas with higher pump fluences states of matter far from equilibrium can be accessed, such as photoinduced superconductivity~\cite{Fausti2011, Hu2014, Kaiser2014, Foerst2014a, Foerst2014b, Khanna2016} or states with renormalized onsite Coulomb interactions~\cite{Baykusheva2022}. Here we report the first pump-probe time- and orbital-resolved investigation of nematicity in a cuprate superconductor. We carefully study the response of both the lattice anisotropy and electronic nematicity as a function of fluence in order to establish a perturbative regime, in which nematicity dynamics can be studied without strongly disturbing related lattice degrees of freedom. 

\vskip 0.5cm
\vskip 0.5cm
\vskip 0.5cm
\vskip 0.5cm
\vskip 0.5cm\vskip 0.5cm
\noindent\textbf{Results}

In this section we describe the results of pump-probe tr-RXS measurements of LESCO, probing at momentum transfer $Q=(001)$. The system is excited using 50~fs pulses of a 1.55~eV (800~nm) Ti:sapphire laser, and then probed at varying time delays with respect to the excitation using $80$~fs soft x-ray pulses. The experiment was performed using the Resonant Soft X-ray Scattering (RSXS) instrument~\cite{Jang2020} at the Pohang Accelerator Laboratory X-ray Free Electron Laser (PAL-XFEL), operating at a 60~Hz repetition rate. In Fig.~\ref{fig:fig1}(c) we plot a representative scan of the $Q=(001)$ peak taken along the (00$L$) direction at 20~K and with probe photons tuned to the planar O~$K$ resonance. In order to visualize the pump-induced peak intensity changes which we discuss throughout this report, we have plotted the corresponding scans taken immediately after ($\sim$~0.5 ps) and at a longer time delay (20~ps) with respect to the excitation for a pump fluence of 50~$\mu$J/cm$^{2}$. Figure~\ref{fig:fig1}(d) schematically depicts the pump-probe geometry with respect to the layer-alternating anisotropy in the low-temperature CDW state.

In order to first identify and isolate the pump-induced response of the lattice, we show in Fig.~\ref{fig:fig2} the $Q=(001)$ tr-RXS intensity measured at the apical O $K$ resonance as a function of pump-probe delay. The choice of the apical O resonance guarantees sensitivity to rotational symmetry breaking driven by the LTT structural distortion, without electronic contributions associated with the CuO$_2$ planes. The LTT structure demonstrates a slow response to optical pumping, characteristic of lattice dynamics, which become vanishingly small for low pump fluences.

\begin{figure*}[ht]
\includegraphics[width=.95\textwidth]{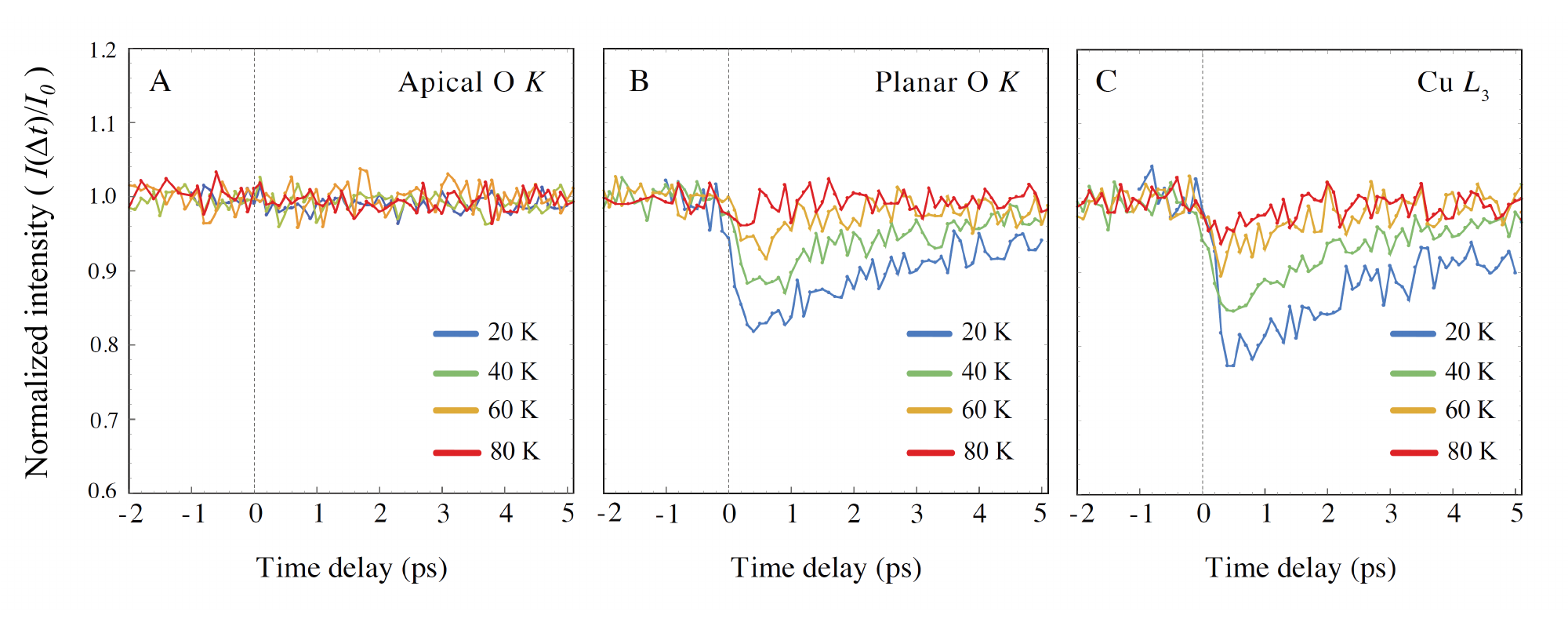}
\caption{\textbf{Temperature series at low fluence.} Normalized tr-RXS intensity ($I(\Delta t)/I_0$) at $Q=(001)$ as a function of pump-probe delay in LESCO ($x=0.15$) with $T_{\text{CDW}}\approx$~75~K~\cite{Fink2009, Achkar2016}. Pump photon energy 1.55 eV and fluence 50~$\mu$J/cm$^2$.  (A) The tr-RXS intensity at the apical O $K$ resonance, sensitive only to structural distortions, is unperturbed by the optical pump at this low fluence. The planar O $K$ resonance (B) and the Cu $L_3$ resonance (C) are sensitive to pump-induced nematicity dynamics in the CDW phase.
\label{fig:fig3}}
\end{figure*}

 When measured at the planar O $K$ or Cu $L_3$ resonances the $Q=(001)$ RXS intensity is sensitive to rotational symmetry breaking associated with both the LTT lattice distortion as well as with electronic nematicity of the CuO$_2$ planes. Based on the fluence-dependent tr-RXS measurements presented in Fig.~\ref{fig:fig2}, we identify a low-fluence regime (approx. $F~<~100$~$\mu$J/cm$^2$) in which the optical pump does not induce a significant response in the lattice. Fig.~\ref{fig:fig3} shows the low fluence (50~$\mu$J/cm$^2$) dynamics observed for incident photons tuned to the apical O $K$, the planar O $K$ and the Cu $L_3$ resonances, and for a series of temperatures below and above $T_{\text{CDW}}$. Whereas the $Q=(001)$ RXS intensity at the apical O $K$ resonance is unperturbed with this low pump fluence, measurements at the planar O $K$ and the Cu $L_3$ resonances detect a large and fast response to pumping, clearly distinct from the LTT structural dynamics. Importantly, the temperature dependence of the fast (001) peak dynamics at the planar O $K$ and the Cu $L_3$ resonances is correlated with the onset of CDW order, indicating that the fast dynamics are associated with electronically driven rotational symmetry breaking.

\begin{figure*}[ht]
\includegraphics[width=\textwidth]{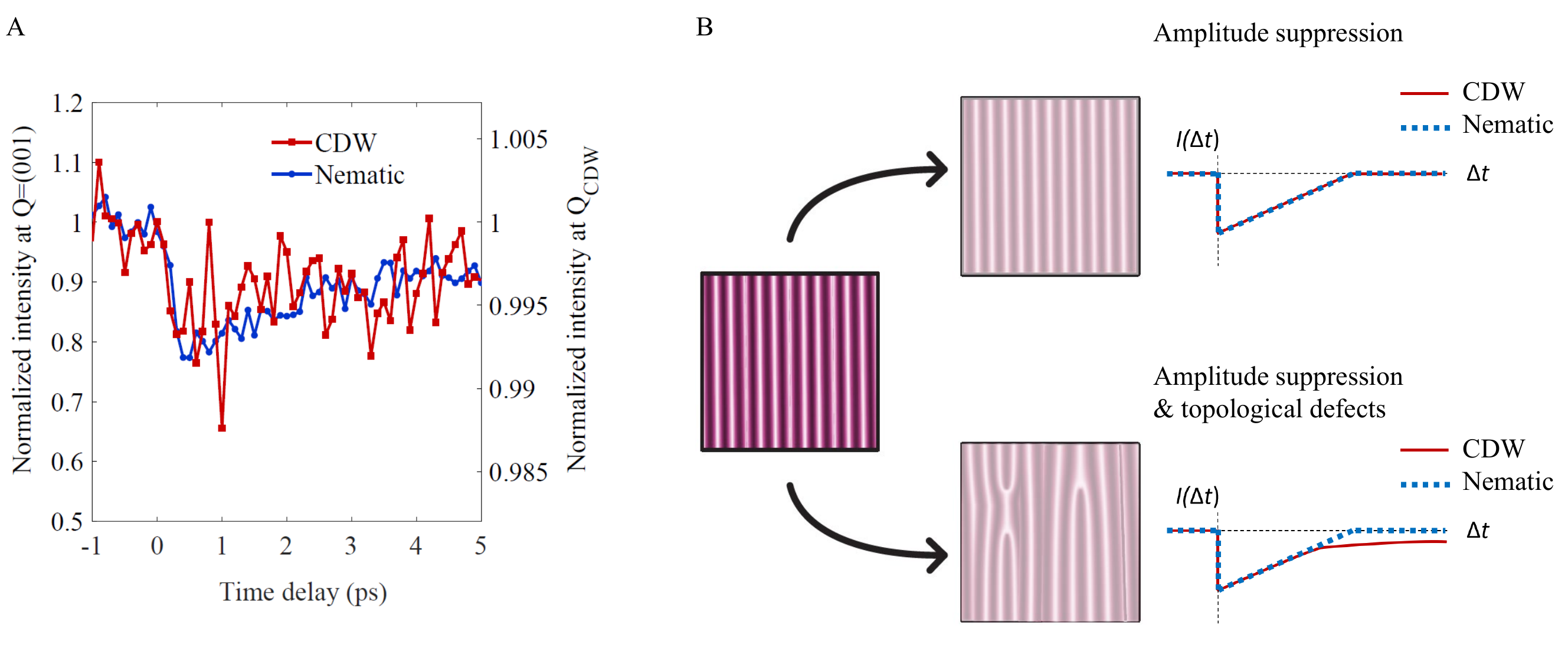}
\caption{\textbf{Comparison of CDW and nematicity dynamics.} (A) CDW translational-symmetry breaking and nematic rotational-symmetry breaking dynamics measured at $Q_{\text{CDW}}=(0.26~0~1.74)$ and $Q=(0 0 1)$ respectively. (B) Schematic depiction of CDW degradation via amplitude suppression or the combination of amplitude suppression and topological defect formation. The corresponding recovery dynamics expected for CDW and nematic order are depicted to the right.
\label{fig:fig4}}
\end{figure*}

Having established the coincident onset of broken rotational and translational symmetry as a function of temperature, it is of interest to ask if and how these broken symmetries remain linked in the photoexcited state. We  address this question by probing the photoexcited dynamics of the CDW translational symmetry breaking at $Q_{\text{CDW}}=(0.26~0~1.74)$. Fig.~\ref{fig:fig4}(a) compares the time-domain response of the tr-RXS intensities associated with both CDW and nematic order under similar pump conditions. Despite the lower signal-to-noise ratio of the $Q_{\text{CDW}}$ measurement, the two orders appear to be suppressed and recover together. The dynamics are also similar to the pump-induced CDW dynamics reported under similar conditions for La$_{2-x}$Ba$_x$CuO$_4$.~\cite{Mitrano2019a, Mitrano2019b}

\begin{figure*}[ht]
\includegraphics[trim={15cm 25cm 25cm 53cm},clip,width=0.8\textwidth]{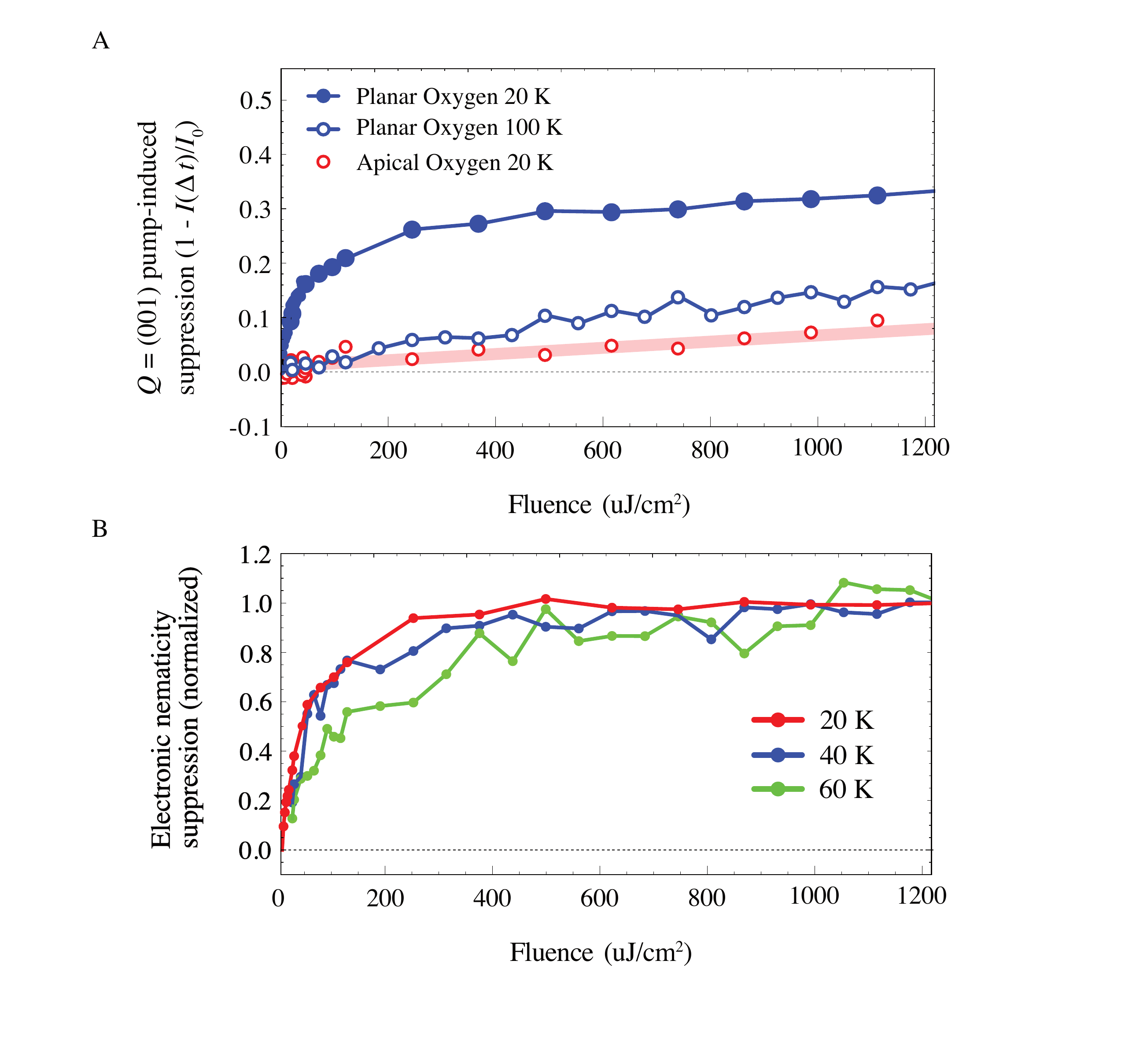}
\caption{\textbf{Fluence dependence.} (A) Pump-induced suppression ($1-I(\Delta t)/I_0$) of the O~$K$ resonance $Q=(001)$ tr-RXS intensity at short time delay as a function of fluence.  The 20~K signal at the planar O~$K$ resonance comprises both lattice and electronic contributions, while the apical resonance is sensitive only to structural anisotropy.  The signal at the planar O~$K$ resonance is also shown at $T=100~\text{K}>T_{\text{CDW}}$ where its fluence dependence is qualitatively similar to that observed at the apical O~$K$ resonance. The shaded red line is a linear fit to the apical O $K$ resonance data. (B)  Fluence-dependent suppression of the electronic nematic signal at $+0.5$~ps (near maximum suppression). The fluence dependent response of the lattice is captured by a linear fit to the 20~K apical O data plotted in red in (A) and then subtracted from fluence-scans collected at the planar O~$K$ resonance inside the CDW phase.
\label{fig:fig5}}
\end{figure*}

At higher fluences, where both the lattice and electronic subsystems are significantly excited, it is more challenging to differentiate contributions to the $Q=(001)$ RXS intensity. In order to do so, we study the fluence dependence of the tr-RXS signal for various temperatures and at each resonance. Fig.~\ref{fig:fig5}(a) shows the fluence dependence of the planar O $K$ resonance $Q=(001)$ suppression at 20~K ($T<T_{\text{CDW}}$) and at 100~K ($T>T_{\text{CDW}}$), immediately following the pump excitation (pump-probe delay $+0.5$~ps). On the same plot we show the corresponding fluence dependence of the apical O $K$ resonance $Q=(001)$ suppression at 20~K (pump-probe delay  +0.7~ps). The apical O measurement captures the fluence-dependent response of the lattice at short time delays. Similarly, the dynamics observed at the planar resonance trend towards a lattice-like response as the temperature is increased beyond $T_{\text{CDW}}$. As the pump fluence is increased, a discrepancy develops between the 20~K data at the apical resonance and the 100~K data at the planar resonance. This discrepancy may result from the temperature-dependent susceptibility of the LTT distortion to optical pumping, whose time-domain response is expected to be a complicated function of both temperature and pump fluence. Alternately, the stronger response seen in the 100~K planar resonance data may correspond to the suppression of high-temperature residual nematicity, associated with either a purely nematic state or with weak short-range CDW correlations.

The apical O $K$ fluence scan in Fig.~\ref{fig:fig5}(a) has been fit with a line (shaded red) to determine the slope of the fluence-dependent suppression of the LTT distortion. This slope is then subtracted from fluence-dependent measurements taken at the planar O $K$ resonance for a series of temperatures. The resulting curves are normalized to the maximum pump-probe effect observed at high fluence and plotted together in Fig.~\ref{fig:fig5}(b). The response of nematicity to optical pumping saturates at a pump fluence which is approximately independent of temperature. In contrast, if the pump-induced effect were driven purely by the post-pump evolution of the transient electronic temperature, one may have expected a decreasing saturation fluence at higher temperatures. Accordingly, this indicates that the suppression of nematic order at short time delays is associated with the establishment of a non-thermal electronic state~\cite{Na2020}.

\vskip 0.5cm
\noindent\textbf{Discussion}

Differentiating translational and rotational-symmetry breaking in CDW-ordered cuprates is difficult but of fundamental interest~\cite{Khosaka07, Lawler2010, Mesaros2011, Blanco-Canosa2014, Comin2015, McMahon2020, Kang2019, Boschini2021, Kim2021}. Of the known stripe-ordered cuprates, La$_{2-x}$Ba$_x$CuO$_4$ with $x\sim~1/8$ demonstrates the most pronounced CDW and has therefore remained a prototype for CDW diffraction studies. Previous tr-RXS studies of La$_{2-x}$Ba$_x$CuO$_4$ have focused on the pump-induced response of the $Q_{\text{CDW}}$ reflection, associated with translation symmetry breaking in the CDW phase, and have treated the $Q=(001)$ reflection purely as a reference probe of the LTT lattice distortion~\cite{Foerst2014a, Khanna2016, Mitrano2019a, Mitrano2019b}. Here we show that the tr-RXS intensity at $Q=(001)$ in the related material LESCO can be an extremely sensitive probe of both LTT lattice dynamics, as well as nematicity dynamics, depending on the choice of probe photon energy.  

Selective measurements of CuO$_{2}$ plane nematicity dynamics demonstrate that rotational symmetry breaking in the photoexcited state remains strongly coupled to the translational symmetry breaking observed at $Q_{\text{CDW}}$. In contrast, the dynamics observed at the apical O resonance and shown in Fig.~\ref{fig:fig2} are typical of the lattice response to photo-excitation in cuprates~\cite{Conte2012, Giannetti2016}, and have no apparent sensitivity to CDW order. Hot electrons excited by the optical pump decay via electron-phonon coupling mediated channels, eventually exciting the entire phonon bath and weakening the LTT distortion over several picoseconds. In contrast, the nematic signal observed below $T_{\text{CDW}}$ and plotted in Fig.~\ref{fig:fig3}(b),(c) is suppressed within hundreds of femtoseconds, consistent with the direct electronic coupling of nematicity to the photoexcited electronic population. Alternately, the fast suppression may involve the participation of phonon modes which are strongly coupled to the CuO$_2$ plane electronic degrees of freedom, such as the transverse acoustic and optical phonon modes, which are known to experience a softening at the CDW wave vector in the CDW phase.~\cite{Tacon2014} The subsequent recovery of the nematic signal unfolds over several picoseconds, comparable to the timescale of the LTT suppression, suggesting a scenario in which the recovery of the electronic nematic ground state occurs via an electron-phonon coupling mediated transfer of energy to the lattice. The recovery of the LTT distortion proceeds over a significantly longer time scale associated with the time it takes for acoustic phonons generated in the pump process to leave the probed region.

Similar to liquid crystals, which can be characterized by both a nematic and a smectic order parameter~\cite{Izzo2020}, the low temperature states of the cuprates appear to be characterized by rotational (Ising nematic) and translational (CDW) symmetry breaking order parameters. These orders are distinct, opening up the possibility that nematic order may survive the melting of CDW order in at least some range of parameter space. A close inspection of Fig. 3(b) and Fig. 3(c) reveals that although the low fluence suppression of nematic order remains largely locked to the onset of CDW order, residual dynamics are observed at 80 K, above the nominal $T_{\text{CDW}}$ determined from equilibrium energy-integrated RXS. This observation may be understood as evidence of nematic order existing in the absence of CDW translational symmetry breaking, similar to observations of nematic order outside the CDW dome in overdoped La$_{1.6-x}$Nd$_{0.4}$Sr$_{x}$CuO$_4$~\cite{Gupta2021}. Alternately, these residual nematicity dynamics may track weak amplitude, short range CDW correlations similar to those observed at high temperatures by resonant inelastic x-ray scattering measurements~\cite{Lee2022, Wang2020a}. Whatever the precise origin, residual nematicity at high-temperatures may be responsible for the discrepancy between the 100~K planar~O and the 20~K apical~O fluence dependences shown in Fig.~\ref{fig:fig5}(a).
 
Comparing the non-equilibrium dynamics of the CDW and nematic orders provides insight into the role of photoexcited topological defects in suppressing CDW order.  While the CDW order parameter is described by a wavevector, an amplitude and a phase, the nematic order parameter is simply described by an Ising-like director and an amplitude. Translational symmetry breaking in the CDW phase may be degraded either by amplitude suppression or by the introduction of topological defects such as discommensuration lines and dislocations, which suppress long-range phase coherence. Previous investigations have observed both the presence~\cite{Zong2018, Mitrano2019a, Wandel2022} and absence~\cite{Lee2022} of topological defects in photo-induced phase transitions of CDW systems. When present, photo-excited topological defects have been shown to decay over a longer time scale than the recovery of the amplitude mode, thus leading to two time scales in the CDW recovery dynamics~\cite{Zong2018}. In contrast nematic order is expected to recover with the amplitude. In Fig.~\ref{fig:fig4}(b) two scenarios are depicted schematically, one in which a defect-free uniaxial CDW is suppressed via pure amplitude suppression, and one in which CDW topological defects are additionally generated by the pump excitation. The observation that CDW and nematic order recover together following photoexcitation (Fig.~\ref{fig:fig4}(a)) is consistent with a scenario in which the pump does not excite a significant number of CDW topological defects. This, however, does not preclude the existence of static CDW defects which are unperturbed at low pump fluences.

Understanding the role of CDW topological defects in the photoinduced dynamics of the cuprates is relevant to understanding the relationship between CDW order and superconductivity. In the cuprate superconductor YBCO, where the uniaxial LTT distortion is absent, the locations of CDW discommensuration lines have been associated with the presence of superconductivity, whose optical melting leads to a strengthening of the CDW phase coherence~\cite{Wandel2022}. In contrast, CDW order in La$_{1.875}$Ba$_{0.125}$CuO$_4$ is insensitive to $T_\text{c}$~\cite{Huecker2011}, and the phase-resolved topology of the CDW domain network remains stable against temperature above both the CDW and LTT transitions melting only for $T>T_{\text{LTO}}$ where four-fold symmetry is restored in the CuO$_2$ plane~\cite{Chen2019a}. It is therefore not surprising that, in the related material LESCO, the 50 $\mu$J/cm$^2$ optical pump used in our experiment does not generate significant changes in the CDW topology. Our result is also reminiscent of the photoinduced CDW dynamics observed in the 214 stripe-ordered nickelate La$_{1.75}$Sr$_{0.25}$NiO$_{4}$, where it was argued that the photoexcited state involves both amplitude and phase fluctuations of the CDW, but that topological defects are not created due to the large energy cost associated with reconfiguring a large number of spins and charges in the coupled charge- and spin-stripe phase~\cite{Lee2012}.

Identifying the local physics associated with rotational symmetry breaking observed via macroscopic probes~\cite{Ando1999, Noda1999, Daou2010, Lubashevsky2014, Wu2019} often proves to be a challenge. Here we have demonstrated an orbital-selective time-resolved probe of electronic nematic order which accompanies the CDW phase in LESCO. The $Q=(001)$ RXS signal averages over the entire $ab$-plane ($Q_a$ = $Q_b$ = 0) while simultaneously being sensitive to intra-unit-cell asymmetries of the Cu~$3d$ and O $2p$ states. The evolution of nematicity dynamics as they are gradually unlocked from the translational symmetry breaking CDW order with additional hole-doping is not fully understood. Future studies may investigate pump-induced nematicity dynamics across the cuprate phase diagram, with a particular focus on the phase region just below the pseudogap critical point, where electronic nematic order was recently reported to persist in the absence of a translational-symmetry breaking CDW~\cite{Gupta2021}. More generally, our results demonstrate the ability to capture the ultrafast behavior of non-equilibrium nematicity dynamics using tr-RXS, thereby paving the way for future tr-RXS studies of $Q=0$ nematic correlations in other systems such as $1T-$TiSe$_2$, which exhibits an ordered stacking of 3 uniaxial CDW's with distinct wavevector orientations~\cite{Ishioka2010}.

\vskip 0.5cm
\noindent\textbf{Methods}

\noindent\textbf{Sample Preparation.} Single crystal La$_{1.65}$Eu$_{0.2}$Sr$_{0.15}$CuO$_4$ was grown using the travelling solvent floating zone technique at the Universit\'e Paris-Saclay. The data presented in the main manuscript was collected on a sample which was cleaved in air to expose a clean surface with an approximate (001)~orientation. The sample dimensions were approximately $2.5$~mm~$\times~1$~mm~$\times~1$~mm.

\noindent\textbf{Equilibrium resonant x-ray scattering.} Equilibrium RXS measurements of both the $Q=(001)$ Bragg peak and the $Q_{\text{CDW}}$ peak were performed  at the REIXS beamline~\cite{Hawthorn2011} of the Canadian Light Source on the same sample studied which was subsequently studied by tr-RXS. A subset of these data are presented in Fig.~\ref{fig:fig1}(b) of the main manuscript. These temperature dependent peak intensities were collected with $\sigma$-polarized photons (polarized parallel to the planar CuO$_2$ bonds). The indicated reciprocal lattice units (r.l.u.) are defined using real-space lattice constants $a~=~b~=~3.79$~\AA~ and $c~=~13.14$~\AA, where the $a$ and $b$ lattice constants are oriented along the planar CuO$_2$ bonds reflecting the high-temperature tetragonal crystal symmetry. 

\noindent\textbf{Time-resolved resonant x-ray scattering.} The tr-RXS measurements presented here were performed at the RSXS endstation of the PAL-XFEL in a vacuum of $2~\times10^{-8}$~mbar. A base temperature of $\sim20$~K was achieved in our experiment using a liquid Helium flow cryostat. Probe XFEL pulses with $\pi$-polarization were produced at a repetition rate of 60~Hz through self-amplified spontaneous emission by passing a 3.15~GeV electron beam through the undulators of the Soft X-ray Scattering and Spectroscopy beamline. A plane grating monochromator was used to achieve monochromatic pulses with a bandwidth of $\sim0.5$~eV, tuned to the relevant soft x-ray resonances: Cu~$L_3$ (931.7~eV), La~$M_5$ pre-edge (832~eV), apical O $K$ (532.4~eV) and planar O $K$ (528.7~eV). The XFEL pulse duration and spot size were approximately 80~fs and 130~$\mu$m~$\times$~250~$\mu$m respectively. Scattered x-rays were detected using an avalanche photodiode with a 1~mm aperture positioned at 150~mm from the sample. 

\noindent\textbf{Pump-probe scheme.} The optical pump pulses were produced using a 1.55~eV Ti:Sapphire laser and delivered to the sample approximately collinear to the trajectory of the probing XFEL pulse. The pump repetition rate was set to 30~Hz such that every second XFEL pulse probes the optically excited system (for positive time delays), while the others probe the equilibrium state. The optical pump spot size ($\sim$650~$\mu$m~$\times$~550~$\mu$m) was significantly larger than the XFEL probe spot size, so as to minimize the spatial inhomogeneity of the pumping observed by the probe. The optical pump was $\sigma$-polarized (parallel to the CuO$_2$ planes and the $\theta$ rotation axis) so as to ensure a constant orientation of the polarization with respect to the crystal axes for all incidence angles $\theta$. A test of the $Q=(001)$ tr-RXS intensity at the Cu $L_3$ resonance indicated that the pump-induced $Q=(001)$ dynamics are independent of the choice of pump polarization (parallel or perpendicular to the CuO$_2$ planes). The pump laser fluence was controlled using a motorized attenuator consisting of a half-wave plate and two broadband thin film polarizers. The reported fluences correspond to the incident fluence after correcting for the different pump (and probe) incidence angles used when probing the (001) Bragg Peak at different resonant photon energies. The duration of the pump pulse was $\sim50$~fs and the delay time between the arrival of the pump and probe pulses was controlled by a mechanical delay stage. Spatial and temporal overlap of the pump and probe spots was achieved using a Ce:YAG crystal (YAG). Absorption of the soft x-ray pulse in the YAG generates free carriers giving rise to a measurable change in its optical properties. After positioning both spots at the same location on the YAG, temporal overlap was achieved by scanning the pump delay stage and monitoring the optical transmission. The overall time resolution achieved at the RSXS endstation is approximately $110$~fs.

\noindent\textbf{Data Analysis.} The normalized tr-RXS intensities plotted in Fig.~\ref{fig:fig2} and Fig.~\ref{fig:fig3} of the main manuscript correspond to the ratio of the tr-RXS intensity $I(\Delta t)$ measured at time delay $\Delta t$ after the pump arrival, and the equilibrium RXS intensity $I_0$ measured before the pump arrival. In turn, $I(\Delta t)$ and $I_0$ are both obtained by normalizing the scattering intensity recorded on the avalanche photodiode by the incident XFEL pulse intensity obtained from a Krypton gas monitor detector located downstream from the grating monochromator. The data in these time-delay scans were binned to the nominal step size of the mechanical delay stage (100~fs for the first 5~ps after excitation). In order to capture the saturating effect of the pump, in the main manuscript Fig.~\ref{fig:fig5}(a) we plot ($1-I(\Delta t)/I_0$), and in  Fig.~\ref{fig:fig5}(b) we plot the same quantity after subtraction of the structural background as described in the text.

\vskip 0.5cm
\vskip 0.5cm
\noindent\textbf{Data availability}

\noindent All data needed to evaluate the conclusions in the paper are present in the paper.


\begin{thebibliography}{79}%
\makeatletter
\providecommand \@ifxundefined [1]{%
 \@ifx{#1\undefined}
}%
\providecommand \@ifnum [1]{%
 \ifnum #1\expandafter \@firstoftwo
 \else \expandafter \@secondoftwo
 \fi
}%
\providecommand \@ifx [1]{%
 \ifx #1\expandafter \@firstoftwo
 \else \expandafter \@secondoftwo
 \fi
}%
\providecommand \natexlab [1]{#1}%
\providecommand \enquote  [1]{``#1''}%
\providecommand \bibnamefont  [1]{#1}%
\providecommand \bibfnamefont [1]{#1}%
\providecommand \citenamefont [1]{#1}%
\providecommand \href@noop [0]{\@secondoftwo}%
\providecommand \href [0]{\begingroup \@sanitize@url \@href}%
\providecommand \@href[1]{\@@startlink{#1}\@@href}%
\providecommand \@@href[1]{\endgroup#1\@@endlink}%
\providecommand \@sanitize@url [0]{\catcode `\\12\catcode `\$12\catcode
  `\&12\catcode `\#12\catcode `\^12\catcode `\_12\catcode `\%12\relax}%
\providecommand \@@startlink[1]{}%
\providecommand \@@endlink[0]{}%
\providecommand \url  [0]{\begingroup\@sanitize@url \@url }%
\providecommand \@url [1]{\endgroup\@href {#1}{\urlprefix }}%
\providecommand \urlprefix  [0]{URL }%
\providecommand \Eprint [0]{\href }%
\providecommand \doibase [0]{https://doi.org/}%
\providecommand \selectlanguage [0]{\@gobble}%
\providecommand \bibinfo  [0]{\@secondoftwo}%
\providecommand \bibfield  [0]{\@secondoftwo}%
\providecommand \translation [1]{[#1]}%
\providecommand \BibitemOpen [0]{}%
\providecommand \bibitemStop [0]{}%
\providecommand \bibitemNoStop [0]{.\EOS\space}%
\providecommand \EOS [0]{\spacefactor3000\relax}%
\providecommand \BibitemShut  [1]{\csname bibitem#1\endcsname}%
\let\auto@bib@innerbib\@empty
\bibitem [{\citenamefont {Keimer}\ and\ \citenamefont
  {Moore}(2017)}]{Keimer2017}%
  \BibitemOpen
  \bibfield  {author} {\bibinfo {author} {\bibfnamefont {B.}~\bibnamefont
  {Keimer}}\ and\ \bibinfo {author} {\bibfnamefont {J.~E.}\ \bibnamefont
  {Moore}},\ }\bibfield  {title} {\bibinfo {title} {The physics of quantum
  materials},\ }\href {https://doi.org/10.1038/nphys4302} {\bibfield  {journal}
  {\bibinfo  {journal} {Nature Physics}\ }\textbf {\bibinfo {volume} {13}},\
  \bibinfo {pages} {1045} (\bibinfo {year} {2017})}\BibitemShut {NoStop}%
\bibitem [{\citenamefont {Fradkin}\ \emph {et~al.}(2015)\citenamefont
  {Fradkin}, \citenamefont {Kivelson},\ and\ \citenamefont
  {Tranquada}}]{Fradkin2015}%
  \BibitemOpen
  \bibfield  {author} {\bibinfo {author} {\bibfnamefont {E.}~\bibnamefont
  {Fradkin}}, \bibinfo {author} {\bibfnamefont {S.~A.}\ \bibnamefont
  {Kivelson}},\ and\ \bibinfo {author} {\bibfnamefont {J.~M.}\ \bibnamefont
  {Tranquada}},\ }\bibfield  {title} {\bibinfo {title} {{Colloquium: Theory of
  intertwined orders in high temperature superconductors}},\ }\href
  {https://doi.org/10.1103/RevModPhys.87.457} {\bibfield  {journal} {\bibinfo
  {journal} {Rev. Mod. Phys.}\ }\textbf {\bibinfo {volume} {87}},\ \bibinfo
  {pages} {457} (\bibinfo {year} {2015})}\BibitemShut {NoStop}%
\bibitem [{\citenamefont {Keimer}\ \emph {et~al.}(2015)\citenamefont {Keimer},
  \citenamefont {Kivelson}, \citenamefont {Norman}, \citenamefont {Uchida},\
  and\ \citenamefont {Zaanen}}]{Keimer2015}%
  \BibitemOpen
  \bibfield  {author} {\bibinfo {author} {\bibfnamefont {B.}~\bibnamefont
  {Keimer}}, \bibinfo {author} {\bibfnamefont {S.~A.}\ \bibnamefont
  {Kivelson}}, \bibinfo {author} {\bibfnamefont {M.~R.}\ \bibnamefont
  {Norman}}, \bibinfo {author} {\bibfnamefont {S.}~\bibnamefont {Uchida}},\
  and\ \bibinfo {author} {\bibfnamefont {J.}~\bibnamefont {Zaanen}},\
  }\bibfield  {title} {\bibinfo {title} {From quantum matter to
  high-temperature superconductivity in copper oxides},\ }\href@noop {}
  {\bibfield  {journal} {\bibinfo  {journal} {Nature}\ }\textbf {\bibinfo
  {volume} {518}},\ \bibinfo {pages} {179} (\bibinfo {year}
  {2015})}\BibitemShut {NoStop}%
\bibitem [{\citenamefont {Comin}\ and\ \citenamefont
  {Damascelli}(2016)}]{Comin2016_CO_review}%
  \BibitemOpen
  \bibfield  {author} {\bibinfo {author} {\bibfnamefont {R.}~\bibnamefont
  {Comin}}\ and\ \bibinfo {author} {\bibfnamefont {A.}~\bibnamefont
  {Damascelli}},\ }\bibfield  {title} {\bibinfo {title} {{Resonant X-ray
  scattering studies of charge order in cuprates}},\ }\href
  {https://doi.org/10.1146/annurev-conmatphys-031115-011401} {\bibfield
  {journal} {\bibinfo  {journal} {Annu. Rev. Conden. Ma. P.}\ }\textbf
  {\bibinfo {volume} {7}},\ \bibinfo {pages} {369} (\bibinfo {year}
  {2016})}\BibitemShut {NoStop}%
\bibitem [{\citenamefont {H\"ucker}\ \emph {et~al.}(2011)\citenamefont
  {H\"ucker}, \citenamefont {v.~Zimmermann}, \citenamefont {Gu}, \citenamefont
  {Xu}, \citenamefont {Wen}, \citenamefont {Xu}, \citenamefont {Kang},
  \citenamefont {Zheludev},\ and\ \citenamefont {Tranquada}}]{Huecker2011}%
  \BibitemOpen
  \bibfield  {author} {\bibinfo {author} {\bibfnamefont {M.}~\bibnamefont
  {H\"ucker}}, \bibinfo {author} {\bibfnamefont {M.}~\bibnamefont
  {v.~Zimmermann}}, \bibinfo {author} {\bibfnamefont {G.~D.}\ \bibnamefont
  {Gu}}, \bibinfo {author} {\bibfnamefont {Z.~J.}\ \bibnamefont {Xu}}, \bibinfo
  {author} {\bibfnamefont {J.~S.}\ \bibnamefont {Wen}}, \bibinfo {author}
  {\bibfnamefont {G.}~\bibnamefont {Xu}}, \bibinfo {author} {\bibfnamefont
  {H.~J.}\ \bibnamefont {Kang}}, \bibinfo {author} {\bibfnamefont
  {A.}~\bibnamefont {Zheludev}},\ and\ \bibinfo {author} {\bibfnamefont
  {J.~M.}\ \bibnamefont {Tranquada}},\ }\bibfield  {title} {\bibinfo {title}
  {Stripe order in superconducting {La$_{2-x}$Ba$_x$CuO$_4$} ($0.095 \leq x
  \leq 0.155$)},\ }\href {https://doi.org/10.1103/PhysRevB.83.104506}
  {\bibfield  {journal} {\bibinfo  {journal} {Phys. Rev. B}\ }\textbf {\bibinfo
  {volume} {83}},\ \bibinfo {pages} {104506} (\bibinfo {year}
  {2011})}\BibitemShut {NoStop}%
\bibitem [{\citenamefont {Ghiringhelli}\ \emph {et~al.}(2012)\citenamefont
  {Ghiringhelli}, \citenamefont {Le~Tacon}, \citenamefont {Minola},
  \citenamefont {Blanco-Canosa}, \citenamefont {Mazzoli}, \citenamefont
  {Brookes}, \citenamefont {De~Luca}, \citenamefont {Frano}, \citenamefont
  {Hawthorn}, \citenamefont {He}, \citenamefont {Loew}, \citenamefont {Sala},
  \citenamefont {Peets}, \citenamefont {Salluzzo}, \citenamefont {Schierle},
  \citenamefont {Sutarto}, \citenamefont {Sawatzky}, \citenamefont {Weschke},
  \citenamefont {Keimer},\ and\ \citenamefont {Braicovich}}]{Ghiringhelli2012}%
  \BibitemOpen
  \bibfield  {author} {\bibinfo {author} {\bibfnamefont {G.}~\bibnamefont
  {Ghiringhelli}}, \bibinfo {author} {\bibfnamefont {M.}~\bibnamefont
  {Le~Tacon}}, \bibinfo {author} {\bibfnamefont {M.}~\bibnamefont {Minola}},
  \bibinfo {author} {\bibfnamefont {S.}~\bibnamefont {Blanco-Canosa}}, \bibinfo
  {author} {\bibfnamefont {C.}~\bibnamefont {Mazzoli}}, \bibinfo {author}
  {\bibfnamefont {N.~B.}\ \bibnamefont {Brookes}}, \bibinfo {author}
  {\bibfnamefont {G.~M.}\ \bibnamefont {De~Luca}}, \bibinfo {author}
  {\bibfnamefont {A.}~\bibnamefont {Frano}}, \bibinfo {author} {\bibfnamefont
  {D.~G.}\ \bibnamefont {Hawthorn}}, \bibinfo {author} {\bibfnamefont
  {F.}~\bibnamefont {He}}, \bibinfo {author} {\bibfnamefont {T.}~\bibnamefont
  {Loew}}, \bibinfo {author} {\bibfnamefont {M.~M.}\ \bibnamefont {Sala}},
  \bibinfo {author} {\bibfnamefont {D.~C.}\ \bibnamefont {Peets}}, \bibinfo
  {author} {\bibfnamefont {M.}~\bibnamefont {Salluzzo}}, \bibinfo {author}
  {\bibfnamefont {E.}~\bibnamefont {Schierle}}, \bibinfo {author}
  {\bibfnamefont {R.}~\bibnamefont {Sutarto}}, \bibinfo {author} {\bibfnamefont
  {G.~A.}\ \bibnamefont {Sawatzky}}, \bibinfo {author} {\bibfnamefont
  {E.}~\bibnamefont {Weschke}}, \bibinfo {author} {\bibfnamefont
  {B.}~\bibnamefont {Keimer}},\ and\ \bibinfo {author} {\bibfnamefont
  {L.}~\bibnamefont {Braicovich}},\ }\bibfield  {title} {\bibinfo {title}
  {{Long-range incommensurate charge fluctuations in
  \\(Y,Nd)Ba$_2$Cu$_3$O$_{6+x}$}},\ }\href
  {https://doi.org/10.1126/science.1223532} {\bibfield  {journal} {\bibinfo
  {journal} {Science}\ }\textbf {\bibinfo {volume} {337}},\ \bibinfo {pages}
  {821} (\bibinfo {year} {2012})}\BibitemShut {NoStop}%
\bibitem [{\citenamefont {Chang}\ \emph {et~al.}(2012)\citenamefont {Chang},
  \citenamefont {Blackburn}, \citenamefont {Holmes}, \citenamefont
  {Christensen}, \citenamefont {Larsen}, \citenamefont {Mesot}, \citenamefont
  {Liang}, \citenamefont {Bonn}, \citenamefont {Hardy}, \citenamefont
  {Watenphul}, \citenamefont {Zimmermann}, \citenamefont {Forgan},\ and\
  \citenamefont {Hayden}}]{Chang2012}%
  \BibitemOpen
  \bibfield  {author} {\bibinfo {author} {\bibfnamefont {J.}~\bibnamefont
  {Chang}}, \bibinfo {author} {\bibfnamefont {E.}~\bibnamefont {Blackburn}},
  \bibinfo {author} {\bibfnamefont {A.~T.}\ \bibnamefont {Holmes}}, \bibinfo
  {author} {\bibfnamefont {N.~B.}\ \bibnamefont {Christensen}}, \bibinfo
  {author} {\bibfnamefont {J.}~\bibnamefont {Larsen}}, \bibinfo {author}
  {\bibfnamefont {J.}~\bibnamefont {Mesot}}, \bibinfo {author} {\bibfnamefont
  {R.}~\bibnamefont {Liang}}, \bibinfo {author} {\bibfnamefont {D.~A.}\
  \bibnamefont {Bonn}}, \bibinfo {author} {\bibfnamefont {W.~N.}\ \bibnamefont
  {Hardy}}, \bibinfo {author} {\bibfnamefont {A.}~\bibnamefont {Watenphul}},
  \bibinfo {author} {\bibfnamefont {M.~v.}\ \bibnamefont {Zimmermann}},
  \bibinfo {author} {\bibfnamefont {E.~M.}\ \bibnamefont {Forgan}},\ and\
  \bibinfo {author} {\bibfnamefont {S.~M.}\ \bibnamefont {Hayden}},\ }\bibfield
   {title} {\bibinfo {title} {{Direct observation of competition between
  superconductivity and charge density wave order in
  YBa$_2$Cu$_3$O$_{6.67}$}},\ }\href@noop {} {\bibfield  {journal} {\bibinfo
  {journal} {Nat. Phys.}\ }\textbf {\bibinfo {volume} {8}},\ \bibinfo {pages}
  {871} (\bibinfo {year} {2012})}\BibitemShut {NoStop}%
\bibitem [{\citenamefont {Chang}\ \emph {et~al.}(2016)\citenamefont {Chang},
  \citenamefont {Blackburn}, \citenamefont {Ivashko}, \citenamefont {Holmes},
  \citenamefont {Christensen}, \citenamefont {H{\"u}cker}, \citenamefont
  {Liang}, \citenamefont {Bonn}, \citenamefont {Hardy}, \citenamefont
  {R{\"u}tt}, \citenamefont {Zimmermann}, \citenamefont {Forgan},\ and\
  \citenamefont {Hayden}}]{Chang2016}%
  \BibitemOpen
  \bibfield  {author} {\bibinfo {author} {\bibfnamefont {J.}~\bibnamefont
  {Chang}}, \bibinfo {author} {\bibfnamefont {E.}~\bibnamefont {Blackburn}},
  \bibinfo {author} {\bibfnamefont {O.}~\bibnamefont {Ivashko}}, \bibinfo
  {author} {\bibfnamefont {A.~T.}\ \bibnamefont {Holmes}}, \bibinfo {author}
  {\bibfnamefont {N.~B.}\ \bibnamefont {Christensen}}, \bibinfo {author}
  {\bibfnamefont {M.}~\bibnamefont {H{\"u}cker}}, \bibinfo {author}
  {\bibfnamefont {R.}~\bibnamefont {Liang}}, \bibinfo {author} {\bibfnamefont
  {D.~A.}\ \bibnamefont {Bonn}}, \bibinfo {author} {\bibfnamefont {W.~N.}\
  \bibnamefont {Hardy}}, \bibinfo {author} {\bibfnamefont {U.}~\bibnamefont
  {R{\"u}tt}}, \bibinfo {author} {\bibfnamefont {M.~v.}\ \bibnamefont
  {Zimmermann}}, \bibinfo {author} {\bibfnamefont {E.~M.}\ \bibnamefont
  {Forgan}},\ and\ \bibinfo {author} {\bibfnamefont {S.~M.}\ \bibnamefont
  {Hayden}},\ }\bibfield  {title} {\bibinfo {title} {{Magnetic field controlled
  charge density wave coupling in underdoped YBa$_2$Cu$_3$O$_{6+x}$}},\
  }\href@noop {} {\bibfield  {journal} {\bibinfo  {journal} {Nat. Commun.}\
  }\textbf {\bibinfo {volume} {7}},\ \bibinfo {pages} {11494} (\bibinfo {year}
  {2016})}\BibitemShut {NoStop}%
\bibitem [{\citenamefont {Jang}\ \emph {et~al.}(2016)\citenamefont {Jang},
  \citenamefont {Lee}, \citenamefont {Nojiri}, \citenamefont {Matsuzawa},
  \citenamefont {Yasumura}, \citenamefont {Nie}, \citenamefont {Maharaj},
  \citenamefont {Gerber}, \citenamefont {Liu}, \citenamefont {Mehta},
  \citenamefont {Bonn}, \citenamefont {Liang}, \citenamefont {Hardy},
  \citenamefont {Burns}, \citenamefont {Islam}, \citenamefont {Song},
  \citenamefont {Hastings}, \citenamefont {Devereaux}, \citenamefont {Shen},
  \citenamefont {Kivelson}, \citenamefont {Kao}, \citenamefont {Zhu},\ and\
  \citenamefont {Lee}}]{Jang2016}%
  \BibitemOpen
  \bibfield  {author} {\bibinfo {author} {\bibfnamefont {H.}~\bibnamefont
  {Jang}}, \bibinfo {author} {\bibfnamefont {W.-S.}\ \bibnamefont {Lee}},
  \bibinfo {author} {\bibfnamefont {H.}~\bibnamefont {Nojiri}}, \bibinfo
  {author} {\bibfnamefont {S.}~\bibnamefont {Matsuzawa}}, \bibinfo {author}
  {\bibfnamefont {H.}~\bibnamefont {Yasumura}}, \bibinfo {author}
  {\bibfnamefont {L.}~\bibnamefont {Nie}}, \bibinfo {author} {\bibfnamefont
  {A.~V.}\ \bibnamefont {Maharaj}}, \bibinfo {author} {\bibfnamefont
  {S.}~\bibnamefont {Gerber}}, \bibinfo {author} {\bibfnamefont {Y.-J.}\
  \bibnamefont {Liu}}, \bibinfo {author} {\bibfnamefont {A.}~\bibnamefont
  {Mehta}}, \bibinfo {author} {\bibfnamefont {D.~A.}\ \bibnamefont {Bonn}},
  \bibinfo {author} {\bibfnamefont {R.}~\bibnamefont {Liang}}, \bibinfo
  {author} {\bibfnamefont {W.~N.}\ \bibnamefont {Hardy}}, \bibinfo {author}
  {\bibfnamefont {C.~A.}\ \bibnamefont {Burns}}, \bibinfo {author}
  {\bibfnamefont {Z.}~\bibnamefont {Islam}}, \bibinfo {author} {\bibfnamefont
  {S.}~\bibnamefont {Song}}, \bibinfo {author} {\bibfnamefont {J.}~\bibnamefont
  {Hastings}}, \bibinfo {author} {\bibfnamefont {T.~P.}\ \bibnamefont
  {Devereaux}}, \bibinfo {author} {\bibfnamefont {Z.-X.}\ \bibnamefont {Shen}},
  \bibinfo {author} {\bibfnamefont {S.~A.}\ \bibnamefont {Kivelson}}, \bibinfo
  {author} {\bibfnamefont {C.-C.}\ \bibnamefont {Kao}}, \bibinfo {author}
  {\bibfnamefont {D.}~\bibnamefont {Zhu}},\ and\ \bibinfo {author}
  {\bibfnamefont {J.-S.}\ \bibnamefont {Lee}},\ }\bibfield  {title} {\bibinfo
  {title} {{Ideal charge-density-wave order in the high-field state of
  superconducting YBCO}},\ }\href {https://doi.org/10.1073/pnas.1612849113}
  {\bibfield  {journal} {\bibinfo  {journal} {Proc. Natl. Acad. Sci. USA}\
  }\textbf {\bibinfo {volume} {113}},\ \bibinfo {pages} {14645} (\bibinfo
  {year} {2016})}\BibitemShut {NoStop}%
\bibitem [{\citenamefont {Kim}\ \emph {et~al.}(2018)\citenamefont {Kim},
  \citenamefont {Souliou}, \citenamefont {Barber}, \citenamefont {Lefran{\c
  c}ois}, \citenamefont {Minola}, \citenamefont {Tortora}, \citenamefont
  {Heid}, \citenamefont {Nandi}, \citenamefont {Borzi}, \citenamefont
  {Garbarino}, \citenamefont {Bosak}, \citenamefont {Porras}, \citenamefont
  {Loew}, \citenamefont {K{\"o}nig}, \citenamefont {Moll}, \citenamefont
  {Mackenzie}, \citenamefont {Keimer}, \citenamefont {Hicks},\ and\
  \citenamefont {Le~Tacon}}]{Kim2018}%
  \BibitemOpen
  \bibfield  {author} {\bibinfo {author} {\bibfnamefont {H.-H.}\ \bibnamefont
  {Kim}}, \bibinfo {author} {\bibfnamefont {S.~M.}\ \bibnamefont {Souliou}},
  \bibinfo {author} {\bibfnamefont {M.~E.}\ \bibnamefont {Barber}}, \bibinfo
  {author} {\bibfnamefont {E.}~\bibnamefont {Lefran{\c c}ois}}, \bibinfo
  {author} {\bibfnamefont {M.}~\bibnamefont {Minola}}, \bibinfo {author}
  {\bibfnamefont {M.}~\bibnamefont {Tortora}}, \bibinfo {author} {\bibfnamefont
  {R.}~\bibnamefont {Heid}}, \bibinfo {author} {\bibfnamefont {N.}~\bibnamefont
  {Nandi}}, \bibinfo {author} {\bibfnamefont {R.~A.}\ \bibnamefont {Borzi}},
  \bibinfo {author} {\bibfnamefont {G.}~\bibnamefont {Garbarino}}, \bibinfo
  {author} {\bibfnamefont {A.}~\bibnamefont {Bosak}}, \bibinfo {author}
  {\bibfnamefont {J.}~\bibnamefont {Porras}}, \bibinfo {author} {\bibfnamefont
  {T.}~\bibnamefont {Loew}}, \bibinfo {author} {\bibfnamefont {M.}~\bibnamefont
  {K{\"o}nig}}, \bibinfo {author} {\bibfnamefont {P.~M.}\ \bibnamefont {Moll}},
  \bibinfo {author} {\bibfnamefont {A.~P.}\ \bibnamefont {Mackenzie}}, \bibinfo
  {author} {\bibfnamefont {B.}~\bibnamefont {Keimer}}, \bibinfo {author}
  {\bibfnamefont {C.~W.}\ \bibnamefont {Hicks}},\ and\ \bibinfo {author}
  {\bibfnamefont {M.}~\bibnamefont {Le~Tacon}},\ }\bibfield  {title} {\bibinfo
  {title} {Uniaxial pressure control of competing orders in a high-temperature
  superconductor},\ }\href {https://doi.org/10.1126/science.aat4708} {\bibfield
   {journal} {\bibinfo  {journal} {Science}\ }\textbf {\bibinfo {volume}
  {362}},\ \bibinfo {pages} {1040} (\bibinfo {year} {2018})}\BibitemShut
  {NoStop}%
\bibitem [{\citenamefont {Tabis}\ \emph {et~al.}(2017)\citenamefont {Tabis},
  \citenamefont {Yu}, \citenamefont {Bialo}, \citenamefont {Bluschke},
  \citenamefont {Kolodziej}, \citenamefont {Kozlowski}, \citenamefont
  {Blackburn}, \citenamefont {Sen}, \citenamefont {Forgan}, \citenamefont
  {Zimmermann}, \citenamefont {Tang}, \citenamefont {Weschke}, \citenamefont
  {Vignolle}, \citenamefont {Hepting}, \citenamefont {Gretarsson},
  \citenamefont {Sutarto}, \citenamefont {He}, \citenamefont {Le~Tacon},
  \citenamefont {Bari\ifmmode \check{s}\else \v{s}\fi{}i\ifmmode~\acute{c}\else
  \'{c}\fi{}}, \citenamefont {Yu},\ and\ \citenamefont {Greven}}]{Tabis2017}%
  \BibitemOpen
  \bibfield  {author} {\bibinfo {author} {\bibfnamefont {W.}~\bibnamefont
  {Tabis}}, \bibinfo {author} {\bibfnamefont {B.}~\bibnamefont {Yu}}, \bibinfo
  {author} {\bibfnamefont {I.}~\bibnamefont {Bialo}}, \bibinfo {author}
  {\bibfnamefont {M.}~\bibnamefont {Bluschke}}, \bibinfo {author}
  {\bibfnamefont {T.}~\bibnamefont {Kolodziej}}, \bibinfo {author}
  {\bibfnamefont {A.}~\bibnamefont {Kozlowski}}, \bibinfo {author}
  {\bibfnamefont {E.}~\bibnamefont {Blackburn}}, \bibinfo {author}
  {\bibfnamefont {K.}~\bibnamefont {Sen}}, \bibinfo {author} {\bibfnamefont
  {E.~M.}\ \bibnamefont {Forgan}}, \bibinfo {author} {\bibfnamefont {M.~v.}\
  \bibnamefont {Zimmermann}}, \bibinfo {author} {\bibfnamefont
  {Y.}~\bibnamefont {Tang}}, \bibinfo {author} {\bibfnamefont {E.}~\bibnamefont
  {Weschke}}, \bibinfo {author} {\bibfnamefont {B.}~\bibnamefont {Vignolle}},
  \bibinfo {author} {\bibfnamefont {M.}~\bibnamefont {Hepting}}, \bibinfo
  {author} {\bibfnamefont {H.}~\bibnamefont {Gretarsson}}, \bibinfo {author}
  {\bibfnamefont {R.}~\bibnamefont {Sutarto}}, \bibinfo {author} {\bibfnamefont
  {F.}~\bibnamefont {He}}, \bibinfo {author} {\bibfnamefont {M.}~\bibnamefont
  {Le~Tacon}}, \bibinfo {author} {\bibfnamefont {N.}~\bibnamefont {Bari\ifmmode
  \check{s}\else \v{s}\fi{}i\ifmmode~\acute{c}\else \'{c}\fi{}}}, \bibinfo
  {author} {\bibfnamefont {G.}~\bibnamefont {Yu}},\ and\ \bibinfo {author}
  {\bibfnamefont {M.}~\bibnamefont {Greven}},\ }\bibfield  {title} {\bibinfo
  {title} {{Synchrotron x-ray scattering study of charge-density-wave order in
  ${\mathrm{HgBa}}_{2}{\mathrm{CuO}}_{4+\ensuremath{\delta}}$}},\ }\href
  {https://doi.org/10.1103/PhysRevB.96.134510} {\bibfield  {journal} {\bibinfo
  {journal} {Phys. Rev. B}\ }\textbf {\bibinfo {volume} {96}},\ \bibinfo
  {pages} {134510} (\bibinfo {year} {2017})}\BibitemShut {NoStop}%
\bibitem [{\citenamefont {Bluschke}\ \emph {et~al.}(2019)\citenamefont
  {Bluschke}, \citenamefont {Yaari}, \citenamefont {Schierle}, \citenamefont
  {Bazalitsky}, \citenamefont {Werner}, \citenamefont {Weschke},\ and\
  \citenamefont {Keren}}]{Bluschke2019}%
  \BibitemOpen
  \bibfield  {author} {\bibinfo {author} {\bibfnamefont {M.}~\bibnamefont
  {Bluschke}}, \bibinfo {author} {\bibfnamefont {M.}~\bibnamefont {Yaari}},
  \bibinfo {author} {\bibfnamefont {E.}~\bibnamefont {Schierle}}, \bibinfo
  {author} {\bibfnamefont {G.}~\bibnamefont {Bazalitsky}}, \bibinfo {author}
  {\bibfnamefont {J.}~\bibnamefont {Werner}}, \bibinfo {author} {\bibfnamefont
  {E.}~\bibnamefont {Weschke}},\ and\ \bibinfo {author} {\bibfnamefont
  {A.}~\bibnamefont {Keren}},\ }\bibfield  {title} {\bibinfo {title} {Adiabatic
  variation of the charge density wave phase diagram in the 123 cuprate
  {(Ca$_x$La$_{1-x}$)(Ba$_{1.75-x}$La$_{0.25+x}$)Cu$_3$O$_y$}},\ }\href
  {https://doi.org/10.1103/PhysRevB.100.035129} {\bibfield  {journal} {\bibinfo
   {journal} {Phys. Rev. B}\ }\textbf {\bibinfo {volume} {100}},\ \bibinfo
  {pages} {035129} (\bibinfo {year} {2019})}\BibitemShut {NoStop}%
\bibitem [{\citenamefont {Wandel}\ \emph {et~al.}(2022)\citenamefont {Wandel},
  \citenamefont {Boschini}, \citenamefont {da~Silva~Neto}, \citenamefont
  {Shen}, \citenamefont {Na}, \citenamefont {Zohar}, \citenamefont {Wang},
  \citenamefont {Welch}, \citenamefont {Seaberg}, \citenamefont {Koralek},
  \citenamefont {Dakovski}, \citenamefont {Hettel}, \citenamefont {Lin},
  \citenamefont {Moeller}, \citenamefont {Schlotter}, \citenamefont {Reid},
  \citenamefont {Minitti}, \citenamefont {Boyle}, \citenamefont {He},
  \citenamefont {Sutarto}, \citenamefont {Liang}, \citenamefont {Bonn},
  \citenamefont {Hardy}, \citenamefont {Kaindl}, \citenamefont {Hawthorn},
  \citenamefont {Lee}, \citenamefont {Kemper}, \citenamefont {Damascelli},
  \citenamefont {Giannetti}, \citenamefont {Turner},\ and\ \citenamefont
  {Coslovich}}]{Wandel2022}%
  \BibitemOpen
  \bibfield  {author} {\bibinfo {author} {\bibfnamefont {S.}~\bibnamefont
  {Wandel}}, \bibinfo {author} {\bibfnamefont {F.}~\bibnamefont {Boschini}},
  \bibinfo {author} {\bibfnamefont {E.~H.}\ \bibnamefont {da~Silva~Neto}},
  \bibinfo {author} {\bibfnamefont {L.}~\bibnamefont {Shen}}, \bibinfo {author}
  {\bibfnamefont {M.~X.}\ \bibnamefont {Na}}, \bibinfo {author} {\bibfnamefont
  {S.}~\bibnamefont {Zohar}}, \bibinfo {author} {\bibfnamefont
  {Y.}~\bibnamefont {Wang}}, \bibinfo {author} {\bibfnamefont {S.~B.}\
  \bibnamefont {Welch}}, \bibinfo {author} {\bibfnamefont {M.~H.}\ \bibnamefont
  {Seaberg}}, \bibinfo {author} {\bibfnamefont {J.~D.}\ \bibnamefont
  {Koralek}}, \bibinfo {author} {\bibfnamefont {G.~L.}\ \bibnamefont
  {Dakovski}}, \bibinfo {author} {\bibfnamefont {W.}~\bibnamefont {Hettel}},
  \bibinfo {author} {\bibfnamefont {M.-F.}\ \bibnamefont {Lin}}, \bibinfo
  {author} {\bibfnamefont {S.~P.}\ \bibnamefont {Moeller}}, \bibinfo {author}
  {\bibfnamefont {W.~F.}\ \bibnamefont {Schlotter}}, \bibinfo {author}
  {\bibfnamefont {A.~H.}\ \bibnamefont {Reid}}, \bibinfo {author}
  {\bibfnamefont {M.~P.}\ \bibnamefont {Minitti}}, \bibinfo {author}
  {\bibfnamefont {T.}~\bibnamefont {Boyle}}, \bibinfo {author} {\bibfnamefont
  {F.}~\bibnamefont {He}}, \bibinfo {author} {\bibfnamefont {R.}~\bibnamefont
  {Sutarto}}, \bibinfo {author} {\bibfnamefont {R.}~\bibnamefont {Liang}},
  \bibinfo {author} {\bibfnamefont {D.}~\bibnamefont {Bonn}}, \bibinfo {author}
  {\bibfnamefont {W.}~\bibnamefont {Hardy}}, \bibinfo {author} {\bibfnamefont
  {R.~A.}\ \bibnamefont {Kaindl}}, \bibinfo {author} {\bibfnamefont {D.~G.}\
  \bibnamefont {Hawthorn}}, \bibinfo {author} {\bibfnamefont {J.-S.}\
  \bibnamefont {Lee}}, \bibinfo {author} {\bibfnamefont {A.~F.}\ \bibnamefont
  {Kemper}}, \bibinfo {author} {\bibfnamefont {A.}~\bibnamefont {Damascelli}},
  \bibinfo {author} {\bibfnamefont {C.}~\bibnamefont {Giannetti}}, \bibinfo
  {author} {\bibfnamefont {J.~J.}\ \bibnamefont {Turner}},\ and\ \bibinfo
  {author} {\bibfnamefont {G.}~\bibnamefont {Coslovich}},\ }\bibfield  {title}
  {\bibinfo {title} {Enhanced charge density wave coherence in a
  light-quenched, high-temperature superconductor},\ }\href
  {https://doi.org/10.1126/science.abd7213} {\bibfield  {journal} {\bibinfo
  {journal} {Science}\ }\textbf {\bibinfo {volume} {376}},\ \bibinfo {pages}
  {860} (\bibinfo {year} {2022})}\BibitemShut {NoStop}%
\bibitem [{\citenamefont {Jang}\ \emph {et~al.}(2022)\citenamefont {Jang},
  \citenamefont {Song}, \citenamefont {Kihara}, \citenamefont {Liu},
  \citenamefont {Lee}, \citenamefont {Park}, \citenamefont {Kim}, \citenamefont
  {Kim}, \citenamefont {Coslovich}, \citenamefont {Nakata}, \citenamefont
  {Kubota}, \citenamefont {Inoue}, \citenamefont {Tamasaku}, \citenamefont
  {Yabashi}, \citenamefont {Lee}, \citenamefont {Song}, \citenamefont {Nojiri},
  \citenamefont {Keimer}, \citenamefont {Kao},\ and\ \citenamefont
  {Lee}}]{Jang2022}%
  \BibitemOpen
  \bibfield  {author} {\bibinfo {author} {\bibfnamefont {H.}~\bibnamefont
  {Jang}}, \bibinfo {author} {\bibfnamefont {S.}~\bibnamefont {Song}}, \bibinfo
  {author} {\bibfnamefont {T.}~\bibnamefont {Kihara}}, \bibinfo {author}
  {\bibfnamefont {Y.}~\bibnamefont {Liu}}, \bibinfo {author} {\bibfnamefont
  {S.-J.}\ \bibnamefont {Lee}}, \bibinfo {author} {\bibfnamefont {S.-Y.}\
  \bibnamefont {Park}}, \bibinfo {author} {\bibfnamefont {M.}~\bibnamefont
  {Kim}}, \bibinfo {author} {\bibfnamefont {H.-D.}\ \bibnamefont {Kim}},
  \bibinfo {author} {\bibfnamefont {G.}~\bibnamefont {Coslovich}}, \bibinfo
  {author} {\bibfnamefont {S.}~\bibnamefont {Nakata}}, \bibinfo {author}
  {\bibfnamefont {Y.}~\bibnamefont {Kubota}}, \bibinfo {author} {\bibfnamefont
  {I.}~\bibnamefont {Inoue}}, \bibinfo {author} {\bibfnamefont
  {K.}~\bibnamefont {Tamasaku}}, \bibinfo {author} {\bibfnamefont
  {M.}~\bibnamefont {Yabashi}}, \bibinfo {author} {\bibfnamefont
  {H.}~\bibnamefont {Lee}}, \bibinfo {author} {\bibfnamefont {C.}~\bibnamefont
  {Song}}, \bibinfo {author} {\bibfnamefont {H.}~\bibnamefont {Nojiri}},
  \bibinfo {author} {\bibfnamefont {B.}~\bibnamefont {Keimer}}, \bibinfo
  {author} {\bibfnamefont {C.-C.}\ \bibnamefont {Kao}},\ and\ \bibinfo {author}
  {\bibfnamefont {J.-S.}\ \bibnamefont {Lee}},\ }\bibfield  {title} {\bibinfo
  {title} {Characterization of photoinduced normal state through charge density
  wave in superconducting {{YBa$_{2}$Cu$_{3}$O$_{6.67}$}}},\ }\href
  {https://doi.org/10.1126/sciadv.abk0832} {\bibfield  {journal} {\bibinfo
  {journal} {Science Advances}\ }\textbf {\bibinfo {volume} {8}},\ \bibinfo
  {pages} {eabk0832} (\bibinfo {year} {2022})}\BibitemShut {NoStop}%
\bibitem [{\citenamefont {Ando}\ \emph {et~al.}(1999)\citenamefont {Ando},
  \citenamefont {Lavrov},\ and\ \citenamefont {Segawa}}]{Ando1999}%
  \BibitemOpen
  \bibfield  {author} {\bibinfo {author} {\bibfnamefont {Y.}~\bibnamefont
  {Ando}}, \bibinfo {author} {\bibfnamefont {A.~N.}\ \bibnamefont {Lavrov}},\
  and\ \bibinfo {author} {\bibfnamefont {K.}~\bibnamefont {Segawa}},\
  }\bibfield  {title} {\bibinfo {title} {Magnetoresistance anomalies in
  antiferromagnetic
  $\mathrm{YBa}{}_{2}\mathrm{Cu}{}_{3}\mathrm{O}{}_{6+\mathit{x}}$:
  Fingerprints of charged stripes},\ }\href
  {https://doi.org/10.1103/PhysRevLett.83.2813} {\bibfield  {journal} {\bibinfo
   {journal} {Phys. Rev. Lett.}\ }\textbf {\bibinfo {volume} {83}},\ \bibinfo
  {pages} {2813} (\bibinfo {year} {1999})}\BibitemShut {NoStop}%
\bibitem [{\citenamefont {Noda}\ \emph {et~al.}(1998)\citenamefont {Noda},
  \citenamefont {Eisaki},\ and\ \citenamefont {Uchida}}]{Noda1999}%
  \BibitemOpen
  \bibfield  {author} {\bibinfo {author} {\bibfnamefont {T.}~\bibnamefont
  {Noda}}, \bibinfo {author} {\bibfnamefont {H.}~\bibnamefont {Eisaki}},\ and\
  \bibinfo {author} {\bibfnamefont {S.}~\bibnamefont {Uchida}},\ }\bibfield
  {title} {\bibinfo {title} {Evidence for one-dimensional charge transport in
  {La$_{2-x-y}$Nd$_{y}$Sr$_x$CuO$_4$}.},\ }\href@noop {} {\bibfield  {journal}
  {\bibinfo  {journal} {Science}\ }\textbf {\bibinfo {volume} {286}},\ \bibinfo
  {pages} {265} (\bibinfo {year} {1998})}\BibitemShut {NoStop}%
\bibitem [{\citenamefont {Mook}\ \emph {et~al.}(2000)\citenamefont {Mook},
  \citenamefont {Dai}, \citenamefont {Dogan},\ and\ \citenamefont
  {Hunt}}]{Mook2000}%
  \BibitemOpen
  \bibfield  {author} {\bibinfo {author} {\bibfnamefont {H.~A.}\ \bibnamefont
  {Mook}}, \bibinfo {author} {\bibfnamefont {P.}~\bibnamefont {Dai}}, \bibinfo
  {author} {\bibfnamefont {F.}~\bibnamefont {Dogan}},\ and\ \bibinfo {author}
  {\bibfnamefont {R.~D.}\ \bibnamefont {Hunt}},\ }\bibfield  {title} {\bibinfo
  {title} {One-dimensional nature of the magnetic fluctuations in
  {YBa$_2$Cu$_3$O$_{6.6}$}},\ }\href {https://doi.org/10.1038/35008005}
  {\bibfield  {journal} {\bibinfo  {journal} {Nature}\ }\textbf {\bibinfo
  {volume} {404}},\ \bibinfo {pages} {729} (\bibinfo {year}
  {2000})}\BibitemShut {NoStop}%
\bibitem [{\citenamefont {Ando}\ \emph {et~al.}(2002)\citenamefont {Ando},
  \citenamefont {Segawa}, \citenamefont {Komiya},\ and\ \citenamefont
  {Lavrov}}]{Ando2002}%
  \BibitemOpen
  \bibfield  {author} {\bibinfo {author} {\bibfnamefont {Y.}~\bibnamefont
  {Ando}}, \bibinfo {author} {\bibfnamefont {K.}~\bibnamefont {Segawa}},
  \bibinfo {author} {\bibfnamefont {S.}~\bibnamefont {Komiya}},\ and\ \bibinfo
  {author} {\bibfnamefont {A.~N.}\ \bibnamefont {Lavrov}},\ }\bibfield  {title}
  {\bibinfo {title} {Electrical resistivity anisotropy from self-organized one
  dimensionality in high-temperature superconductors},\ }\href
  {https://doi.org/10.1103/PhysRevLett.88.137005} {\bibfield  {journal}
  {\bibinfo  {journal} {Phys. Rev. Lett.}\ }\textbf {\bibinfo {volume} {88}},\
  \bibinfo {pages} {137005} (\bibinfo {year} {2002})}\BibitemShut {NoStop}%
\bibitem [{\citenamefont {Stock}\ \emph {et~al.}(2004)\citenamefont {Stock},
  \citenamefont {Buyers}, \citenamefont {Liang}, \citenamefont {Peets},
  \citenamefont {Tun}, \citenamefont {Bonn}, \citenamefont {Hardy},\ and\
  \citenamefont {Birgeneau}}]{Stock2004}%
  \BibitemOpen
  \bibfield  {author} {\bibinfo {author} {\bibfnamefont {C.}~\bibnamefont
  {Stock}}, \bibinfo {author} {\bibfnamefont {W.~J.~L.}\ \bibnamefont
  {Buyers}}, \bibinfo {author} {\bibfnamefont {R.}~\bibnamefont {Liang}},
  \bibinfo {author} {\bibfnamefont {D.}~\bibnamefont {Peets}}, \bibinfo
  {author} {\bibfnamefont {Z.}~\bibnamefont {Tun}}, \bibinfo {author}
  {\bibfnamefont {D.}~\bibnamefont {Bonn}}, \bibinfo {author} {\bibfnamefont
  {W.~N.}\ \bibnamefont {Hardy}},\ and\ \bibinfo {author} {\bibfnamefont
  {R.~J.}\ \bibnamefont {Birgeneau}},\ }\bibfield  {title} {\bibinfo {title}
  {Dynamic stripes and resonance in the superconducting and normal phases of
  {${\mathrm{YBa}}_{2}{\mathrm{Cu}}_{3}{\mathrm{O}}_{6.5}$} ortho-{II}
  superconductor},\ }\href {https://doi.org/10.1103/PhysRevB.69.014502}
  {\bibfield  {journal} {\bibinfo  {journal} {Phys. Rev. B}\ }\textbf {\bibinfo
  {volume} {69}},\ \bibinfo {pages} {014502} (\bibinfo {year}
  {2004})}\BibitemShut {NoStop}%
\bibitem [{\citenamefont {Hinkov}\ \emph {et~al.}(2007)\citenamefont {Hinkov},
  \citenamefont {Bourges}, \citenamefont {Pailh{\`e}s}, \citenamefont {Sidis},
  \citenamefont {Ivanov}, \citenamefont {Frost}, \citenamefont {Perring},
  \citenamefont {Lin}, \citenamefont {Chen},\ and\ \citenamefont
  {Keimer}}]{Hinkov2007}%
  \BibitemOpen
  \bibfield  {author} {\bibinfo {author} {\bibfnamefont {V.}~\bibnamefont
  {Hinkov}}, \bibinfo {author} {\bibfnamefont {P.}~\bibnamefont {Bourges}},
  \bibinfo {author} {\bibfnamefont {S.}~\bibnamefont {Pailh{\`e}s}}, \bibinfo
  {author} {\bibfnamefont {Y.}~\bibnamefont {Sidis}}, \bibinfo {author}
  {\bibfnamefont {A.}~\bibnamefont {Ivanov}}, \bibinfo {author} {\bibfnamefont
  {C.~D.}\ \bibnamefont {Frost}}, \bibinfo {author} {\bibfnamefont {T.~G.}\
  \bibnamefont {Perring}}, \bibinfo {author} {\bibfnamefont {C.~T.}\
  \bibnamefont {Lin}}, \bibinfo {author} {\bibfnamefont {D.~P.}\ \bibnamefont
  {Chen}},\ and\ \bibinfo {author} {\bibfnamefont {B.}~\bibnamefont {Keimer}},\
  }\bibfield  {title} {\bibinfo {title} {Spin dynamics in the pseudogap state
  of a high-temperature superconductor},\ }\href@noop {} {\bibfield  {journal}
  {\bibinfo  {journal} {Nature Physics}\ }\textbf {\bibinfo {volume} {3}},\
  \bibinfo {pages} {780} (\bibinfo {year} {2007})}\BibitemShut {NoStop}%
\bibitem [{\citenamefont {Hinkov}\ \emph {et~al.}(2008)\citenamefont {Hinkov},
  \citenamefont {Haug}, \citenamefont {Fauqu{\'e}}, \citenamefont {Bourges},
  \citenamefont {Sidis}, \citenamefont {Ivanov}, \citenamefont {Bernhard},
  \citenamefont {Lin},\ and\ \citenamefont {Keimer}}]{Hinkov2008}%
  \BibitemOpen
  \bibfield  {author} {\bibinfo {author} {\bibfnamefont {V.}~\bibnamefont
  {Hinkov}}, \bibinfo {author} {\bibfnamefont {D.}~\bibnamefont {Haug}},
  \bibinfo {author} {\bibfnamefont {B.}~\bibnamefont {Fauqu{\'e}}}, \bibinfo
  {author} {\bibfnamefont {P.}~\bibnamefont {Bourges}}, \bibinfo {author}
  {\bibfnamefont {Y.}~\bibnamefont {Sidis}}, \bibinfo {author} {\bibfnamefont
  {A.}~\bibnamefont {Ivanov}}, \bibinfo {author} {\bibfnamefont
  {C.}~\bibnamefont {Bernhard}}, \bibinfo {author} {\bibfnamefont {C.~T.}\
  \bibnamefont {Lin}},\ and\ \bibinfo {author} {\bibfnamefont {B.}~\bibnamefont
  {Keimer}},\ }\bibfield  {title} {\bibinfo {title} {Electronic liquid crystal
  state in the high-temperature superconductor {YBa$_2$Cu$_3$O$_{6.45}$}},\
  }\href {https://doi.org/10.1126/science.1152309} {\bibfield  {journal}
  {\bibinfo  {journal} {Science}\ }\textbf {\bibinfo {volume} {319}},\ \bibinfo
  {pages} {597} (\bibinfo {year} {2008})}\BibitemShut {NoStop}%
\bibitem [{\citenamefont {Daou}\ \emph {et~al.}(2010)\citenamefont {Daou},
  \citenamefont {Chang}, \citenamefont {LeBoeuf}, \citenamefont
  {Cyr-Choini{\`e}re}, \citenamefont {Lalibert{\'e}}, \citenamefont
  {Doiron-Leyraud}, \citenamefont {Ramshaw}, \citenamefont {Liang},
  \citenamefont {Bonn}, \citenamefont {Hardy},\ and\ \citenamefont
  {Taillefer}}]{Daou2010}%
  \BibitemOpen
  \bibfield  {author} {\bibinfo {author} {\bibfnamefont {R.}~\bibnamefont
  {Daou}}, \bibinfo {author} {\bibfnamefont {J.}~\bibnamefont {Chang}},
  \bibinfo {author} {\bibfnamefont {D.}~\bibnamefont {LeBoeuf}}, \bibinfo
  {author} {\bibfnamefont {O.}~\bibnamefont {Cyr-Choini{\`e}re}}, \bibinfo
  {author} {\bibfnamefont {F.}~\bibnamefont {Lalibert{\'e}}}, \bibinfo {author}
  {\bibfnamefont {N.}~\bibnamefont {Doiron-Leyraud}}, \bibinfo {author}
  {\bibfnamefont {B.~J.}\ \bibnamefont {Ramshaw}}, \bibinfo {author}
  {\bibfnamefont {R.}~\bibnamefont {Liang}}, \bibinfo {author} {\bibfnamefont
  {D.~A.}\ \bibnamefont {Bonn}}, \bibinfo {author} {\bibfnamefont {W.~N.}\
  \bibnamefont {Hardy}},\ and\ \bibinfo {author} {\bibfnamefont
  {L.}~\bibnamefont {Taillefer}},\ }\bibfield  {title} {\bibinfo {title}
  {Broken rotational symmetry in the pseudogap phase of a high-{$T_\text{c}$}
  superconductor},\ }\href@noop {} {\bibfield  {journal} {\bibinfo  {journal}
  {Nature}\ }\textbf {\bibinfo {volume} {463}},\ \bibinfo {pages} {519}
  (\bibinfo {year} {2010})}\BibitemShut {NoStop}%
\bibitem [{\citenamefont {Lawler}\ \emph {et~al.}(2010)\citenamefont {Lawler},
  \citenamefont {Fujita}, \citenamefont {Lee}, \citenamefont {Schmidt},
  \citenamefont {Kohsaka}, \citenamefont {Kim}, \citenamefont {Eisaki},
  \citenamefont {Uchida}, \citenamefont {Davis}, \citenamefont {Sethna},\ and\
  \citenamefont {Kim}}]{Lawler2010}%
  \BibitemOpen
  \bibfield  {author} {\bibinfo {author} {\bibfnamefont {M.~J.}\ \bibnamefont
  {Lawler}}, \bibinfo {author} {\bibfnamefont {K.}~\bibnamefont {Fujita}},
  \bibinfo {author} {\bibfnamefont {J.}~\bibnamefont {Lee}}, \bibinfo {author}
  {\bibfnamefont {A.~R.}\ \bibnamefont {Schmidt}}, \bibinfo {author}
  {\bibfnamefont {Y.}~\bibnamefont {Kohsaka}}, \bibinfo {author} {\bibfnamefont
  {C.~K.}\ \bibnamefont {Kim}}, \bibinfo {author} {\bibfnamefont
  {H.}~\bibnamefont {Eisaki}}, \bibinfo {author} {\bibfnamefont
  {S.}~\bibnamefont {Uchida}}, \bibinfo {author} {\bibfnamefont {J.~C.}\
  \bibnamefont {Davis}}, \bibinfo {author} {\bibfnamefont {J.~P.}\ \bibnamefont
  {Sethna}},\ and\ \bibinfo {author} {\bibfnamefont {E.-A.}\ \bibnamefont
  {Kim}},\ }\bibfield  {title} {\bibinfo {title} {Intra-unit-cell electronic
  nematicity of the high-${T}_\text{c}$ copper-oxide pseudogap states},\
  }\href@noop {} {\bibfield  {journal} {\bibinfo  {journal} {Nature}\ }\textbf
  {\bibinfo {volume} {466}},\ \bibinfo {pages} {347} (\bibinfo {year}
  {2010})}\BibitemShut {NoStop}%
\bibitem [{\citenamefont {Mesaros}\ \emph {et~al.}(2011)\citenamefont
  {Mesaros}, \citenamefont {Fujita}, \citenamefont {Eisaki}, \citenamefont
  {Uchida}, \citenamefont {Davis}, \citenamefont {Sachdev}, \citenamefont
  {Zaanen}, \citenamefont {Lawler},\ and\ \citenamefont {Kim}}]{Mesaros2011}%
  \BibitemOpen
  \bibfield  {author} {\bibinfo {author} {\bibfnamefont {A.}~\bibnamefont
  {Mesaros}}, \bibinfo {author} {\bibfnamefont {K.}~\bibnamefont {Fujita}},
  \bibinfo {author} {\bibfnamefont {H.}~\bibnamefont {Eisaki}}, \bibinfo
  {author} {\bibfnamefont {S.}~\bibnamefont {Uchida}}, \bibinfo {author}
  {\bibfnamefont {J.~C.}\ \bibnamefont {Davis}}, \bibinfo {author}
  {\bibfnamefont {S.}~\bibnamefont {Sachdev}}, \bibinfo {author} {\bibfnamefont
  {J.}~\bibnamefont {Zaanen}}, \bibinfo {author} {\bibfnamefont {M.~J.}\
  \bibnamefont {Lawler}},\ and\ \bibinfo {author} {\bibfnamefont {E.-A.}\
  \bibnamefont {Kim}},\ }\bibfield  {title} {\bibinfo {title} {Topological
  defects coupling smectic modulations to intra{\textendash}unit-cell
  nematicity in cuprates},\ }\href {https://doi.org/10.1126/science.1201082}
  {\bibfield  {journal} {\bibinfo  {journal} {Science}\ }\textbf {\bibinfo
  {volume} {333}},\ \bibinfo {pages} {426} (\bibinfo {year}
  {2011})}\BibitemShut {NoStop}%
\bibitem [{\citenamefont {Wu}\ \emph {et~al.}(2019)\citenamefont {Wu},
  \citenamefont {Bollinger}, \citenamefont {He}, \citenamefont {Gu},
  \citenamefont {Miao}, \citenamefont {Dean}, \citenamefont {Robinson},\ and\
  \citenamefont {Bo{\v{z}}ovi{\'{c}}}}]{Wu2019}%
  \BibitemOpen
  \bibfield  {author} {\bibinfo {author} {\bibfnamefont {J.}~\bibnamefont
  {Wu}}, \bibinfo {author} {\bibfnamefont {A.~T.}\ \bibnamefont {Bollinger}},
  \bibinfo {author} {\bibfnamefont {X.}~\bibnamefont {He}}, \bibinfo {author}
  {\bibfnamefont {G.~D.}\ \bibnamefont {Gu}}, \bibinfo {author} {\bibfnamefont
  {H.}~\bibnamefont {Miao}}, \bibinfo {author} {\bibfnamefont {M.~P.~M.}\
  \bibnamefont {Dean}}, \bibinfo {author} {\bibfnamefont {I.~K.}\ \bibnamefont
  {Robinson}},\ and\ \bibinfo {author} {\bibfnamefont {I.}~\bibnamefont
  {Bo{\v{z}}ovi{\'{c}}}},\ }\bibfield  {title} {\bibinfo {title}
  {Angle-resolved transport measurements reveal electronic nematicity in
  cuprate superconductors},\ }\href
  {https://doi.org/10.1007/s10948-019-05222-5} {\bibfield  {journal} {\bibinfo
  {journal} {Journal of Superconductivity and Novel Magnetism}\ }\textbf
  {\bibinfo {volume} {33}},\ \bibinfo {pages} {87} (\bibinfo {year}
  {2019})}\BibitemShut {NoStop}%
\bibitem [{\citenamefont {Auvray}\ \emph {et~al.}(2019)\citenamefont {Auvray},
  \citenamefont {Loret}, \citenamefont {Benhabib}, \citenamefont {Cazayous},
  \citenamefont {Zhong}, \citenamefont {Schneeloch}, \citenamefont {Gu},
  \citenamefont {Forget}, \citenamefont {Colson}, \citenamefont {Paul},
  \citenamefont {Sacuto},\ and\ \citenamefont {Gallais}}]{Auvray2019}%
  \BibitemOpen
  \bibfield  {author} {\bibinfo {author} {\bibfnamefont {N.}~\bibnamefont
  {Auvray}}, \bibinfo {author} {\bibfnamefont {B.}~\bibnamefont {Loret}},
  \bibinfo {author} {\bibfnamefont {S.}~\bibnamefont {Benhabib}}, \bibinfo
  {author} {\bibfnamefont {M.}~\bibnamefont {Cazayous}}, \bibinfo {author}
  {\bibfnamefont {R.~D.}\ \bibnamefont {Zhong}}, \bibinfo {author}
  {\bibfnamefont {J.}~\bibnamefont {Schneeloch}}, \bibinfo {author}
  {\bibfnamefont {G.~D.}\ \bibnamefont {Gu}}, \bibinfo {author} {\bibfnamefont
  {A.}~\bibnamefont {Forget}}, \bibinfo {author} {\bibfnamefont
  {D.}~\bibnamefont {Colson}}, \bibinfo {author} {\bibfnamefont
  {I.}~\bibnamefont {Paul}}, \bibinfo {author} {\bibfnamefont {A.}~\bibnamefont
  {Sacuto}},\ and\ \bibinfo {author} {\bibfnamefont {Y.}~\bibnamefont
  {Gallais}},\ }\bibfield  {title} {\bibinfo {title} {Nematic fluctuations in
  the cuprate superconductor {Bi$_2$Sr$_2$CaCu$_2$O$_{8+\delta}$}},\
  }\href@noop {} {\bibfield  {journal} {\bibinfo  {journal} {Nature
  Communications}\ }\textbf {\bibinfo {volume} {10}} (\bibinfo {year}
  {2019})}\BibitemShut {NoStop}%
\bibitem [{\citenamefont {Wang}\ \emph {et~al.}()\citenamefont {Wang},
  \citenamefont {Kennedy}, \citenamefont {Fujita}, \citenamefont {Uchida},
  \citenamefont {Eisaki}, \citenamefont {Johnson}, \citenamefont {Davis},\ and\
  \citenamefont {O'Mahony}}]{Wang2023}%
  \BibitemOpen
  \bibfield  {author} {\bibinfo {author} {\bibfnamefont {S.}~\bibnamefont
  {Wang}}, \bibinfo {author} {\bibfnamefont {N.}~\bibnamefont {Kennedy}},
  \bibinfo {author} {\bibfnamefont {K.}~\bibnamefont {Fujita}}, \bibinfo
  {author} {\bibfnamefont {S.-I.}\ \bibnamefont {Uchida}}, \bibinfo {author}
  {\bibfnamefont {H.}~\bibnamefont {Eisaki}}, \bibinfo {author} {\bibfnamefont
  {P.~D.}\ \bibnamefont {Johnson}}, \bibinfo {author} {\bibfnamefont
  {J.~C.~S.}\ \bibnamefont {Davis}},\ and\ \bibinfo {author} {\bibfnamefont
  {S.~M.}\ \bibnamefont {O'Mahony}},\ }\href@noop {} {\bibinfo {title}
  {Discovery of orbital ordering in {Bi$_2$Sr$_2$CaCu$_2$O$_{8+x}$.}}},\
  \bibinfo {note} {\textit{preprint available at
  https://arxiv.org/abs/2302.14855} (2023)}\BibitemShut {NoStop}%
\bibitem [{\citenamefont {Kivelson}\ \emph {et~al.}(1998)\citenamefont
  {Kivelson}, \citenamefont {Fradkin},\ and\ \citenamefont
  {Emery}}]{Kivelson1998}%
  \BibitemOpen
  \bibfield  {author} {\bibinfo {author} {\bibfnamefont {S.~A.}\ \bibnamefont
  {Kivelson}}, \bibinfo {author} {\bibfnamefont {E.}~\bibnamefont {Fradkin}},\
  and\ \bibinfo {author} {\bibfnamefont {V.~J.}\ \bibnamefont {Emery}},\
  }\bibfield  {title} {\bibinfo {title} {Electronic liquid-crystal phases of a
  doped {M}ott insulator},\ }\href@noop {} {\bibfield  {journal} {\bibinfo
  {journal} {Nature}\ }\textbf {\bibinfo {volume} {393}},\ \bibinfo {pages}
  {550} (\bibinfo {year} {1998})}\BibitemShut {NoStop}%
\bibitem [{\citenamefont {Fradkin}\ \emph {et~al.}(2010)\citenamefont
  {Fradkin}, \citenamefont {Kivelson}, \citenamefont {Lawler}, \citenamefont
  {Eisenstein},\ and\ \citenamefont {Mackenzie}}]{Fradkin2010}%
  \BibitemOpen
  \bibfield  {author} {\bibinfo {author} {\bibfnamefont {E.}~\bibnamefont
  {Fradkin}}, \bibinfo {author} {\bibfnamefont {S.~A.}\ \bibnamefont
  {Kivelson}}, \bibinfo {author} {\bibfnamefont {M.~J.}\ \bibnamefont
  {Lawler}}, \bibinfo {author} {\bibfnamefont {J.~P.}\ \bibnamefont
  {Eisenstein}},\ and\ \bibinfo {author} {\bibfnamefont {A.~P.}\ \bibnamefont
  {Mackenzie}},\ }\bibfield  {title} {\bibinfo {title} {Nematic {F}ermi fluids
  in condensed matter physics},\ }\href
  {https://doi.org/10.1146/annurev-conmatphys-070909-103925} {\bibfield
  {journal} {\bibinfo  {journal} {Annual Review of Condensed Matter Physics}\
  }\textbf {\bibinfo {volume} {1}},\ \bibinfo {pages} {153} (\bibinfo {year}
  {2010})}\BibitemShut {NoStop}%
\bibitem [{\citenamefont {Zaanen}\ and\ \citenamefont
  {Gunnarsson}(1989)}]{Zaanen1989}%
  \BibitemOpen
  \bibfield  {author} {\bibinfo {author} {\bibfnamefont {J.}~\bibnamefont
  {Zaanen}}\ and\ \bibinfo {author} {\bibfnamefont {O.}~\bibnamefont
  {Gunnarsson}},\ }\bibfield  {title} {\bibinfo {title} {Charged magnetic
  domain lines and the magnetism of high-${T}_{c}$ oxides},\ }\href
  {https://doi.org/10.1103/PhysRevB.40.7391} {\bibfield  {journal} {\bibinfo
  {journal} {Phys. Rev. B}\ }\textbf {\bibinfo {volume} {40}},\ \bibinfo
  {pages} {7391} (\bibinfo {year} {1989})}\BibitemShut {NoStop}%
\bibitem [{\citenamefont {Machida}(1989)}]{Machida1989}%
  \BibitemOpen
  \bibfield  {author} {\bibinfo {author} {\bibfnamefont {K.}~\bibnamefont
  {Machida}},\ }\bibfield  {title} {\bibinfo {title} {Magnetism in
  {L}a$_2${C}u{O}$_4$ based compounds},\ }\href
  {https://doi.org/https://doi.org/10.1016/0921-4534(89)90316-X} {\bibfield
  {journal} {\bibinfo  {journal} {Physica C: Superconductivity}\ }\textbf
  {\bibinfo {volume} {158}},\ \bibinfo {pages} {192} (\bibinfo {year}
  {1989})}\BibitemShut {NoStop}%
\bibitem [{\citenamefont {Kato}\ \emph {et~al.}(1990)\citenamefont {Kato},
  \citenamefont {Machida}, \citenamefont {Nakanishi},\ and\ \citenamefont
  {Fujita}}]{Kato1990}%
  \BibitemOpen
  \bibfield  {author} {\bibinfo {author} {\bibfnamefont {M.}~\bibnamefont
  {Kato}}, \bibinfo {author} {\bibfnamefont {K.}~\bibnamefont {Machida}},
  \bibinfo {author} {\bibfnamefont {H.}~\bibnamefont {Nakanishi}},\ and\
  \bibinfo {author} {\bibfnamefont {M.}~\bibnamefont {Fujita}},\ }\bibfield
  {title} {\bibinfo {title} {Soliton lattice modulation of incommensurate spin
  density wave in two dimensional hubbard model - {A} mean field study},\
  }\href {https://doi.org/10.1143/jpsj.59.1047} {\bibfield  {journal} {\bibinfo
   {journal} {Journal of the Physical Society of Japan}\ }\textbf {\bibinfo
  {volume} {59}},\ \bibinfo {pages} {1047} (\bibinfo {year}
  {1990})}\BibitemShut {NoStop}%
\bibitem [{\citenamefont {Emery}\ and\ \citenamefont
  {Kivelson}(1993)}]{Emery1993}%
  \BibitemOpen
  \bibfield  {author} {\bibinfo {author} {\bibfnamefont {V.}~\bibnamefont
  {Emery}}\ and\ \bibinfo {author} {\bibfnamefont {S.}~\bibnamefont
  {Kivelson}},\ }\bibfield  {title} {\bibinfo {title} {Frustrated electronic
  phase separation and high-temperature superconductors},\ }\href
  {https://doi.org/https://doi.org/10.1016/0921-4534(93)90581-A} {\bibfield
  {journal} {\bibinfo  {journal} {Physica C: Superconductivity}\ }\textbf
  {\bibinfo {volume} {209}},\ \bibinfo {pages} {597} (\bibinfo {year}
  {1993})}\BibitemShut {NoStop}%
\bibitem [{\citenamefont {White}\ and\ \citenamefont
  {Scalapino}(1998)}]{White1998}%
  \BibitemOpen
  \bibfield  {author} {\bibinfo {author} {\bibfnamefont {S.~R.}\ \bibnamefont
  {White}}\ and\ \bibinfo {author} {\bibfnamefont {D.~J.}\ \bibnamefont
  {Scalapino}},\ }\bibfield  {title} {\bibinfo {title} {Density matrix
  renormalization group study of the striped phase in the 2d
  $\mathit{t}\ensuremath{-}\mathit{J}$ model},\ }\href
  {https://doi.org/10.1103/PhysRevLett.80.1272} {\bibfield  {journal} {\bibinfo
   {journal} {Phys. Rev. Lett.}\ }\textbf {\bibinfo {volume} {80}},\ \bibinfo
  {pages} {1272} (\bibinfo {year} {1998})}\BibitemShut {NoStop}%
\bibitem [{\citenamefont {Halboth}\ and\ \citenamefont
  {Metzner}(2000)}]{Halboth2000}%
  \BibitemOpen
  \bibfield  {author} {\bibinfo {author} {\bibfnamefont {C.~J.}\ \bibnamefont
  {Halboth}}\ and\ \bibinfo {author} {\bibfnamefont {W.}~\bibnamefont
  {Metzner}},\ }\bibfield  {title} {\bibinfo {title} {$\mathit{d}$-wave
  superconductivity and pomeranchuk instability in the two-dimensional hubbard
  model},\ }\href {https://doi.org/10.1103/PhysRevLett.85.5162} {\bibfield
  {journal} {\bibinfo  {journal} {Phys. Rev. Lett.}\ }\textbf {\bibinfo
  {volume} {85}},\ \bibinfo {pages} {5162} (\bibinfo {year}
  {2000})}\BibitemShut {NoStop}%
\bibitem [{\citenamefont {Lorenzana}\ and\ \citenamefont
  {Seibold}(2002)}]{Lorenzana2002}%
  \BibitemOpen
  \bibfield  {author} {\bibinfo {author} {\bibfnamefont {J.}~\bibnamefont
  {Lorenzana}}\ and\ \bibinfo {author} {\bibfnamefont {G.}~\bibnamefont
  {Seibold}},\ }\bibfield  {title} {\bibinfo {title} {Metallic mean-field
  stripes, incommensurability, and chemical potential in cuprates},\ }\href
  {https://doi.org/10.1103/physrevlett.89.136401} {\bibfield  {journal}
  {\bibinfo  {journal} {Physical Review Letters}\ }\textbf {\bibinfo {volume}
  {89}},\ \bibinfo {pages} {136401} (\bibinfo {year} {2002})}\BibitemShut
  {NoStop}%
\bibitem [{\citenamefont {Kivelson}\ \emph {et~al.}(2003)\citenamefont
  {Kivelson}, \citenamefont {Bindloss}, \citenamefont {Fradkin}, \citenamefont
  {Oganesyan}, \citenamefont {Tranquada}, \citenamefont {Kapitulnik},\ and\
  \citenamefont {Howald}}]{Kivelson2003}%
  \BibitemOpen
  \bibfield  {author} {\bibinfo {author} {\bibfnamefont {S.~A.}\ \bibnamefont
  {Kivelson}}, \bibinfo {author} {\bibfnamefont {I.~P.}\ \bibnamefont
  {Bindloss}}, \bibinfo {author} {\bibfnamefont {E.}~\bibnamefont {Fradkin}},
  \bibinfo {author} {\bibfnamefont {V.}~\bibnamefont {Oganesyan}}, \bibinfo
  {author} {\bibfnamefont {J.~M.}\ \bibnamefont {Tranquada}}, \bibinfo {author}
  {\bibfnamefont {A.}~\bibnamefont {Kapitulnik}},\ and\ \bibinfo {author}
  {\bibfnamefont {C.}~\bibnamefont {Howald}},\ }\bibfield  {title} {\bibinfo
  {title} {How to detect fluctuating stripes in the high-temperature
  superconductors},\ }\href {https://doi.org/10.1103/revmodphys.75.1201}
  {\bibfield  {journal} {\bibinfo  {journal} {Reviews of Modern Physics}\
  }\textbf {\bibinfo {volume} {75}},\ \bibinfo {pages} {1201} (\bibinfo {year}
  {2003})}\BibitemShut {NoStop}%
\bibitem [{\citenamefont {Gerber}\ \emph {et~al.}(2015)\citenamefont {Gerber},
  \citenamefont {Jang}, \citenamefont {Nojiri}, \citenamefont {Matsuzawa},
  \citenamefont {Yasumura}, \citenamefont {Bonn}, \citenamefont {Liang},
  \citenamefont {Hardy}, \citenamefont {Islam}, \citenamefont {Mehta},
  \citenamefont {Song}, \citenamefont {Sikorski}, \citenamefont {Stefanescu},
  \citenamefont {Feng}, \citenamefont {Kivelson}, \citenamefont {Devereaux},
  \citenamefont {Shen}, \citenamefont {Kao}, \citenamefont {Lee}, \citenamefont
  {Zhu},\ and\ \citenamefont {Lee}}]{Gerber2015}%
  \BibitemOpen
  \bibfield  {author} {\bibinfo {author} {\bibfnamefont {S.}~\bibnamefont
  {Gerber}}, \bibinfo {author} {\bibfnamefont {H.}~\bibnamefont {Jang}},
  \bibinfo {author} {\bibfnamefont {H.}~\bibnamefont {Nojiri}}, \bibinfo
  {author} {\bibfnamefont {S.}~\bibnamefont {Matsuzawa}}, \bibinfo {author}
  {\bibfnamefont {H.}~\bibnamefont {Yasumura}}, \bibinfo {author}
  {\bibfnamefont {D.~A.}\ \bibnamefont {Bonn}}, \bibinfo {author}
  {\bibfnamefont {R.}~\bibnamefont {Liang}}, \bibinfo {author} {\bibfnamefont
  {W.~N.}\ \bibnamefont {Hardy}}, \bibinfo {author} {\bibfnamefont
  {Z.}~\bibnamefont {Islam}}, \bibinfo {author} {\bibfnamefont
  {A.}~\bibnamefont {Mehta}}, \bibinfo {author} {\bibfnamefont
  {S.}~\bibnamefont {Song}}, \bibinfo {author} {\bibfnamefont {M.}~\bibnamefont
  {Sikorski}}, \bibinfo {author} {\bibfnamefont {D.}~\bibnamefont
  {Stefanescu}}, \bibinfo {author} {\bibfnamefont {Y.}~\bibnamefont {Feng}},
  \bibinfo {author} {\bibfnamefont {S.~A.}\ \bibnamefont {Kivelson}}, \bibinfo
  {author} {\bibfnamefont {T.~P.}\ \bibnamefont {Devereaux}}, \bibinfo {author}
  {\bibfnamefont {Z.-X.}\ \bibnamefont {Shen}}, \bibinfo {author}
  {\bibfnamefont {C.-C.}\ \bibnamefont {Kao}}, \bibinfo {author} {\bibfnamefont
  {W.-S.}\ \bibnamefont {Lee}}, \bibinfo {author} {\bibfnamefont
  {D.}~\bibnamefont {Zhu}},\ and\ \bibinfo {author} {\bibfnamefont {J.-S.}\
  \bibnamefont {Lee}},\ }\bibfield  {title} {\bibinfo {title}
  {Three-dimensional charge density wave order in
  {{YBa$_{2}$Cu$_{3}$O$_{6.67}$}} at high magnetic fields},\ }\href
  {https://doi.org/10.1126/science.aac6257} {\bibfield  {journal} {\bibinfo
  {journal} {Science}\ }\textbf {\bibinfo {volume} {350}},\ \bibinfo {pages}
  {949} (\bibinfo {year} {2015})}\BibitemShut {NoStop}%
\bibitem [{\citenamefont {Kim}\ \emph {et~al.}(2021)\citenamefont {Kim},
  \citenamefont {Lefran{\c{c}}ois}, \citenamefont {Kummer}, \citenamefont
  {Fumagalli}, \citenamefont {Brookes}, \citenamefont {Betto}, \citenamefont
  {Nakata}, \citenamefont {Tortora}, \citenamefont {Porras}, \citenamefont
  {Loew}, \citenamefont {Barber}, \citenamefont {Braicovich}, \citenamefont
  {Mackenzie}, \citenamefont {Hicks}, \citenamefont {Keimer}, \citenamefont
  {Minola},\ and\ \citenamefont {Tacon}}]{Kim2021}%
  \BibitemOpen
  \bibfield  {author} {\bibinfo {author} {\bibfnamefont {H.-H.}\ \bibnamefont
  {Kim}}, \bibinfo {author} {\bibfnamefont {E.}~\bibnamefont
  {Lefran{\c{c}}ois}}, \bibinfo {author} {\bibfnamefont {K.}~\bibnamefont
  {Kummer}}, \bibinfo {author} {\bibfnamefont {R.}~\bibnamefont {Fumagalli}},
  \bibinfo {author} {\bibfnamefont {N.}~\bibnamefont {Brookes}}, \bibinfo
  {author} {\bibfnamefont {D.}~\bibnamefont {Betto}}, \bibinfo {author}
  {\bibfnamefont {S.}~\bibnamefont {Nakata}}, \bibinfo {author} {\bibfnamefont
  {M.}~\bibnamefont {Tortora}}, \bibinfo {author} {\bibfnamefont
  {J.}~\bibnamefont {Porras}}, \bibinfo {author} {\bibfnamefont
  {T.}~\bibnamefont {Loew}}, \bibinfo {author} {\bibfnamefont {M.}~\bibnamefont
  {Barber}}, \bibinfo {author} {\bibfnamefont {L.}~\bibnamefont {Braicovich}},
  \bibinfo {author} {\bibfnamefont {A.}~\bibnamefont {Mackenzie}}, \bibinfo
  {author} {\bibfnamefont {C.}~\bibnamefont {Hicks}}, \bibinfo {author}
  {\bibfnamefont {B.}~\bibnamefont {Keimer}}, \bibinfo {author} {\bibfnamefont
  {M.}~\bibnamefont {Minola}},\ and\ \bibinfo {author} {\bibfnamefont {M.~L.}\
  \bibnamefont {Tacon}},\ }\bibfield  {title} {\bibinfo {title} {Charge density
  waves in {YBa$_2$Cu$_3$O$_{6.67}$} probed by resonant x-ray scattering under
  uniaxial compression},\ }\href
  {https://doi.org/10.1103/physrevlett.126.037002} {\bibfield  {journal}
  {\bibinfo  {journal} {Physical Review Letters}\ }\textbf {\bibinfo {volume}
  {126}},\ \bibinfo {pages} {037002} (\bibinfo {year} {2021})}\BibitemShut
  {NoStop}%
\bibitem [{\citenamefont {Kivelson}\ \emph {et~al.}(2004)\citenamefont
  {Kivelson}, \citenamefont {Fradkin},\ and\ \citenamefont
  {Geballe}}]{Kivelson2004}%
  \BibitemOpen
  \bibfield  {author} {\bibinfo {author} {\bibfnamefont {S.~A.}\ \bibnamefont
  {Kivelson}}, \bibinfo {author} {\bibfnamefont {E.}~\bibnamefont {Fradkin}},\
  and\ \bibinfo {author} {\bibfnamefont {T.~H.}\ \bibnamefont {Geballe}},\
  }\bibfield  {title} {\bibinfo {title} {Quasi-one-dimensional dynamics and
  nematic phases in the two-dimensional {E}mery model},\ }\href@noop {}
  {\bibfield  {journal} {\bibinfo  {journal} {Physical Review B}\ }\textbf
  {\bibinfo {volume} {69}},\ \bibinfo {pages} {144505} (\bibinfo {year}
  {2004})}\BibitemShut {NoStop}%
\bibitem [{\citenamefont {Fischer}\ \emph {et~al.}(2014)\citenamefont
  {Fischer}, \citenamefont {Wu}, \citenamefont {Lawler}, \citenamefont
  {Paramekanti},\ and\ \citenamefont {Kim}}]{Fischer2014}%
  \BibitemOpen
  \bibfield  {author} {\bibinfo {author} {\bibfnamefont {M.~H.}\ \bibnamefont
  {Fischer}}, \bibinfo {author} {\bibfnamefont {S.}~\bibnamefont {Wu}},
  \bibinfo {author} {\bibfnamefont {M.}~\bibnamefont {Lawler}}, \bibinfo
  {author} {\bibfnamefont {A.}~\bibnamefont {Paramekanti}},\ and\ \bibinfo
  {author} {\bibfnamefont {E.-A.}\ \bibnamefont {Kim}},\ }\bibfield  {title}
  {\bibinfo {title} {Nematic and spin-charge orders driven by hole-doping a
  charge-transfer insulator},\ }\href@noop {} {\bibfield  {journal} {\bibinfo
  {journal} {New Journal of Physics}\ }\textbf {\bibinfo {volume} {16}},\
  \bibinfo {pages} {093057} (\bibinfo {year} {2014})}\BibitemShut {NoStop}%
\bibitem [{\citenamefont {Yamase}\ and\ \citenamefont
  {Kohno}(2000)}]{Yamase2000}%
  \BibitemOpen
  \bibfield  {author} {\bibinfo {author} {\bibfnamefont {H.}~\bibnamefont
  {Yamase}}\ and\ \bibinfo {author} {\bibfnamefont {H.}~\bibnamefont {Kohno}},\
  }\bibfield  {title} {\bibinfo {title} {Instability toward formation of
  quasi-one-dimensional {F}ermi surface in two-dimensional {t-J} model},\
  }\href {https://doi.org/10.1143/JPSJ.69.2151} {\bibfield  {journal} {\bibinfo
   {journal} {Journal of the Physical Society of Japan}\ }\textbf {\bibinfo
  {volume} {69}},\ \bibinfo {pages} {2151} (\bibinfo {year}
  {2000})}\BibitemShut {NoStop}%
\bibitem [{\citenamefont {Yamase}(2021)}]{Yamase2021}%
  \BibitemOpen
  \bibfield  {author} {\bibinfo {author} {\bibfnamefont {H.}~\bibnamefont
  {Yamase}},\ }\bibfield  {title} {\bibinfo {title} {Theoretical insights into
  electronic nematic order, bond-charge orders, and plasmons in cuprate
  superconductors},\ }\href {https://doi.org/10.7566/JPSJ.90.111011} {\bibfield
   {journal} {\bibinfo  {journal} {Journal of the Physical Society of Japan}\
  }\textbf {\bibinfo {volume} {90}},\ \bibinfo {pages} {111011} (\bibinfo
  {year} {2021})}\BibitemShut {NoStop}%
\bibitem [{\citenamefont {Achkar}\ \emph {et~al.}(2016)\citenamefont {Achkar},
  \citenamefont {Zwiebler}, \citenamefont {McMahon}, \citenamefont {He},
  \citenamefont {Sutarto}, \citenamefont {Djianto}, \citenamefont {Hao},
  \citenamefont {Gingras}, \citenamefont {Hücker}, \citenamefont {Gu},
  \citenamefont {Revcolevschi}, \citenamefont {Zhang}, \citenamefont {Kim},
  \citenamefont {Geck},\ and\ \citenamefont {Hawthorn}}]{Achkar2016}%
  \BibitemOpen
  \bibfield  {author} {\bibinfo {author} {\bibfnamefont {A.~J.}\ \bibnamefont
  {Achkar}}, \bibinfo {author} {\bibfnamefont {M.}~\bibnamefont {Zwiebler}},
  \bibinfo {author} {\bibfnamefont {C.}~\bibnamefont {McMahon}}, \bibinfo
  {author} {\bibfnamefont {F.}~\bibnamefont {He}}, \bibinfo {author}
  {\bibfnamefont {R.}~\bibnamefont {Sutarto}}, \bibinfo {author} {\bibfnamefont
  {I.}~\bibnamefont {Djianto}}, \bibinfo {author} {\bibfnamefont
  {Z.}~\bibnamefont {Hao}}, \bibinfo {author} {\bibfnamefont {M.~J.~P.}\
  \bibnamefont {Gingras}}, \bibinfo {author} {\bibfnamefont {M.}~\bibnamefont
  {Hücker}}, \bibinfo {author} {\bibfnamefont {G.~D.}\ \bibnamefont {Gu}},
  \bibinfo {author} {\bibfnamefont {A.}~\bibnamefont {Revcolevschi}}, \bibinfo
  {author} {\bibfnamefont {H.}~\bibnamefont {Zhang}}, \bibinfo {author}
  {\bibfnamefont {Y.-J.}\ \bibnamefont {Kim}}, \bibinfo {author} {\bibfnamefont
  {J.}~\bibnamefont {Geck}},\ and\ \bibinfo {author} {\bibfnamefont {D.~G.}\
  \bibnamefont {Hawthorn}},\ }\bibfield  {title} {\bibinfo {title} {Nematicity
  in stripe-ordered cuprates probed via resonant x-ray scattering},\ }\href
  {https://doi.org/10.1126/science.aad1824} {\bibfield  {journal} {\bibinfo
  {journal} {Science}\ }\textbf {\bibinfo {volume} {351}},\ \bibinfo {pages}
  {576} (\bibinfo {year} {2016})}\BibitemShut {NoStop}%
\bibitem [{\citenamefont {Gupta}\ \emph {et~al.}(2021)\citenamefont {Gupta},
  \citenamefont {McMahon}, \citenamefont {Sutarto}, \citenamefont {Shi},
  \citenamefont {Gong}, \citenamefont {Wei}, \citenamefont {Shen},
  \citenamefont {He}, \citenamefont {Ma}, \citenamefont {Dragomir},
  \citenamefont {Gaulin},\ and\ \citenamefont {Hawthorn}}]{Gupta2021}%
  \BibitemOpen
  \bibfield  {author} {\bibinfo {author} {\bibfnamefont {N.~K.}\ \bibnamefont
  {Gupta}}, \bibinfo {author} {\bibfnamefont {C.}~\bibnamefont {McMahon}},
  \bibinfo {author} {\bibfnamefont {R.}~\bibnamefont {Sutarto}}, \bibinfo
  {author} {\bibfnamefont {T.}~\bibnamefont {Shi}}, \bibinfo {author}
  {\bibfnamefont {R.}~\bibnamefont {Gong}}, \bibinfo {author} {\bibfnamefont
  {H.~I.}\ \bibnamefont {Wei}}, \bibinfo {author} {\bibfnamefont {K.~M.}\
  \bibnamefont {Shen}}, \bibinfo {author} {\bibfnamefont {F.}~\bibnamefont
  {He}}, \bibinfo {author} {\bibfnamefont {Q.}~\bibnamefont {Ma}}, \bibinfo
  {author} {\bibfnamefont {M.}~\bibnamefont {Dragomir}}, \bibinfo {author}
  {\bibfnamefont {B.~D.}\ \bibnamefont {Gaulin}},\ and\ \bibinfo {author}
  {\bibfnamefont {D.~G.}\ \bibnamefont {Hawthorn}},\ }\bibfield  {title}
  {\bibinfo {title} {Vanishing nematic order beyond the pseudogap phase in
  overdoped cuprate superconductors},\ }\href
  {https://doi.org/10.1073/pnas.2106881118} {\bibfield  {journal} {\bibinfo
  {journal} {Proceedings of the National Academy of Sciences}\ }\textbf
  {\bibinfo {volume} {118}},\ \bibinfo {pages} {e2106881118} (\bibinfo {year}
  {2021})}\BibitemShut {NoStop}%
\bibitem [{\citenamefont {Gedik}\ \emph {et~al.}(2005)\citenamefont {Gedik},
  \citenamefont {Langner}, \citenamefont {Orenstein}, \citenamefont {Ono},
  \citenamefont {Abe},\ and\ \citenamefont {Ando}}]{Gedik2005}%
  \BibitemOpen
  \bibfield  {author} {\bibinfo {author} {\bibfnamefont {N.}~\bibnamefont
  {Gedik}}, \bibinfo {author} {\bibfnamefont {M.}~\bibnamefont {Langner}},
  \bibinfo {author} {\bibfnamefont {J.}~\bibnamefont {Orenstein}}, \bibinfo
  {author} {\bibfnamefont {S.}~\bibnamefont {Ono}}, \bibinfo {author}
  {\bibfnamefont {Y.}~\bibnamefont {Abe}},\ and\ \bibinfo {author}
  {\bibfnamefont {Y.}~\bibnamefont {Ando}},\ }\bibfield  {title} {\bibinfo
  {title} {Abrupt transition in quasiparticle dynamics at optimal doping in a
  cuprate superconductor system},\ }\href
  {https://doi.org/10.1103/PhysRevLett.95.117005} {\bibfield  {journal}
  {\bibinfo  {journal} {Phys. Rev. Lett.}\ }\textbf {\bibinfo {volume} {95}},\
  \bibinfo {pages} {117005} (\bibinfo {year} {2005})}\BibitemShut {NoStop}%
\bibitem [{\citenamefont {Torchinsky}\ \emph {et~al.}(2013)\citenamefont
  {Torchinsky}, \citenamefont {Mahmood}, \citenamefont {Bollinger},
  \citenamefont {Bo{\v{z}}ovi{\'{c}}},\ and\ \citenamefont
  {Gedik}}]{Torchinsky2013}%
  \BibitemOpen
  \bibfield  {author} {\bibinfo {author} {\bibfnamefont {D.~H.}\ \bibnamefont
  {Torchinsky}}, \bibinfo {author} {\bibfnamefont {F.}~\bibnamefont {Mahmood}},
  \bibinfo {author} {\bibfnamefont {A.~T.}\ \bibnamefont {Bollinger}}, \bibinfo
  {author} {\bibfnamefont {I.}~\bibnamefont {Bo{\v{z}}ovi{\'{c}}}},\ and\
  \bibinfo {author} {\bibfnamefont {N.}~\bibnamefont {Gedik}},\ }\bibfield
  {title} {\bibinfo {title} {Fluctuating charge-density waves in a cuprate
  superconductor},\ }\href {https://doi.org/10.1038/nmat3571} {\bibfield
  {journal} {\bibinfo  {journal} {Nature Materials}\ }\textbf {\bibinfo
  {volume} {12}},\ \bibinfo {pages} {387} (\bibinfo {year} {2013})}\BibitemShut
  {NoStop}%
\bibitem [{\citenamefont {Dakovski}\ \emph {et~al.}(2015)\citenamefont
  {Dakovski}, \citenamefont {Lee}, \citenamefont {Hawthorn}, \citenamefont
  {Garner}, \citenamefont {Bonn}, \citenamefont {Hardy}, \citenamefont {Liang},
  \citenamefont {Hoffmann},\ and\ \citenamefont {Turner}}]{Dakovski2015a}%
  \BibitemOpen
  \bibfield  {author} {\bibinfo {author} {\bibfnamefont {G.~L.}\ \bibnamefont
  {Dakovski}}, \bibinfo {author} {\bibfnamefont {W.-S.}\ \bibnamefont {Lee}},
  \bibinfo {author} {\bibfnamefont {D.~G.}\ \bibnamefont {Hawthorn}}, \bibinfo
  {author} {\bibfnamefont {N.}~\bibnamefont {Garner}}, \bibinfo {author}
  {\bibfnamefont {D.}~\bibnamefont {Bonn}}, \bibinfo {author} {\bibfnamefont
  {W.}~\bibnamefont {Hardy}}, \bibinfo {author} {\bibfnamefont
  {R.}~\bibnamefont {Liang}}, \bibinfo {author} {\bibfnamefont {M.~C.}\
  \bibnamefont {Hoffmann}},\ and\ \bibinfo {author} {\bibfnamefont {J.~J.}\
  \bibnamefont {Turner}},\ }\bibfield  {title} {\bibinfo {title} {Enhanced
  coherent oscillations in the superconducting state of underdoped
  {YBa$_2$Cu$_3$O$_{6+x}$} induced via ultrafast terahertz excitation},\ }\href
  {https://doi.org/10.1103/physrevb.91.220506} {\bibfield  {journal} {\bibinfo
  {journal} {Physical Review B}\ }\textbf {\bibinfo {volume} {91}},\ \bibinfo
  {pages} {220506} (\bibinfo {year} {2015})}\BibitemShut {NoStop}%
\bibitem [{\citenamefont {Boschini}\ \emph {et~al.}(2018)\citenamefont
  {Boschini}, \citenamefont {da~Silva~Neto}, \citenamefont {Razzoli},
  \citenamefont {Zonno}, \citenamefont {Peli}, \citenamefont {Day},
  \citenamefont {Michiardi}, \citenamefont {Schneider}, \citenamefont
  {Zwartsenberg}, \citenamefont {Nigge}, \citenamefont {Zhong}, \citenamefont
  {Schneeloch}, \citenamefont {Gu}, \citenamefont {Zhdanovich}, \citenamefont
  {Mills}, \citenamefont {Levy}, \citenamefont {Jones}, \citenamefont
  {Giannetti},\ and\ \citenamefont {Damascelli}}]{Boschini2018}%
  \BibitemOpen
  \bibfield  {author} {\bibinfo {author} {\bibfnamefont {F.}~\bibnamefont
  {Boschini}}, \bibinfo {author} {\bibfnamefont {E.~H.}\ \bibnamefont
  {da~Silva~Neto}}, \bibinfo {author} {\bibfnamefont {E.}~\bibnamefont
  {Razzoli}}, \bibinfo {author} {\bibfnamefont {M.}~\bibnamefont {Zonno}},
  \bibinfo {author} {\bibfnamefont {S.}~\bibnamefont {Peli}}, \bibinfo {author}
  {\bibfnamefont {R.~P.}\ \bibnamefont {Day}}, \bibinfo {author} {\bibfnamefont
  {M.}~\bibnamefont {Michiardi}}, \bibinfo {author} {\bibfnamefont
  {M.}~\bibnamefont {Schneider}}, \bibinfo {author} {\bibfnamefont
  {B.}~\bibnamefont {Zwartsenberg}}, \bibinfo {author} {\bibfnamefont
  {P.}~\bibnamefont {Nigge}}, \bibinfo {author} {\bibfnamefont {R.~D.}\
  \bibnamefont {Zhong}}, \bibinfo {author} {\bibfnamefont {J.}~\bibnamefont
  {Schneeloch}}, \bibinfo {author} {\bibfnamefont {G.~D.}\ \bibnamefont {Gu}},
  \bibinfo {author} {\bibfnamefont {S.}~\bibnamefont {Zhdanovich}}, \bibinfo
  {author} {\bibfnamefont {A.~K.}\ \bibnamefont {Mills}}, \bibinfo {author}
  {\bibfnamefont {G.}~\bibnamefont {Levy}}, \bibinfo {author} {\bibfnamefont
  {D.~J.}\ \bibnamefont {Jones}}, \bibinfo {author} {\bibfnamefont
  {C.}~\bibnamefont {Giannetti}},\ and\ \bibinfo {author} {\bibfnamefont
  {A.}~\bibnamefont {Damascelli}},\ }\bibfield  {title} {\bibinfo {title}
  {Collapse of superconductivity in cuprates via ultrafast quenching of phase
  coherence},\ }\href {https://doi.org/10.1038/s41563-018-0045-1} {\bibfield
  {journal} {\bibinfo  {journal} {Nature Materials}\ }\textbf {\bibinfo
  {volume} {17}},\ \bibinfo {pages} {416} (\bibinfo {year} {2018})}\BibitemShut
  {NoStop}%
\bibitem [{\citenamefont {Mitrano}\ \emph
  {et~al.}(2019{\natexlab{a}})\citenamefont {Mitrano}, \citenamefont {Lee},
  \citenamefont {Husain}, \citenamefont {Delacretaz}, \citenamefont {Zhu},
  \citenamefont {de~la Pe{\~{n}}a~Munoz}, \citenamefont {Sun}, \citenamefont
  {Joe}, \citenamefont {Reid}, \citenamefont {Wandel}, \citenamefont
  {Coslovich}, \citenamefont {Schlotter}, \citenamefont {van Driel},
  \citenamefont {Schneeloch}, \citenamefont {Gu}, \citenamefont {Hartnoll},
  \citenamefont {Goldenfeld},\ and\ \citenamefont {Abbamonte}}]{Mitrano2019a}%
  \BibitemOpen
  \bibfield  {author} {\bibinfo {author} {\bibfnamefont {M.}~\bibnamefont
  {Mitrano}}, \bibinfo {author} {\bibfnamefont {S.}~\bibnamefont {Lee}},
  \bibinfo {author} {\bibfnamefont {A.~A.}\ \bibnamefont {Husain}}, \bibinfo
  {author} {\bibfnamefont {L.}~\bibnamefont {Delacretaz}}, \bibinfo {author}
  {\bibfnamefont {M.}~\bibnamefont {Zhu}}, \bibinfo {author} {\bibfnamefont
  {G.}~\bibnamefont {de~la Pe{\~{n}}a~Munoz}}, \bibinfo {author} {\bibfnamefont
  {S.~X.-L.}\ \bibnamefont {Sun}}, \bibinfo {author} {\bibfnamefont {Y.~I.}\
  \bibnamefont {Joe}}, \bibinfo {author} {\bibfnamefont {A.~H.}\ \bibnamefont
  {Reid}}, \bibinfo {author} {\bibfnamefont {S.~F.}\ \bibnamefont {Wandel}},
  \bibinfo {author} {\bibfnamefont {G.}~\bibnamefont {Coslovich}}, \bibinfo
  {author} {\bibfnamefont {W.}~\bibnamefont {Schlotter}}, \bibinfo {author}
  {\bibfnamefont {T.}~\bibnamefont {van Driel}}, \bibinfo {author}
  {\bibfnamefont {J.}~\bibnamefont {Schneeloch}}, \bibinfo {author}
  {\bibfnamefont {G.~D.}\ \bibnamefont {Gu}}, \bibinfo {author} {\bibfnamefont
  {S.}~\bibnamefont {Hartnoll}}, \bibinfo {author} {\bibfnamefont
  {N.}~\bibnamefont {Goldenfeld}},\ and\ \bibinfo {author} {\bibfnamefont
  {P.}~\bibnamefont {Abbamonte}},\ }\bibfield  {title} {\bibinfo {title}
  {Ultrafast time-resolved x-ray scattering reveals diffusive charge order
  dynamics in {La$_{2-x}$Ba$_{x}$CuO$_{4}$}},\ }\href@noop {} {\bibfield
  {journal} {\bibinfo  {journal} {Science Advances}\ }\textbf {\bibinfo
  {volume} {5}},\ \bibinfo {pages} {eaax3346} (\bibinfo {year}
  {2019}{\natexlab{a}})}\BibitemShut {NoStop}%
\bibitem [{\citenamefont {Mitrano}\ \emph
  {et~al.}(2019{\natexlab{b}})\citenamefont {Mitrano}, \citenamefont {Lee},
  \citenamefont {Husain}, \citenamefont {Zhu}, \citenamefont {de~la
  Pe{\~{n}}a~Munoz}, \citenamefont {Sun}, \citenamefont {Joe}, \citenamefont
  {Reid}, \citenamefont {Wandel}, \citenamefont {Coslovich}, \citenamefont
  {Schlotter}, \citenamefont {van Driel}, \citenamefont {Schneeloch},
  \citenamefont {Gu}, \citenamefont {Goldenfeld},\ and\ \citenamefont
  {Abbamonte}}]{Mitrano2019b}%
  \BibitemOpen
  \bibfield  {author} {\bibinfo {author} {\bibfnamefont {M.}~\bibnamefont
  {Mitrano}}, \bibinfo {author} {\bibfnamefont {S.}~\bibnamefont {Lee}},
  \bibinfo {author} {\bibfnamefont {A.~A.}\ \bibnamefont {Husain}}, \bibinfo
  {author} {\bibfnamefont {M.}~\bibnamefont {Zhu}}, \bibinfo {author}
  {\bibfnamefont {G.}~\bibnamefont {de~la Pe{\~{n}}a~Munoz}}, \bibinfo {author}
  {\bibfnamefont {S.~X.-L.}\ \bibnamefont {Sun}}, \bibinfo {author}
  {\bibfnamefont {Y.~I.}\ \bibnamefont {Joe}}, \bibinfo {author} {\bibfnamefont
  {A.~H.}\ \bibnamefont {Reid}}, \bibinfo {author} {\bibfnamefont {S.~F.}\
  \bibnamefont {Wandel}}, \bibinfo {author} {\bibfnamefont {G.}~\bibnamefont
  {Coslovich}}, \bibinfo {author} {\bibfnamefont {W.}~\bibnamefont
  {Schlotter}}, \bibinfo {author} {\bibfnamefont {T.}~\bibnamefont {van
  Driel}}, \bibinfo {author} {\bibfnamefont {J.}~\bibnamefont {Schneeloch}},
  \bibinfo {author} {\bibfnamefont {G.~D.}\ \bibnamefont {Gu}}, \bibinfo
  {author} {\bibfnamefont {N.}~\bibnamefont {Goldenfeld}},\ and\ \bibinfo
  {author} {\bibfnamefont {P.}~\bibnamefont {Abbamonte}},\ }\bibfield  {title}
  {\bibinfo {title} {Evidence for photoinduced sliding of the charge-order
  condensate in {La$_{1.875}$Ba$_{0.125}$CuO$_{4}$}},\ }\href@noop {}
  {\bibfield  {journal} {\bibinfo  {journal} {Physical Review B}\ }\textbf
  {\bibinfo {volume} {100}} (\bibinfo {year} {2019}{\natexlab{b}})}\BibitemShut
  {NoStop}%
\bibitem [{\citenamefont {Fausti}\ \emph {et~al.}(2011)\citenamefont {Fausti},
  \citenamefont {Tobey}, \citenamefont {Dean}, \citenamefont {Kaiser},
  \citenamefont {Dienst}, \citenamefont {Hoffmann}, \citenamefont {Pyon},
  \citenamefont {Takayama}, \citenamefont {Takagi},\ and\ \citenamefont
  {Cavalleri}}]{Fausti2011}%
  \BibitemOpen
  \bibfield  {author} {\bibinfo {author} {\bibfnamefont {D.}~\bibnamefont
  {Fausti}}, \bibinfo {author} {\bibfnamefont {R.~I.}\ \bibnamefont {Tobey}},
  \bibinfo {author} {\bibfnamefont {N.}~\bibnamefont {Dean}}, \bibinfo {author}
  {\bibfnamefont {S.}~\bibnamefont {Kaiser}}, \bibinfo {author} {\bibfnamefont
  {A.}~\bibnamefont {Dienst}}, \bibinfo {author} {\bibfnamefont {M.~C.}\
  \bibnamefont {Hoffmann}}, \bibinfo {author} {\bibfnamefont {S.}~\bibnamefont
  {Pyon}}, \bibinfo {author} {\bibfnamefont {T.}~\bibnamefont {Takayama}},
  \bibinfo {author} {\bibfnamefont {H.}~\bibnamefont {Takagi}},\ and\ \bibinfo
  {author} {\bibfnamefont {A.}~\bibnamefont {Cavalleri}},\ }\bibfield  {title}
  {\bibinfo {title} {Light-induced superconductivity in a stripe-ordered
  cuprate},\ }\href {https://doi.org/10.1126/science.1197294} {\bibfield
  {journal} {\bibinfo  {journal} {Science}\ }\textbf {\bibinfo {volume}
  {331}},\ \bibinfo {pages} {189} (\bibinfo {year} {2011})}\BibitemShut
  {NoStop}%
\bibitem [{\citenamefont {Hu}\ \emph {et~al.}(2014)\citenamefont {Hu},
  \citenamefont {Kaiser}, \citenamefont {Nicoletti}, \citenamefont {Hunt},
  \citenamefont {Gierz}, \citenamefont {Hoffmann}, \citenamefont {Tacon},
  \citenamefont {Loew}, \citenamefont {Keimer},\ and\ \citenamefont
  {Cavalleri}}]{Hu2014}%
  \BibitemOpen
  \bibfield  {author} {\bibinfo {author} {\bibfnamefont {W.}~\bibnamefont
  {Hu}}, \bibinfo {author} {\bibfnamefont {S.}~\bibnamefont {Kaiser}}, \bibinfo
  {author} {\bibfnamefont {D.}~\bibnamefont {Nicoletti}}, \bibinfo {author}
  {\bibfnamefont {C.~R.}\ \bibnamefont {Hunt}}, \bibinfo {author}
  {\bibfnamefont {I.}~\bibnamefont {Gierz}}, \bibinfo {author} {\bibfnamefont
  {M.~C.}\ \bibnamefont {Hoffmann}}, \bibinfo {author} {\bibfnamefont {M.~L.}\
  \bibnamefont {Tacon}}, \bibinfo {author} {\bibfnamefont {T.}~\bibnamefont
  {Loew}}, \bibinfo {author} {\bibfnamefont {B.}~\bibnamefont {Keimer}},\ and\
  \bibinfo {author} {\bibfnamefont {A.}~\bibnamefont {Cavalleri}},\ }\bibfield
  {title} {\bibinfo {title} {Optically enhanced coherent transport in
  {{YBa$_{2}$Cu$_{3}$O$_{6.5}$}} by ultrafast redistribution of interlayer
  coupling},\ }\href@noop {} {\bibfield  {journal} {\bibinfo  {journal} {Nature
  Materials}\ }\textbf {\bibinfo {volume} {13}},\ \bibinfo {pages} {705}
  (\bibinfo {year} {2014})}\BibitemShut {NoStop}%
\bibitem [{\citenamefont {Kaiser}\ \emph {et~al.}(2014)\citenamefont {Kaiser},
  \citenamefont {Hunt}, \citenamefont {Nicoletti}, \citenamefont {Hu},
  \citenamefont {Gierz}, \citenamefont {Liu}, \citenamefont {Tacon},
  \citenamefont {Loew}, \citenamefont {Haug}, \citenamefont {Keimer},\ and\
  \citenamefont {Cavalleri}}]{Kaiser2014}%
  \BibitemOpen
  \bibfield  {author} {\bibinfo {author} {\bibfnamefont {S.}~\bibnamefont
  {Kaiser}}, \bibinfo {author} {\bibfnamefont {C.~R.}\ \bibnamefont {Hunt}},
  \bibinfo {author} {\bibfnamefont {D.}~\bibnamefont {Nicoletti}}, \bibinfo
  {author} {\bibfnamefont {W.}~\bibnamefont {Hu}}, \bibinfo {author}
  {\bibfnamefont {I.}~\bibnamefont {Gierz}}, \bibinfo {author} {\bibfnamefont
  {H.~Y.}\ \bibnamefont {Liu}}, \bibinfo {author} {\bibfnamefont {M.~L.}\
  \bibnamefont {Tacon}}, \bibinfo {author} {\bibfnamefont {T.}~\bibnamefont
  {Loew}}, \bibinfo {author} {\bibfnamefont {D.}~\bibnamefont {Haug}}, \bibinfo
  {author} {\bibfnamefont {B.}~\bibnamefont {Keimer}},\ and\ \bibinfo {author}
  {\bibfnamefont {A.}~\bibnamefont {Cavalleri}},\ }\bibfield  {title} {\bibinfo
  {title} {Optically induced coherent transport far above ${T}_{\text{c}}$ in
  underdoped {{YBa$_{2}$Cu$_{3}$O$_{6}$}}.},\ }\href@noop {} {\bibfield
  {journal} {\bibinfo  {journal} {Physical Review B}\ }\textbf {\bibinfo
  {volume} {89}} (\bibinfo {year} {2014})}\BibitemShut {NoStop}%
\bibitem [{\citenamefont {Först}\ \emph
  {et~al.}(2014{\natexlab{a}})\citenamefont {Först}, \citenamefont {Tobey},
  \citenamefont {Bromberger}, \citenamefont {Wilkins}, \citenamefont {Khanna},
  \citenamefont {Caviglia}, \citenamefont {Chuang}, \citenamefont {Lee},
  \citenamefont {Schlotter}, \citenamefont {Turner}, \citenamefont {Minitti},
  \citenamefont {Krupin}, \citenamefont {Xu}, \citenamefont {Wen},
  \citenamefont {Gu}, \citenamefont {Dhesi}, \citenamefont {Cavalleri},\ and\
  \citenamefont {Hill}}]{Foerst2014a}%
  \BibitemOpen
  \bibfield  {author} {\bibinfo {author} {\bibfnamefont {M.}~\bibnamefont
  {Först}}, \bibinfo {author} {\bibfnamefont {R.}~\bibnamefont {Tobey}},
  \bibinfo {author} {\bibfnamefont {H.}~\bibnamefont {Bromberger}}, \bibinfo
  {author} {\bibfnamefont {S.}~\bibnamefont {Wilkins}}, \bibinfo {author}
  {\bibfnamefont {V.}~\bibnamefont {Khanna}}, \bibinfo {author} {\bibfnamefont
  {A.}~\bibnamefont {Caviglia}}, \bibinfo {author} {\bibfnamefont {Y.-D.}\
  \bibnamefont {Chuang}}, \bibinfo {author} {\bibfnamefont {W.}~\bibnamefont
  {Lee}}, \bibinfo {author} {\bibfnamefont {W.}~\bibnamefont {Schlotter}},
  \bibinfo {author} {\bibfnamefont {J.}~\bibnamefont {Turner}}, \bibinfo
  {author} {\bibfnamefont {M.}~\bibnamefont {Minitti}}, \bibinfo {author}
  {\bibfnamefont {O.}~\bibnamefont {Krupin}}, \bibinfo {author} {\bibfnamefont
  {Z.}~\bibnamefont {Xu}}, \bibinfo {author} {\bibfnamefont {J.}~\bibnamefont
  {Wen}}, \bibinfo {author} {\bibfnamefont {G.}~\bibnamefont {Gu}}, \bibinfo
  {author} {\bibfnamefont {S.}~\bibnamefont {Dhesi}}, \bibinfo {author}
  {\bibfnamefont {A.}~\bibnamefont {Cavalleri}},\ and\ \bibinfo {author}
  {\bibfnamefont {J.}~\bibnamefont {Hill}},\ }\bibfield  {title} {\bibinfo
  {title} {Melting of charge stripes in vibrationally driven
  {La$_{1.875}$Ba$_{0.125}$CuO$_{4}$}: Assessing the respective roles of
  electronic and lattice order in frustrated superconductors},\ }\href
  {https://doi.org/10.1103/physrevlett.112.157002} {\bibfield  {journal}
  {\bibinfo  {journal} {Physical Review Letters}\ }\textbf {\bibinfo {volume}
  {112}},\ \bibinfo {pages} {157002} (\bibinfo {year}
  {2014}{\natexlab{a}})}\BibitemShut {NoStop}%
\bibitem [{\citenamefont {Först}\ \emph
  {et~al.}(2014{\natexlab{b}})\citenamefont {Först}, \citenamefont {Frano},
  \citenamefont {Kaiser}, \citenamefont {Mankowsky}, \citenamefont {Hunt},
  \citenamefont {Turner}, \citenamefont {Dakovski}, \citenamefont {Minitti},
  \citenamefont {Robinson}, \citenamefont {Loew}, \citenamefont {Tacon},
  \citenamefont {Keimer}, \citenamefont {Hill}, \citenamefont {Cavalleri},\
  and\ \citenamefont {Dhesi}}]{Foerst2014b}%
  \BibitemOpen
  \bibfield  {author} {\bibinfo {author} {\bibfnamefont {M.}~\bibnamefont
  {Först}}, \bibinfo {author} {\bibfnamefont {A.}~\bibnamefont {Frano}},
  \bibinfo {author} {\bibfnamefont {S.}~\bibnamefont {Kaiser}}, \bibinfo
  {author} {\bibfnamefont {R.}~\bibnamefont {Mankowsky}}, \bibinfo {author}
  {\bibfnamefont {C.~R.}\ \bibnamefont {Hunt}}, \bibinfo {author}
  {\bibfnamefont {J.~J.}\ \bibnamefont {Turner}}, \bibinfo {author}
  {\bibfnamefont {G.~L.}\ \bibnamefont {Dakovski}}, \bibinfo {author}
  {\bibfnamefont {M.~P.}\ \bibnamefont {Minitti}}, \bibinfo {author}
  {\bibfnamefont {J.}~\bibnamefont {Robinson}}, \bibinfo {author}
  {\bibfnamefont {T.}~\bibnamefont {Loew}}, \bibinfo {author} {\bibfnamefont
  {M.~L.}\ \bibnamefont {Tacon}}, \bibinfo {author} {\bibfnamefont
  {B.}~\bibnamefont {Keimer}}, \bibinfo {author} {\bibfnamefont {J.~P.}\
  \bibnamefont {Hill}}, \bibinfo {author} {\bibfnamefont {A.}~\bibnamefont
  {Cavalleri}},\ and\ \bibinfo {author} {\bibfnamefont {S.~S.}\ \bibnamefont
  {Dhesi}},\ }\bibfield  {title} {\bibinfo {title} {Femtosecond x rays link
  melting of charge-density wave correlations and light-enhanced coherent
  transport in {{YBa$_{2}$Cu$_{3}$O$_{6.6}$}}.},\ }\href
  {https://doi.org/10.1103/physrevb.90.184514} {\bibfield  {journal} {\bibinfo
  {journal} {Physical Review B}\ }\textbf {\bibinfo {volume} {90}},\ \bibinfo
  {pages} {184514} (\bibinfo {year} {2014}{\natexlab{b}})}\BibitemShut
  {NoStop}%
\bibitem [{\citenamefont {Khanna}\ \emph {et~al.}(2016)\citenamefont {Khanna},
  \citenamefont {Mankowsky}, \citenamefont {Petrich}, \citenamefont
  {Bromberger}, \citenamefont {Cavill}, \citenamefont {M\"ohr-Vorobeva},
  \citenamefont {Nicoletti}, \citenamefont {Laplace}, \citenamefont {Gu},
  \citenamefont {Hill}, \citenamefont {F\"orst}, \citenamefont {Cavalleri},\
  and\ \citenamefont {Dhesi}}]{Khanna2016}%
  \BibitemOpen
  \bibfield  {author} {\bibinfo {author} {\bibfnamefont {V.}~\bibnamefont
  {Khanna}}, \bibinfo {author} {\bibfnamefont {R.}~\bibnamefont {Mankowsky}},
  \bibinfo {author} {\bibfnamefont {M.}~\bibnamefont {Petrich}}, \bibinfo
  {author} {\bibfnamefont {H.}~\bibnamefont {Bromberger}}, \bibinfo {author}
  {\bibfnamefont {S.~A.}\ \bibnamefont {Cavill}}, \bibinfo {author}
  {\bibfnamefont {E.}~\bibnamefont {M\"ohr-Vorobeva}}, \bibinfo {author}
  {\bibfnamefont {D.}~\bibnamefont {Nicoletti}}, \bibinfo {author}
  {\bibfnamefont {Y.}~\bibnamefont {Laplace}}, \bibinfo {author} {\bibfnamefont
  {G.~D.}\ \bibnamefont {Gu}}, \bibinfo {author} {\bibfnamefont {J.~P.}\
  \bibnamefont {Hill}}, \bibinfo {author} {\bibfnamefont {M.}~\bibnamefont
  {F\"orst}}, \bibinfo {author} {\bibfnamefont {A.}~\bibnamefont {Cavalleri}},\
  and\ \bibinfo {author} {\bibfnamefont {S.~S.}\ \bibnamefont {Dhesi}},\
  }\bibfield  {title} {\bibinfo {title} {Restoring interlayer {J}osephson
  coupling in {${\mathrm{La}}_{1.885}{\mathrm{Ba}}_{0.115}{\mathrm{CuO}}_{4}$}
  by charge transfer melting of stripe order},\ }\href
  {https://doi.org/10.1103/PhysRevB.93.224522} {\bibfield  {journal} {\bibinfo
  {journal} {Phys. Rev. B}\ }\textbf {\bibinfo {volume} {93}},\ \bibinfo
  {pages} {224522} (\bibinfo {year} {2016})}\BibitemShut {NoStop}%
\bibitem [{\citenamefont {Baykusheva}\ \emph {et~al.}(2022)\citenamefont
  {Baykusheva}, \citenamefont {Jang}, \citenamefont {Husain}, \citenamefont
  {Lee}, \citenamefont {TenHuisen}, \citenamefont {Zhou}, \citenamefont {Park},
  \citenamefont {Kim}, \citenamefont {Kim}, \citenamefont {Kim}, \citenamefont
  {Kim}, \citenamefont {Park}, \citenamefont {Abbamonte}, \citenamefont {Kim},
  \citenamefont {Gu}, \citenamefont {Wang},\ and\ \citenamefont
  {Mitrano}}]{Baykusheva2022}%
  \BibitemOpen
  \bibfield  {author} {\bibinfo {author} {\bibfnamefont {D.~R.}\ \bibnamefont
  {Baykusheva}}, \bibinfo {author} {\bibfnamefont {H.}~\bibnamefont {Jang}},
  \bibinfo {author} {\bibfnamefont {A.~A.}\ \bibnamefont {Husain}}, \bibinfo
  {author} {\bibfnamefont {S.}~\bibnamefont {Lee}}, \bibinfo {author}
  {\bibfnamefont {S.~F.}\ \bibnamefont {TenHuisen}}, \bibinfo {author}
  {\bibfnamefont {P.}~\bibnamefont {Zhou}}, \bibinfo {author} {\bibfnamefont
  {S.}~\bibnamefont {Park}}, \bibinfo {author} {\bibfnamefont {H.}~\bibnamefont
  {Kim}}, \bibinfo {author} {\bibfnamefont {J.-K.}\ \bibnamefont {Kim}},
  \bibinfo {author} {\bibfnamefont {H.-D.}\ \bibnamefont {Kim}}, \bibinfo
  {author} {\bibfnamefont {M.}~\bibnamefont {Kim}}, \bibinfo {author}
  {\bibfnamefont {S.-Y.}\ \bibnamefont {Park}}, \bibinfo {author}
  {\bibfnamefont {P.}~\bibnamefont {Abbamonte}}, \bibinfo {author}
  {\bibfnamefont {B.}~\bibnamefont {Kim}}, \bibinfo {author} {\bibfnamefont
  {G.}~\bibnamefont {Gu}}, \bibinfo {author} {\bibfnamefont {Y.}~\bibnamefont
  {Wang}},\ and\ \bibinfo {author} {\bibfnamefont {M.}~\bibnamefont
  {Mitrano}},\ }\bibfield  {title} {\bibinfo {title} {Ultrafast renormalization
  of the on-site coulomb repulsion in a cuprate superconductor},\ }\href
  {https://doi.org/10.1103/physrevx.12.011013} {\bibfield  {journal} {\bibinfo
  {journal} {Physical Review X}\ }\textbf {\bibinfo {volume} {12}},\ \bibinfo
  {pages} {011013} (\bibinfo {year} {2022})}\BibitemShut {NoStop}%
\bibitem [{\citenamefont {Jang}\ \emph {et~al.}(2020)\citenamefont {Jang},
  \citenamefont {Kim}, \citenamefont {Kim}, \citenamefont {Park}, \citenamefont
  {Kwon}, \citenamefont {Lee}, \citenamefont {Park}, \citenamefont {Park},
  \citenamefont {Kim}, \citenamefont {Hyun}, \citenamefont {Hwang},
  \citenamefont {Lee}, \citenamefont {Lim}, \citenamefont {Gang}, \citenamefont
  {Kim}, \citenamefont {Heo}, \citenamefont {Kim}, \citenamefont {Jung},
  \citenamefont {Kim}, \citenamefont {Park}, \citenamefont {Kim}, \citenamefont
  {Shin}, \citenamefont {Park}, \citenamefont {Koo}, \citenamefont {Shin},
  \citenamefont {Heo}, \citenamefont {Kim}, \citenamefont {Min}, \citenamefont
  {Han}, \citenamefont {Kang}, \citenamefont {Lee}, \citenamefont {Kim},
  \citenamefont {Eom},\ and\ \citenamefont {Rah}}]{Jang2020}%
  \BibitemOpen
  \bibfield  {author} {\bibinfo {author} {\bibfnamefont {H.}~\bibnamefont
  {Jang}}, \bibinfo {author} {\bibfnamefont {H.-D.}\ \bibnamefont {Kim}},
  \bibinfo {author} {\bibfnamefont {M.}~\bibnamefont {Kim}}, \bibinfo {author}
  {\bibfnamefont {S.~H.}\ \bibnamefont {Park}}, \bibinfo {author}
  {\bibfnamefont {S.}~\bibnamefont {Kwon}}, \bibinfo {author} {\bibfnamefont
  {J.~Y.}\ \bibnamefont {Lee}}, \bibinfo {author} {\bibfnamefont {S.-Y.}\
  \bibnamefont {Park}}, \bibinfo {author} {\bibfnamefont {G.}~\bibnamefont
  {Park}}, \bibinfo {author} {\bibfnamefont {S.}~\bibnamefont {Kim}}, \bibinfo
  {author} {\bibfnamefont {H.}~\bibnamefont {Hyun}}, \bibinfo {author}
  {\bibfnamefont {S.}~\bibnamefont {Hwang}}, \bibinfo {author} {\bibfnamefont
  {C.-S.}\ \bibnamefont {Lee}}, \bibinfo {author} {\bibfnamefont {C.-Y.}\
  \bibnamefont {Lim}}, \bibinfo {author} {\bibfnamefont {W.}~\bibnamefont
  {Gang}}, \bibinfo {author} {\bibfnamefont {M.}~\bibnamefont {Kim}}, \bibinfo
  {author} {\bibfnamefont {S.}~\bibnamefont {Heo}}, \bibinfo {author}
  {\bibfnamefont {J.}~\bibnamefont {Kim}}, \bibinfo {author} {\bibfnamefont
  {G.}~\bibnamefont {Jung}}, \bibinfo {author} {\bibfnamefont {S.}~\bibnamefont
  {Kim}}, \bibinfo {author} {\bibfnamefont {J.}~\bibnamefont {Park}}, \bibinfo
  {author} {\bibfnamefont {J.}~\bibnamefont {Kim}}, \bibinfo {author}
  {\bibfnamefont {H.}~\bibnamefont {Shin}}, \bibinfo {author} {\bibfnamefont
  {J.}~\bibnamefont {Park}}, \bibinfo {author} {\bibfnamefont {T.-Y.}\
  \bibnamefont {Koo}}, \bibinfo {author} {\bibfnamefont {H.-J.}\ \bibnamefont
  {Shin}}, \bibinfo {author} {\bibfnamefont {H.}~\bibnamefont {Heo}}, \bibinfo
  {author} {\bibfnamefont {C.}~\bibnamefont {Kim}}, \bibinfo {author}
  {\bibfnamefont {C.-K.}\ \bibnamefont {Min}}, \bibinfo {author} {\bibfnamefont
  {J.-H.}\ \bibnamefont {Han}}, \bibinfo {author} {\bibfnamefont {H.-S.}\
  \bibnamefont {Kang}}, \bibinfo {author} {\bibfnamefont {H.-S.}\ \bibnamefont
  {Lee}}, \bibinfo {author} {\bibfnamefont {K.~S.}\ \bibnamefont {Kim}},
  \bibinfo {author} {\bibfnamefont {I.}~\bibnamefont {Eom}},\ and\ \bibinfo
  {author} {\bibfnamefont {S.}~\bibnamefont {Rah}},\ }\bibfield  {title}
  {\bibinfo {title} {Time-resolved resonant elastic soft x-ray scattering at
  pohang accelerator laboratory x-ray free electron laser},\ }\href
  {https://doi.org/10.1063/5.0016414} {\bibfield  {journal} {\bibinfo
  {journal} {Review of Scientific Instruments}\ }\textbf {\bibinfo {volume}
  {91}},\ \bibinfo {pages} {083904} (\bibinfo {year} {2020})}\BibitemShut
  {NoStop}%
\bibitem [{\citenamefont {Fink}\ \emph {et~al.}(2009)\citenamefont {Fink},
  \citenamefont {Schierle}, \citenamefont {Weschke}, \citenamefont {Geck},
  \citenamefont {Hawthorn}, \citenamefont {Soltwisch}, \citenamefont {Wadati},
  \citenamefont {Wu}, \citenamefont {D\"urr}, \citenamefont {Wizent},
  \citenamefont {B\"uchner},\ and\ \citenamefont {Sawatzky}}]{Fink2009}%
  \BibitemOpen
  \bibfield  {author} {\bibinfo {author} {\bibfnamefont {J.}~\bibnamefont
  {Fink}}, \bibinfo {author} {\bibfnamefont {E.}~\bibnamefont {Schierle}},
  \bibinfo {author} {\bibfnamefont {E.}~\bibnamefont {Weschke}}, \bibinfo
  {author} {\bibfnamefont {J.}~\bibnamefont {Geck}}, \bibinfo {author}
  {\bibfnamefont {D.}~\bibnamefont {Hawthorn}}, \bibinfo {author}
  {\bibfnamefont {V.}~\bibnamefont {Soltwisch}}, \bibinfo {author}
  {\bibfnamefont {H.}~\bibnamefont {Wadati}}, \bibinfo {author} {\bibfnamefont
  {H.-H.}\ \bibnamefont {Wu}}, \bibinfo {author} {\bibfnamefont {H.~A.}\
  \bibnamefont {D\"urr}}, \bibinfo {author} {\bibfnamefont {N.}~\bibnamefont
  {Wizent}}, \bibinfo {author} {\bibfnamefont {B.}~\bibnamefont {B\"uchner}},\
  and\ \bibinfo {author} {\bibfnamefont {G.~A.}\ \bibnamefont {Sawatzky}},\
  }\bibfield  {title} {\bibinfo {title} {Charge ordering in
  {${\text{La}}_{1.8\ensuremath{-}x}{\text{Eu}}_{0.2}{\text{Sr}}_{x}{\text{CuO}}_{4}$}
  studied by resonant soft x-ray diffraction},\ }\href
  {https://doi.org/10.1103/PhysRevB.79.100502} {\bibfield  {journal} {\bibinfo
  {journal} {Phys. Rev. B}\ }\textbf {\bibinfo {volume} {79}},\ \bibinfo
  {pages} {100502} (\bibinfo {year} {2009})}\BibitemShut {NoStop}%
\bibitem [{\citenamefont {Na}\ \emph {et~al.}(2020)\citenamefont {Na},
  \citenamefont {Boschini}, \citenamefont {Mills}, \citenamefont {Michiardi},
  \citenamefont {Day}, \citenamefont {Zwartsenberg}, \citenamefont {Levy},
  \citenamefont {Zhdanovich}, \citenamefont {Kemper}, \citenamefont {Jones},\
  and\ \citenamefont {Damascelli}}]{Na2020}%
  \BibitemOpen
  \bibfield  {author} {\bibinfo {author} {\bibfnamefont {M.~X.}\ \bibnamefont
  {Na}}, \bibinfo {author} {\bibfnamefont {F.}~\bibnamefont {Boschini}},
  \bibinfo {author} {\bibfnamefont {A.~K.}\ \bibnamefont {Mills}}, \bibinfo
  {author} {\bibfnamefont {M.}~\bibnamefont {Michiardi}}, \bibinfo {author}
  {\bibfnamefont {R.~P.}\ \bibnamefont {Day}}, \bibinfo {author} {\bibfnamefont
  {B.}~\bibnamefont {Zwartsenberg}}, \bibinfo {author} {\bibfnamefont
  {G.}~\bibnamefont {Levy}}, \bibinfo {author} {\bibfnamefont {S.}~\bibnamefont
  {Zhdanovich}}, \bibinfo {author} {\bibfnamefont {A.~F.}\ \bibnamefont
  {Kemper}}, \bibinfo {author} {\bibfnamefont {D.~J.}\ \bibnamefont {Jones}},\
  and\ \bibinfo {author} {\bibfnamefont {A.}~\bibnamefont {Damascelli}},\
  }\bibfield  {title} {\bibinfo {title} {Establishing nonthermal regimes in
  pump-probe electron relaxation dynamics},\ }\href@noop {} {\bibfield
  {journal} {\bibinfo  {journal} {Phys. Rev. B}\ }\textbf {\bibinfo {volume}
  {102}},\ \bibinfo {pages} {184307} (\bibinfo {year} {2020})}\BibitemShut
  {NoStop}%
\bibitem [{\citenamefont {Kohsaka}\ \emph {et~al.}(2007)\citenamefont
  {Kohsaka}, \citenamefont {Taylor}, \citenamefont {Fujita}, \citenamefont
  {Schmidt}, \citenamefont {Lupien}, \citenamefont {Hanaguri}, \citenamefont
  {Azuma}, \citenamefont {Takano}, \citenamefont {Eisaki}, \citenamefont
  {Takagi}, \citenamefont {Uchida},\ and\ \citenamefont {Davis}}]{Khosaka07}%
  \BibitemOpen
  \bibfield  {author} {\bibinfo {author} {\bibfnamefont {Y.}~\bibnamefont
  {Kohsaka}}, \bibinfo {author} {\bibfnamefont {C.}~\bibnamefont {Taylor}},
  \bibinfo {author} {\bibfnamefont {K.}~\bibnamefont {Fujita}}, \bibinfo
  {author} {\bibfnamefont {A.}~\bibnamefont {Schmidt}}, \bibinfo {author}
  {\bibfnamefont {C.}~\bibnamefont {Lupien}}, \bibinfo {author} {\bibfnamefont
  {T.}~\bibnamefont {Hanaguri}}, \bibinfo {author} {\bibfnamefont
  {M.}~\bibnamefont {Azuma}}, \bibinfo {author} {\bibfnamefont
  {M.}~\bibnamefont {Takano}}, \bibinfo {author} {\bibfnamefont
  {H.}~\bibnamefont {Eisaki}}, \bibinfo {author} {\bibfnamefont
  {H.}~\bibnamefont {Takagi}}, \bibinfo {author} {\bibfnamefont
  {S.}~\bibnamefont {Uchida}},\ and\ \bibinfo {author} {\bibfnamefont {J.~C.}\
  \bibnamefont {Davis}},\ }\bibfield  {title} {\bibinfo {title} {An intrinsic
  bond-centered electronic glass with unidirectional domains in underdoped
  cuprates},\ }\href {https://doi.org/10.1126/science.1138584} {\bibfield
  {journal} {\bibinfo  {journal} {Science}\ }\textbf {\bibinfo {volume}
  {315}},\ \bibinfo {pages} {1380} (\bibinfo {year} {2007})}\BibitemShut
  {NoStop}%
\bibitem [{\citenamefont {Blanco-Canosa}\ \emph {et~al.}(2014)\citenamefont
  {Blanco-Canosa}, \citenamefont {Frano}, \citenamefont {Schierle},
  \citenamefont {Porras}, \citenamefont {Loew}, \citenamefont {Minola},
  \citenamefont {Bluschke}, \citenamefont {Weschke}, \citenamefont {Keimer},\
  and\ \citenamefont {Le~Tacon}}]{Blanco-Canosa2014}%
  \BibitemOpen
  \bibfield  {author} {\bibinfo {author} {\bibfnamefont {S.}~\bibnamefont
  {Blanco-Canosa}}, \bibinfo {author} {\bibfnamefont {A.}~\bibnamefont
  {Frano}}, \bibinfo {author} {\bibfnamefont {E.}~\bibnamefont {Schierle}},
  \bibinfo {author} {\bibfnamefont {J.}~\bibnamefont {Porras}}, \bibinfo
  {author} {\bibfnamefont {T.}~\bibnamefont {Loew}}, \bibinfo {author}
  {\bibfnamefont {M.}~\bibnamefont {Minola}}, \bibinfo {author} {\bibfnamefont
  {M.}~\bibnamefont {Bluschke}}, \bibinfo {author} {\bibfnamefont
  {E.}~\bibnamefont {Weschke}}, \bibinfo {author} {\bibfnamefont
  {B.}~\bibnamefont {Keimer}},\ and\ \bibinfo {author} {\bibfnamefont
  {M.}~\bibnamefont {Le~Tacon}},\ }\bibfield  {title} {\bibinfo {title}
  {{Resonant X-ray scattering study of charge-density wave correlations in
  {YBa$_{2}$Cu$_{3}$O$_{6+x}$}}},\ }\href
  {https://doi.org/10.1103/PhysRevB.90.054513} {\bibfield  {journal} {\bibinfo
  {journal} {Phys. Rev. B}\ }\textbf {\bibinfo {volume} {90}},\ \bibinfo
  {pages} {054513} (\bibinfo {year} {2014})}\BibitemShut {NoStop}%
\bibitem [{\citenamefont {Comin}\ \emph {et~al.}(2015)\citenamefont {Comin},
  \citenamefont {Sutarto}, \citenamefont {da~Silva~Neto}, \citenamefont
  {Chauviere}, \citenamefont {Liang}, \citenamefont {Hardy}, \citenamefont
  {Bonn}, \citenamefont {He}, \citenamefont {Sawatzky},\ and\ \citenamefont
  {Damascelli}}]{Comin2015}%
  \BibitemOpen
  \bibfield  {author} {\bibinfo {author} {\bibfnamefont {R.}~\bibnamefont
  {Comin}}, \bibinfo {author} {\bibfnamefont {R.}~\bibnamefont {Sutarto}},
  \bibinfo {author} {\bibfnamefont {E.~H.}\ \bibnamefont {da~Silva~Neto}},
  \bibinfo {author} {\bibfnamefont {L.}~\bibnamefont {Chauviere}}, \bibinfo
  {author} {\bibfnamefont {R.}~\bibnamefont {Liang}}, \bibinfo {author}
  {\bibfnamefont {W.~N.}\ \bibnamefont {Hardy}}, \bibinfo {author}
  {\bibfnamefont {D.~A.}\ \bibnamefont {Bonn}}, \bibinfo {author}
  {\bibfnamefont {F.}~\bibnamefont {He}}, \bibinfo {author} {\bibfnamefont
  {G.~A.}\ \bibnamefont {Sawatzky}},\ and\ \bibinfo {author} {\bibfnamefont
  {A.}~\bibnamefont {Damascelli}},\ }\bibfield  {title} {\bibinfo {title}
  {Broken translational and rotational symmetry via charge stripe order in
  underdoped {YBa$_2$Cu$_3$O$_{6+y}$}},\ }\href
  {https://doi.org/10.1126/science.1258399} {\bibfield  {journal} {\bibinfo
  {journal} {Science}\ }\textbf {\bibinfo {volume} {347}},\ \bibinfo {pages}
  {1335} (\bibinfo {year} {2015})}\BibitemShut {NoStop}%
\bibitem [{\citenamefont {McMahon}\ \emph {et~al.}(2020)\citenamefont
  {McMahon}, \citenamefont {Achkar}, \citenamefont {da~Silva~Neto},
  \citenamefont {Djianto}, \citenamefont {Menard}, \citenamefont {He},
  \citenamefont {Sutarto}, \citenamefont {Comin}, \citenamefont {Liang},
  \citenamefont {Bonn}, \citenamefont {Hardy}, \citenamefont {Damascelli},\
  and\ \citenamefont {Hawthorn}}]{McMahon2020}%
  \BibitemOpen
  \bibfield  {author} {\bibinfo {author} {\bibfnamefont {C.}~\bibnamefont
  {McMahon}}, \bibinfo {author} {\bibfnamefont {A.~J.}\ \bibnamefont {Achkar}},
  \bibinfo {author} {\bibfnamefont {E.~H.}\ \bibnamefont {da~Silva~Neto}},
  \bibinfo {author} {\bibfnamefont {I.}~\bibnamefont {Djianto}}, \bibinfo
  {author} {\bibfnamefont {J.}~\bibnamefont {Menard}}, \bibinfo {author}
  {\bibfnamefont {F.}~\bibnamefont {He}}, \bibinfo {author} {\bibfnamefont
  {R.}~\bibnamefont {Sutarto}}, \bibinfo {author} {\bibfnamefont
  {R.}~\bibnamefont {Comin}}, \bibinfo {author} {\bibfnamefont
  {R.}~\bibnamefont {Liang}}, \bibinfo {author} {\bibfnamefont {D.~A.}\
  \bibnamefont {Bonn}}, \bibinfo {author} {\bibfnamefont {W.~N.}\ \bibnamefont
  {Hardy}}, \bibinfo {author} {\bibfnamefont {A.}~\bibnamefont {Damascelli}},\
  and\ \bibinfo {author} {\bibfnamefont {D.~G.}\ \bibnamefont {Hawthorn}},\
  }\bibfield  {title} {\bibinfo {title} {Orbital symmetries of charge density
  wave order in {YBa$_2$Cu$_3$O$_{6+x}$}},\ }\href@noop {} {\bibfield
  {journal} {\bibinfo  {journal} {Science Advances}\ }\textbf {\bibinfo
  {volume} {6}},\ \bibinfo {pages} {eaay0345} (\bibinfo {year}
  {2020})}\BibitemShut {NoStop}%
\bibitem [{\citenamefont {Kang}\ \emph {et~al.}(2019)\citenamefont {Kang},
  \citenamefont {Pelliciari}, \citenamefont {Frano}, \citenamefont {Breznay},
  \citenamefont {Schierle}, \citenamefont {Weschke}, \citenamefont {Sutarto},
  \citenamefont {He}, \citenamefont {Shafer}, \citenamefont {Arenholz},
  \citenamefont {Chen}, \citenamefont {Zhang}, \citenamefont {Ruiz},
  \citenamefont {Hao}, \citenamefont {Lewin}, \citenamefont {Analytis},
  \citenamefont {Krockenberger}, \citenamefont {Yamamoto}, \citenamefont
  {Das},\ and\ \citenamefont {Comin}}]{Kang2019}%
  \BibitemOpen
  \bibfield  {author} {\bibinfo {author} {\bibfnamefont {M.}~\bibnamefont
  {Kang}}, \bibinfo {author} {\bibfnamefont {J.}~\bibnamefont {Pelliciari}},
  \bibinfo {author} {\bibfnamefont {A.}~\bibnamefont {Frano}}, \bibinfo
  {author} {\bibfnamefont {N.}~\bibnamefont {Breznay}}, \bibinfo {author}
  {\bibfnamefont {E.}~\bibnamefont {Schierle}}, \bibinfo {author}
  {\bibfnamefont {E.}~\bibnamefont {Weschke}}, \bibinfo {author} {\bibfnamefont
  {R.}~\bibnamefont {Sutarto}}, \bibinfo {author} {\bibfnamefont
  {F.}~\bibnamefont {He}}, \bibinfo {author} {\bibfnamefont {P.}~\bibnamefont
  {Shafer}}, \bibinfo {author} {\bibfnamefont {E.}~\bibnamefont {Arenholz}},
  \bibinfo {author} {\bibfnamefont {M.}~\bibnamefont {Chen}}, \bibinfo {author}
  {\bibfnamefont {K.}~\bibnamefont {Zhang}}, \bibinfo {author} {\bibfnamefont
  {A.}~\bibnamefont {Ruiz}}, \bibinfo {author} {\bibfnamefont {Z.}~\bibnamefont
  {Hao}}, \bibinfo {author} {\bibfnamefont {S.}~\bibnamefont {Lewin}}, \bibinfo
  {author} {\bibfnamefont {J.}~\bibnamefont {Analytis}}, \bibinfo {author}
  {\bibfnamefont {Y.}~\bibnamefont {Krockenberger}}, \bibinfo {author}
  {\bibfnamefont {H.}~\bibnamefont {Yamamoto}}, \bibinfo {author}
  {\bibfnamefont {T.}~\bibnamefont {Das}},\ and\ \bibinfo {author}
  {\bibfnamefont {R.}~\bibnamefont {Comin}},\ }\bibfield  {title} {\bibinfo
  {title} {Evolution of charge order topology across a magnetic phase
  transition in cuprate superconductors},\ }\href@noop {} {\bibfield  {journal}
  {\bibinfo  {journal} {Nature Physics}\ }\textbf {\bibinfo {volume} {15}}
  (\bibinfo {year} {2019})}\BibitemShut {NoStop}%
\bibitem [{\citenamefont {Boschini}\ \emph {et~al.}(2021)\citenamefont
  {Boschini}, \citenamefont {Minola}, \citenamefont {Sutarto}, \citenamefont
  {Schierle}, \citenamefont {Bluschke}, \citenamefont {Das}, \citenamefont
  {Yang}, \citenamefont {Michiardi}, \citenamefont {Shao}, \citenamefont
  {Feng}, \citenamefont {Ono}, \citenamefont {Zhong}, \citenamefont
  {Schneeloch}, \citenamefont {Gu}, \citenamefont {Weschke}, \citenamefont
  {He}, \citenamefont {Chuang}, \citenamefont {Keimer}, \citenamefont
  {Damascelli}, \citenamefont {Frano},\ and\ \citenamefont
  {da~Silva~Neto}}]{Boschini2021}%
  \BibitemOpen
  \bibfield  {author} {\bibinfo {author} {\bibfnamefont {F.}~\bibnamefont
  {Boschini}}, \bibinfo {author} {\bibfnamefont {M.}~\bibnamefont {Minola}},
  \bibinfo {author} {\bibfnamefont {R.}~\bibnamefont {Sutarto}}, \bibinfo
  {author} {\bibfnamefont {E.}~\bibnamefont {Schierle}}, \bibinfo {author}
  {\bibfnamefont {M.}~\bibnamefont {Bluschke}}, \bibinfo {author}
  {\bibfnamefont {S.}~\bibnamefont {Das}}, \bibinfo {author} {\bibfnamefont
  {Y.}~\bibnamefont {Yang}}, \bibinfo {author} {\bibfnamefont {M.}~\bibnamefont
  {Michiardi}}, \bibinfo {author} {\bibfnamefont {Y.~C.}\ \bibnamefont {Shao}},
  \bibinfo {author} {\bibfnamefont {X.}~\bibnamefont {Feng}}, \bibinfo {author}
  {\bibfnamefont {S.}~\bibnamefont {Ono}}, \bibinfo {author} {\bibfnamefont
  {R.~D.}\ \bibnamefont {Zhong}}, \bibinfo {author} {\bibfnamefont {J.~A.}\
  \bibnamefont {Schneeloch}}, \bibinfo {author} {\bibfnamefont {G.~D.}\
  \bibnamefont {Gu}}, \bibinfo {author} {\bibfnamefont {E.}~\bibnamefont
  {Weschke}}, \bibinfo {author} {\bibfnamefont {F.}~\bibnamefont {He}},
  \bibinfo {author} {\bibfnamefont {Y.~D.}\ \bibnamefont {Chuang}}, \bibinfo
  {author} {\bibfnamefont {B.}~\bibnamefont {Keimer}}, \bibinfo {author}
  {\bibfnamefont {A.}~\bibnamefont {Damascelli}}, \bibinfo {author}
  {\bibfnamefont {A.}~\bibnamefont {Frano}},\ and\ \bibinfo {author}
  {\bibfnamefont {E.~H.}\ \bibnamefont {da~Silva~Neto}},\ }\bibfield  {title}
  {\bibinfo {title} {Dynamic electron correlations with charge order wavelength
  along all directions in the copper oxide plane},\ }\href
  {https://doi.org/10.1038/s41467-020-20824-7} {\bibfield  {journal} {\bibinfo
  {journal} {Nature Communications}\ }\textbf {\bibinfo {volume} {12}},\
  \bibinfo {pages} {597} (\bibinfo {year} {2021})}\BibitemShut {NoStop}%
\bibitem [{\citenamefont {Conte}\ \emph {et~al.}(2012)\citenamefont {Conte},
  \citenamefont {Giannetti}, \citenamefont {Coslovich}, \citenamefont
  {Cilento}, \citenamefont {Bossini}, \citenamefont {Abebaw}, \citenamefont
  {Banfi}, \citenamefont {Ferrini}, \citenamefont {Eisaki}, \citenamefont
  {Greven}, \citenamefont {Damascelli}, \citenamefont {van~der Marel},\ and\
  \citenamefont {Parmigiani}}]{Conte2012}%
  \BibitemOpen
  \bibfield  {author} {\bibinfo {author} {\bibfnamefont {S.~D.}\ \bibnamefont
  {Conte}}, \bibinfo {author} {\bibfnamefont {C.}~\bibnamefont {Giannetti}},
  \bibinfo {author} {\bibfnamefont {G.}~\bibnamefont {Coslovich}}, \bibinfo
  {author} {\bibfnamefont {F.}~\bibnamefont {Cilento}}, \bibinfo {author}
  {\bibfnamefont {D.}~\bibnamefont {Bossini}}, \bibinfo {author} {\bibfnamefont
  {T.}~\bibnamefont {Abebaw}}, \bibinfo {author} {\bibfnamefont
  {F.}~\bibnamefont {Banfi}}, \bibinfo {author} {\bibfnamefont
  {G.}~\bibnamefont {Ferrini}}, \bibinfo {author} {\bibfnamefont
  {H.}~\bibnamefont {Eisaki}}, \bibinfo {author} {\bibfnamefont
  {M.}~\bibnamefont {Greven}}, \bibinfo {author} {\bibfnamefont
  {A.}~\bibnamefont {Damascelli}}, \bibinfo {author} {\bibfnamefont
  {D.}~\bibnamefont {van~der Marel}},\ and\ \bibinfo {author} {\bibfnamefont
  {F.}~\bibnamefont {Parmigiani}},\ }\bibfield  {title} {\bibinfo {title}
  {Disentangling the electronic and phononic glue in a high-{$T_{\text{c}}$}
  superconductor},\ }\href {https://doi.org/10.1126/science.1216765} {\bibfield
   {journal} {\bibinfo  {journal} {Science}\ }\textbf {\bibinfo {volume}
  {335}},\ \bibinfo {pages} {1600} (\bibinfo {year} {2012})}\BibitemShut
  {NoStop}%
\bibitem [{\citenamefont {Giannetti}\ \emph {et~al.}(2016)\citenamefont
  {Giannetti}, \citenamefont {Capone}, \citenamefont {Fausti}, \citenamefont
  {Fabrizio}, \citenamefont {Parmigiani},\ and\ \citenamefont
  {Mihailovic}}]{Giannetti2016}%
  \BibitemOpen
  \bibfield  {author} {\bibinfo {author} {\bibfnamefont {C.}~\bibnamefont
  {Giannetti}}, \bibinfo {author} {\bibfnamefont {M.}~\bibnamefont {Capone}},
  \bibinfo {author} {\bibfnamefont {D.}~\bibnamefont {Fausti}}, \bibinfo
  {author} {\bibfnamefont {M.}~\bibnamefont {Fabrizio}}, \bibinfo {author}
  {\bibfnamefont {F.}~\bibnamefont {Parmigiani}},\ and\ \bibinfo {author}
  {\bibfnamefont {D.}~\bibnamefont {Mihailovic}},\ }\bibfield  {title}
  {\bibinfo {title} {Ultrafast optical spectroscopy of strongly correlated
  materials and high-temperature superconductors: a non-equilibrium approach},\
  }\href {https://doi.org/10.1080/00018732.2016.1194044} {\bibfield  {journal}
  {\bibinfo  {journal} {Advances in Physics}\ }\textbf {\bibinfo {volume}
  {65}},\ \bibinfo {pages} {58} (\bibinfo {year} {2016})}\BibitemShut {NoStop}%
\bibitem [{\citenamefont {Tacon}\ \emph {et~al.}(2013)\citenamefont {Tacon},
  \citenamefont {Bosak}, \citenamefont {Souliou}, \citenamefont {Dellea},
  \citenamefont {Loew}, \citenamefont {Heid}, \citenamefont {Bohnen},
  \citenamefont {Ghiringhelli}, \citenamefont {Krisch},\ and\ \citenamefont
  {Keimer}}]{Tacon2014}%
  \BibitemOpen
  \bibfield  {author} {\bibinfo {author} {\bibfnamefont {M.~L.}\ \bibnamefont
  {Tacon}}, \bibinfo {author} {\bibfnamefont {A.}~\bibnamefont {Bosak}},
  \bibinfo {author} {\bibfnamefont {S.~M.}\ \bibnamefont {Souliou}}, \bibinfo
  {author} {\bibfnamefont {G.}~\bibnamefont {Dellea}}, \bibinfo {author}
  {\bibfnamefont {T.}~\bibnamefont {Loew}}, \bibinfo {author} {\bibfnamefont
  {R.}~\bibnamefont {Heid}}, \bibinfo {author} {\bibfnamefont {K.-P.}\
  \bibnamefont {Bohnen}}, \bibinfo {author} {\bibfnamefont {G.}~\bibnamefont
  {Ghiringhelli}}, \bibinfo {author} {\bibfnamefont {M.}~\bibnamefont
  {Krisch}},\ and\ \bibinfo {author} {\bibfnamefont {B.}~\bibnamefont
  {Keimer}},\ }\bibfield  {title} {\bibinfo {title} {Inelastic x-ray scattering
  in {YBa}$_2${Cu}$_3${O}$_{6.6}$ reveals giant phonon anomalies and elastic
  central peak due to charge-density-wave formation},\ }\href
  {https://doi.org/10.1038/nphys2805} {\bibfield  {journal} {\bibinfo
  {journal} {Nature Physics}\ }\textbf {\bibinfo {volume} {10}},\ \bibinfo
  {pages} {52} (\bibinfo {year} {2013})}\BibitemShut {NoStop}%
\bibitem [{\citenamefont {Izzo}\ and\ \citenamefont
  {Oliveira}(2020)}]{Izzo2020}%
  \BibitemOpen
  \bibfield  {author} {\bibinfo {author} {\bibfnamefont {D.}~\bibnamefont
  {Izzo}}\ and\ \bibinfo {author} {\bibfnamefont {M.~J.~D.}\ \bibnamefont
  {Oliveira}},\ }\bibfield  {title} {\bibinfo {title} {Landau theory for
  isotropic, nematic, smectic-a, and smectic-c phases},\ }\href
  {https://doi.org/10.1080/02678292.2019.1631968} {\bibfield  {journal}
  {\bibinfo  {journal} {Liquid Crystals}\ }\textbf {\bibinfo {volume} {47}},\
  \bibinfo {pages} {99} (\bibinfo {year} {2020})}\BibitemShut {NoStop}%
\bibitem [{\citenamefont {Lee}\ \emph {et~al.}(2022)\citenamefont {Lee},
  \citenamefont {Huang}, \citenamefont {Johnson}, \citenamefont {Guo},
  \citenamefont {Husain}, \citenamefont {Mitrano}, \citenamefont {Lu},
  \citenamefont {Zakrzewski}, \citenamefont {de~la Pe{\~{n}}a}, \citenamefont
  {Peng}, \citenamefont {Huang}, \citenamefont {Lee}, \citenamefont {Jang},
  \citenamefont {Lee}, \citenamefont {Joe}, \citenamefont {Doriese},
  \citenamefont {Szypryt}, \citenamefont {Swetz}, \citenamefont {Chi},
  \citenamefont {Aczel}, \citenamefont {MacDougall}, \citenamefont {Kivelson},
  \citenamefont {Fradkin},\ and\ \citenamefont {Abbamonte}}]{Lee2022}%
  \BibitemOpen
  \bibfield  {author} {\bibinfo {author} {\bibfnamefont {S.}~\bibnamefont
  {Lee}}, \bibinfo {author} {\bibfnamefont {E.~W.}\ \bibnamefont {Huang}},
  \bibinfo {author} {\bibfnamefont {T.~A.}\ \bibnamefont {Johnson}}, \bibinfo
  {author} {\bibfnamefont {X.}~\bibnamefont {Guo}}, \bibinfo {author}
  {\bibfnamefont {A.~A.}\ \bibnamefont {Husain}}, \bibinfo {author}
  {\bibfnamefont {M.}~\bibnamefont {Mitrano}}, \bibinfo {author} {\bibfnamefont
  {K.}~\bibnamefont {Lu}}, \bibinfo {author} {\bibfnamefont {A.~V.}\
  \bibnamefont {Zakrzewski}}, \bibinfo {author} {\bibfnamefont {G.~A.}\
  \bibnamefont {de~la Pe{\~{n}}a}}, \bibinfo {author} {\bibfnamefont
  {Y.}~\bibnamefont {Peng}}, \bibinfo {author} {\bibfnamefont {H.}~\bibnamefont
  {Huang}}, \bibinfo {author} {\bibfnamefont {S.-J.}\ \bibnamefont {Lee}},
  \bibinfo {author} {\bibfnamefont {H.}~\bibnamefont {Jang}}, \bibinfo {author}
  {\bibfnamefont {J.-S.}\ \bibnamefont {Lee}}, \bibinfo {author} {\bibfnamefont
  {Y.~I.}\ \bibnamefont {Joe}}, \bibinfo {author} {\bibfnamefont {W.~B.}\
  \bibnamefont {Doriese}}, \bibinfo {author} {\bibfnamefont {P.}~\bibnamefont
  {Szypryt}}, \bibinfo {author} {\bibfnamefont {D.~S.}\ \bibnamefont {Swetz}},
  \bibinfo {author} {\bibfnamefont {S.}~\bibnamefont {Chi}}, \bibinfo {author}
  {\bibfnamefont {A.~A.}\ \bibnamefont {Aczel}}, \bibinfo {author}
  {\bibfnamefont {G.~J.}\ \bibnamefont {MacDougall}}, \bibinfo {author}
  {\bibfnamefont {S.~A.}\ \bibnamefont {Kivelson}}, \bibinfo {author}
  {\bibfnamefont {E.}~\bibnamefont {Fradkin}},\ and\ \bibinfo {author}
  {\bibfnamefont {P.}~\bibnamefont {Abbamonte}},\ }\bibfield  {title} {\bibinfo
  {title} {Generic character of charge and spin density waves in
  superconducting cuprates},\ }\href@noop {} {\bibfield  {journal} {\bibinfo
  {journal} {Proceedings of the National Academy of Sciences}\ }\textbf
  {\bibinfo {volume} {119}} (\bibinfo {year} {2022})}\BibitemShut {NoStop}%
\bibitem [{\citenamefont {Wang}\ \emph {et~al.}(2020)\citenamefont {Wang},
  \citenamefont {Horio}, \citenamefont {von Arx}, \citenamefont {Shen},
  \citenamefont {Mukkattukavil}, \citenamefont {Sassa}, \citenamefont
  {Ivashko}, \citenamefont {Matt}, \citenamefont {Pyon}, \citenamefont
  {Takayama}, \citenamefont {Takagi}, \citenamefont {Kurosawa}, \citenamefont
  {Momono}, \citenamefont {Oda}, \citenamefont {Adachi}, \citenamefont
  {Haidar}, \citenamefont {Koike}, \citenamefont {Tseng}, \citenamefont
  {Zhang}, \citenamefont {Zhao}, \citenamefont {Kummer}, \citenamefont
  {Garcia-Fernandez}, \citenamefont {Zhou}, \citenamefont {Christensen},
  \citenamefont {R{\o}nnow}, \citenamefont {Schmitt},\ and\ \citenamefont
  {Chang}}]{Wang2020a}%
  \BibitemOpen
  \bibfield  {author} {\bibinfo {author} {\bibfnamefont {Q.}~\bibnamefont
  {Wang}}, \bibinfo {author} {\bibfnamefont {M.}~\bibnamefont {Horio}},
  \bibinfo {author} {\bibfnamefont {K.}~\bibnamefont {von Arx}}, \bibinfo
  {author} {\bibfnamefont {Y.}~\bibnamefont {Shen}}, \bibinfo {author}
  {\bibfnamefont {D.~J.}\ \bibnamefont {Mukkattukavil}}, \bibinfo {author}
  {\bibfnamefont {Y.}~\bibnamefont {Sassa}}, \bibinfo {author} {\bibfnamefont
  {O.}~\bibnamefont {Ivashko}}, \bibinfo {author} {\bibfnamefont
  {C.}~\bibnamefont {Matt}}, \bibinfo {author} {\bibfnamefont {S.}~\bibnamefont
  {Pyon}}, \bibinfo {author} {\bibfnamefont {T.}~\bibnamefont {Takayama}},
  \bibinfo {author} {\bibfnamefont {H.}~\bibnamefont {Takagi}}, \bibinfo
  {author} {\bibfnamefont {T.}~\bibnamefont {Kurosawa}}, \bibinfo {author}
  {\bibfnamefont {N.}~\bibnamefont {Momono}}, \bibinfo {author} {\bibfnamefont
  {M.}~\bibnamefont {Oda}}, \bibinfo {author} {\bibfnamefont {T.}~\bibnamefont
  {Adachi}}, \bibinfo {author} {\bibfnamefont {S.}~\bibnamefont {Haidar}},
  \bibinfo {author} {\bibfnamefont {Y.}~\bibnamefont {Koike}}, \bibinfo
  {author} {\bibfnamefont {Y.}~\bibnamefont {Tseng}}, \bibinfo {author}
  {\bibfnamefont {W.}~\bibnamefont {Zhang}}, \bibinfo {author} {\bibfnamefont
  {J.}~\bibnamefont {Zhao}}, \bibinfo {author} {\bibfnamefont {K.}~\bibnamefont
  {Kummer}}, \bibinfo {author} {\bibfnamefont {M.}~\bibnamefont
  {Garcia-Fernandez}}, \bibinfo {author} {\bibfnamefont {K.-J.}\ \bibnamefont
  {Zhou}}, \bibinfo {author} {\bibfnamefont {N.}~\bibnamefont {Christensen}},
  \bibinfo {author} {\bibfnamefont {H.}~\bibnamefont {R{\o}nnow}}, \bibinfo
  {author} {\bibfnamefont {T.}~\bibnamefont {Schmitt}},\ and\ \bibinfo {author}
  {\bibfnamefont {J.}~\bibnamefont {Chang}},\ }\bibfield  {title} {\bibinfo
  {title} {High-temperature charge-stripe correlations in
  {${\text{La}}_{1.675\ensuremath{-}x}{\text{Eu}}_{0.2}{\text{Sr}}_{0.125}{\text{CuO}}_{4}$}},\
  }\href {https://doi.org/10.1103/physrevlett.124.187002} {\bibfield  {journal}
  {\bibinfo  {journal} {Physical Review Letters}\ }\textbf {\bibinfo {volume}
  {124}},\ \bibinfo {pages} {187002} (\bibinfo {year} {2020})}\BibitemShut
  {NoStop}%
\bibitem [{\citenamefont {Zong}\ \emph {et~al.}(2018)\citenamefont {Zong},
  \citenamefont {Kogar}, \citenamefont {Bie}, \citenamefont {Rohwer},
  \citenamefont {Lee}, \citenamefont {Baldini}, \citenamefont {Erge{\c{c}}en},
  \citenamefont {Yilmaz}, \citenamefont {Freelon}, \citenamefont {Sie},
  \citenamefont {Zhou}, \citenamefont {Straquadine}, \citenamefont {Walmsley},
  \citenamefont {Dolgirev}, \citenamefont {Rozhkov}, \citenamefont {Fisher},
  \citenamefont {Jarillo-Herrero}, \citenamefont {Fine},\ and\ \citenamefont
  {Gedik}}]{Zong2018}%
  \BibitemOpen
  \bibfield  {author} {\bibinfo {author} {\bibfnamefont {A.}~\bibnamefont
  {Zong}}, \bibinfo {author} {\bibfnamefont {A.}~\bibnamefont {Kogar}},
  \bibinfo {author} {\bibfnamefont {Y.-Q.}\ \bibnamefont {Bie}}, \bibinfo
  {author} {\bibfnamefont {T.}~\bibnamefont {Rohwer}}, \bibinfo {author}
  {\bibfnamefont {C.}~\bibnamefont {Lee}}, \bibinfo {author} {\bibfnamefont
  {E.}~\bibnamefont {Baldini}}, \bibinfo {author} {\bibfnamefont
  {E.}~\bibnamefont {Erge{\c{c}}en}}, \bibinfo {author} {\bibfnamefont {M.~B.}\
  \bibnamefont {Yilmaz}}, \bibinfo {author} {\bibfnamefont {B.}~\bibnamefont
  {Freelon}}, \bibinfo {author} {\bibfnamefont {E.~J.}\ \bibnamefont {Sie}},
  \bibinfo {author} {\bibfnamefont {H.}~\bibnamefont {Zhou}}, \bibinfo {author}
  {\bibfnamefont {J.}~\bibnamefont {Straquadine}}, \bibinfo {author}
  {\bibfnamefont {P.}~\bibnamefont {Walmsley}}, \bibinfo {author}
  {\bibfnamefont {P.~E.}\ \bibnamefont {Dolgirev}}, \bibinfo {author}
  {\bibfnamefont {A.~V.}\ \bibnamefont {Rozhkov}}, \bibinfo {author}
  {\bibfnamefont {I.~R.}\ \bibnamefont {Fisher}}, \bibinfo {author}
  {\bibfnamefont {P.}~\bibnamefont {Jarillo-Herrero}}, \bibinfo {author}
  {\bibfnamefont {B.~V.}\ \bibnamefont {Fine}},\ and\ \bibinfo {author}
  {\bibfnamefont {N.}~\bibnamefont {Gedik}},\ }\bibfield  {title} {\bibinfo
  {title} {Evidence for topological defects in a photoinduced phase
  transition},\ }\href {https://doi.org/10.1038/s41567-018-0311-9} {\bibfield
  {journal} {\bibinfo  {journal} {Nature Physics}\ }\textbf {\bibinfo {volume}
  {15}},\ \bibinfo {pages} {27} (\bibinfo {year} {2018})}\BibitemShut {NoStop}%
\bibitem [{\citenamefont {Chen}\ \emph {et~al.}(2019)\citenamefont {Chen},
  \citenamefont {Mazzoli}, \citenamefont {Cao}, \citenamefont {Thampy},
  \citenamefont {Barbour}, \citenamefont {Hu}, \citenamefont {Lu},
  \citenamefont {Assefa}, \citenamefont {Miao}, \citenamefont {Fabbris},
  \citenamefont {Gu}, \citenamefont {Tranquada}, \citenamefont {Dean},
  \citenamefont {Wilkins},\ and\ \citenamefont {Robinson}}]{Chen2019a}%
  \BibitemOpen
  \bibfield  {author} {\bibinfo {author} {\bibfnamefont {X.~M.}\ \bibnamefont
  {Chen}}, \bibinfo {author} {\bibfnamefont {C.}~\bibnamefont {Mazzoli}},
  \bibinfo {author} {\bibfnamefont {Y.}~\bibnamefont {Cao}}, \bibinfo {author}
  {\bibfnamefont {V.}~\bibnamefont {Thampy}}, \bibinfo {author} {\bibfnamefont
  {A.~M.}\ \bibnamefont {Barbour}}, \bibinfo {author} {\bibfnamefont
  {W.}~\bibnamefont {Hu}}, \bibinfo {author} {\bibfnamefont {M.}~\bibnamefont
  {Lu}}, \bibinfo {author} {\bibfnamefont {T.~A.}\ \bibnamefont {Assefa}},
  \bibinfo {author} {\bibfnamefont {H.}~\bibnamefont {Miao}}, \bibinfo {author}
  {\bibfnamefont {G.}~\bibnamefont {Fabbris}}, \bibinfo {author} {\bibfnamefont
  {G.~D.}\ \bibnamefont {Gu}}, \bibinfo {author} {\bibfnamefont {J.~M.}\
  \bibnamefont {Tranquada}}, \bibinfo {author} {\bibfnamefont {M.~P.~M.}\
  \bibnamefont {Dean}}, \bibinfo {author} {\bibfnamefont {S.~B.}\ \bibnamefont
  {Wilkins}},\ and\ \bibinfo {author} {\bibfnamefont {I.~K.}\ \bibnamefont
  {Robinson}},\ }\bibfield  {title} {\bibinfo {title} {Charge density wave
  memory in a cuprate superconductor},\ }\href@noop {} {\bibfield  {journal}
  {\bibinfo  {journal} {Nature Communications}\ }\textbf {\bibinfo {volume}
  {10}} (\bibinfo {year} {2019})}\BibitemShut {NoStop}%
\bibitem [{\citenamefont {Lee}\ \emph {et~al.}(2012)\citenamefont {Lee},
  \citenamefont {Chuang}, \citenamefont {Moore}, \citenamefont {Zhu},
  \citenamefont {Patthey}, \citenamefont {Trigo}, \citenamefont {Lu},
  \citenamefont {Kirchmann}, \citenamefont {Krupin}, \citenamefont {Yi},
  \citenamefont {Langner}, \citenamefont {Huse}, \citenamefont {Robinson},
  \citenamefont {Chen}, \citenamefont {Zhou}, \citenamefont {Coslovich},
  \citenamefont {Huber}, \citenamefont {Reis}, \citenamefont {Kaindl},
  \citenamefont {Schoenlein}, \citenamefont {Doering}, \citenamefont {Denes},
  \citenamefont {Schlotter}, \citenamefont {Turner}, \citenamefont {Johnson},
  \citenamefont {Först}, \citenamefont {Sasagawa}, \citenamefont {Kung},
  \citenamefont {Sorini}, \citenamefont {Kemper}, \citenamefont {Moritz},
  \citenamefont {Devereaux}, \citenamefont {Lee}, \citenamefont {Shen},\ and\
  \citenamefont {Hussain}}]{Lee2012}%
  \BibitemOpen
  \bibfield  {author} {\bibinfo {author} {\bibfnamefont {W.}~\bibnamefont
  {Lee}}, \bibinfo {author} {\bibfnamefont {Y.}~\bibnamefont {Chuang}},
  \bibinfo {author} {\bibfnamefont {R.}~\bibnamefont {Moore}}, \bibinfo
  {author} {\bibfnamefont {Y.}~\bibnamefont {Zhu}}, \bibinfo {author}
  {\bibfnamefont {L.}~\bibnamefont {Patthey}}, \bibinfo {author} {\bibfnamefont
  {M.}~\bibnamefont {Trigo}}, \bibinfo {author} {\bibfnamefont
  {D.}~\bibnamefont {Lu}}, \bibinfo {author} {\bibfnamefont {P.}~\bibnamefont
  {Kirchmann}}, \bibinfo {author} {\bibfnamefont {O.}~\bibnamefont {Krupin}},
  \bibinfo {author} {\bibfnamefont {M.}~\bibnamefont {Yi}}, \bibinfo {author}
  {\bibfnamefont {M.}~\bibnamefont {Langner}}, \bibinfo {author} {\bibfnamefont
  {N.}~\bibnamefont {Huse}}, \bibinfo {author} {\bibfnamefont {J.}~\bibnamefont
  {Robinson}}, \bibinfo {author} {\bibfnamefont {Y.}~\bibnamefont {Chen}},
  \bibinfo {author} {\bibfnamefont {S.}~\bibnamefont {Zhou}}, \bibinfo {author}
  {\bibfnamefont {G.}~\bibnamefont {Coslovich}}, \bibinfo {author}
  {\bibfnamefont {B.}~\bibnamefont {Huber}}, \bibinfo {author} {\bibfnamefont
  {D.}~\bibnamefont {Reis}}, \bibinfo {author} {\bibfnamefont {R.}~\bibnamefont
  {Kaindl}}, \bibinfo {author} {\bibfnamefont {R.}~\bibnamefont {Schoenlein}},
  \bibinfo {author} {\bibfnamefont {D.}~\bibnamefont {Doering}}, \bibinfo
  {author} {\bibfnamefont {P.}~\bibnamefont {Denes}}, \bibinfo {author}
  {\bibfnamefont {W.}~\bibnamefont {Schlotter}}, \bibinfo {author}
  {\bibfnamefont {J.}~\bibnamefont {Turner}}, \bibinfo {author} {\bibfnamefont
  {S.}~\bibnamefont {Johnson}}, \bibinfo {author} {\bibfnamefont
  {M.}~\bibnamefont {Först}}, \bibinfo {author} {\bibfnamefont
  {T.}~\bibnamefont {Sasagawa}}, \bibinfo {author} {\bibfnamefont
  {Y.}~\bibnamefont {Kung}}, \bibinfo {author} {\bibfnamefont {A.}~\bibnamefont
  {Sorini}}, \bibinfo {author} {\bibfnamefont {A.}~\bibnamefont {Kemper}},
  \bibinfo {author} {\bibfnamefont {B.}~\bibnamefont {Moritz}}, \bibinfo
  {author} {\bibfnamefont {T.}~\bibnamefont {Devereaux}}, \bibinfo {author}
  {\bibfnamefont {D.-H.}\ \bibnamefont {Lee}}, \bibinfo {author} {\bibfnamefont
  {Z.}~\bibnamefont {Shen}},\ and\ \bibinfo {author} {\bibfnamefont
  {Z.}~\bibnamefont {Hussain}},\ }\bibfield  {title} {\bibinfo {title} {Phase
  fluctuations and the absence of topological defects in a photo-excited
  charge-ordered nickelate},\ }\href@noop {} {\bibfield  {journal} {\bibinfo
  {journal} {Nature Communications}\ }\textbf {\bibinfo {volume} {3}} (\bibinfo
  {year} {2012})}\BibitemShut {NoStop}%
\bibitem [{\citenamefont {Lubashevsky}\ \emph {et~al.}(2014)\citenamefont
  {Lubashevsky}, \citenamefont {Pan}, \citenamefont {Kirzhner}, \citenamefont
  {Koren},\ and\ \citenamefont {Armitage}}]{Lubashevsky2014}%
  \BibitemOpen
  \bibfield  {author} {\bibinfo {author} {\bibfnamefont {Y.}~\bibnamefont
  {Lubashevsky}}, \bibinfo {author} {\bibfnamefont {L.}~\bibnamefont {Pan}},
  \bibinfo {author} {\bibfnamefont {T.}~\bibnamefont {Kirzhner}}, \bibinfo
  {author} {\bibfnamefont {G.}~\bibnamefont {Koren}},\ and\ \bibinfo {author}
  {\bibfnamefont {N.~P.}\ \bibnamefont {Armitage}},\ }\bibfield  {title}
  {\bibinfo {title} {Optical birefringence and dichroism of cuprate
  superconductors in the {THz} regime},\ }\href
  {https://doi.org/10.1103/PhysRevLett.112.147001} {\bibfield  {journal}
  {\bibinfo  {journal} {Phys. Rev. Lett.}\ }\textbf {\bibinfo {volume} {112}},\
  \bibinfo {pages} {147001} (\bibinfo {year} {2014})}\BibitemShut {NoStop}%
\bibitem [{\citenamefont {Ishioka}\ \emph {et~al.}(2010)\citenamefont
  {Ishioka}, \citenamefont {Liu}, \citenamefont {Shimatake}, \citenamefont
  {Kurosawa}, \citenamefont {Ichimura}, \citenamefont {Toda}, \citenamefont
  {Oda},\ and\ \citenamefont {Tanda}}]{Ishioka2010}%
  \BibitemOpen
  \bibfield  {author} {\bibinfo {author} {\bibfnamefont {J.}~\bibnamefont
  {Ishioka}}, \bibinfo {author} {\bibfnamefont {Y.~H.}\ \bibnamefont {Liu}},
  \bibinfo {author} {\bibfnamefont {K.}~\bibnamefont {Shimatake}}, \bibinfo
  {author} {\bibfnamefont {T.}~\bibnamefont {Kurosawa}}, \bibinfo {author}
  {\bibfnamefont {K.}~\bibnamefont {Ichimura}}, \bibinfo {author}
  {\bibfnamefont {Y.}~\bibnamefont {Toda}}, \bibinfo {author} {\bibfnamefont
  {M.}~\bibnamefont {Oda}},\ and\ \bibinfo {author} {\bibfnamefont
  {S.}~\bibnamefont {Tanda}},\ }\bibfield  {title} {\bibinfo {title} {Chiral
  charge-density waves},\ }\href
  {https://doi.org/10.1103/physrevlett.105.176401} {\bibfield  {journal}
  {\bibinfo  {journal} {Physical Review Letters}\ }\textbf {\bibinfo {volume}
  {105}},\ \bibinfo {pages} {176401} (\bibinfo {year} {2010})}\BibitemShut
  {NoStop}%
\bibitem [{\citenamefont {Hawthorn}\ \emph {et~al.}(2011)\citenamefont
  {Hawthorn}, \citenamefont {He}, \citenamefont {Venema}, \citenamefont
  {Davis}, \citenamefont {Achkar}, \citenamefont {Zhang}, \citenamefont
  {Sutarto}, \citenamefont {Wadati}, \citenamefont {Radi}, \citenamefont
  {Wilson}, \citenamefont {Wright}, \citenamefont {Shen}, \citenamefont {Geck},
  \citenamefont {Zhang}, \citenamefont {Novák},\ and\ \citenamefont
  {Sawatzky}}]{Hawthorn2011}%
  \BibitemOpen
  \bibfield  {author} {\bibinfo {author} {\bibfnamefont {D.~G.}\ \bibnamefont
  {Hawthorn}}, \bibinfo {author} {\bibfnamefont {F.}~\bibnamefont {He}},
  \bibinfo {author} {\bibfnamefont {L.}~\bibnamefont {Venema}}, \bibinfo
  {author} {\bibfnamefont {H.}~\bibnamefont {Davis}}, \bibinfo {author}
  {\bibfnamefont {A.~J.}\ \bibnamefont {Achkar}}, \bibinfo {author}
  {\bibfnamefont {J.}~\bibnamefont {Zhang}}, \bibinfo {author} {\bibfnamefont
  {R.}~\bibnamefont {Sutarto}}, \bibinfo {author} {\bibfnamefont
  {H.}~\bibnamefont {Wadati}}, \bibinfo {author} {\bibfnamefont
  {A.}~\bibnamefont {Radi}}, \bibinfo {author} {\bibfnamefont {T.}~\bibnamefont
  {Wilson}}, \bibinfo {author} {\bibfnamefont {G.}~\bibnamefont {Wright}},
  \bibinfo {author} {\bibfnamefont {K.~M.}\ \bibnamefont {Shen}}, \bibinfo
  {author} {\bibfnamefont {J.}~\bibnamefont {Geck}}, \bibinfo {author}
  {\bibfnamefont {H.}~\bibnamefont {Zhang}}, \bibinfo {author} {\bibfnamefont
  {V.}~\bibnamefont {Novák}},\ and\ \bibinfo {author} {\bibfnamefont {G.~A.}\
  \bibnamefont {Sawatzky}},\ }\bibfield  {title} {\bibinfo {title} {An
  in-vacuum diffractometer for resonant elastic soft x-ray scattering},\ }\href
  {https://doi.org/10.1063/1.3607438} {\bibfield  {journal} {\bibinfo
  {journal} {Review of Scientific Instruments}\ }\textbf {\bibinfo {volume}
  {82}},\ \bibinfo {pages} {073104} (\bibinfo {year} {2011})}\BibitemShut
  {NoStop}%
\end{thebibliography}
%

\vskip 0.5cm
\vskip 0.5cm
\vskip 0.5cm
\noindent\textbf{Acknowledgements}

\noindent We gratefully acknowledge Fabio~Boschini, Marta~Zonno, Eduardo~da~Silva~Neto, Alex~Frano, Steve~Dodge, and Vincent~Esposito for useful discussions. We also acknowledge assistance from Hiruy Hale and Andrew Dube from Science Technical Services in machining sample holders suitable for RSXS and tr-RSXS experiments. 

\noindent Funding for this research has been provided by the Alexander von Humboldt Foundation in the form of a Feodor Lynen Research Fellowship (M.B.). This research is funded in part by a QuantEmX grant from ICAM and the Gordon and Betty Moore Foundation through Grant GBMF5305 to Dr.~Martin~Bluschke. N. K.~Gupta acknowledges support from the Waterloo Institute of Nanotechnology (WIN), Mike \& Ophelia Lazaridis Quantum-Nano Centre at University of Waterloo, Canada. The tr-RXS experiments were performed at the SSS-RSXS endstation (proposal number: 2022-1st-SSS-028) of the PAL-XFEL funded by the Korea government (MSIT). H.~Jang acknowledges the support by the National Research Foundation grant funded by the Korea government (MSIT) (grant no. 2019R1F1A1060295). This work was supported by GSDC and KREONET provided by the Korea Institute of Science and Technology Information (KISTI). This research was undertaken thanks in part to funding from the Max Planck-UBC-UTokyo Center for Quantum Materials and the Canada First Research Excellence Fund, Quantum Materials and Future Technologies Program. This project is also funded by the Gordon and Betty Moore Foundation’s EPiQS Initiative, Grant GMBF4779 (A. Damascelli); the Killam, Alfred P. Sloan, and Natural Sciences and Engineering Research Council of Canada’s (NSERC’s) Steacie Memorial Fellowships (A.D.); the Alexander von Humboldt Foundation (A.D.); the Canada Research Chairs Program (A.D.); NSERC, Canada Foundation for Innovation (CFI); the Department of National Defence (DND); the British Columbia Knowledge Development Fund (BCKDF); and the CIFAR Quantum Materials Program. The work at Max Planck POSTECH/Korea Research Initiative was supported by the National Research Foundation of Korea funded by the Ministry of Science and ICT, Grant No. 2022M3H4A1A04074153 and 2020M3H4A2084417. Part of the research described in this paper was performed at the Canadian Light Source, a national research facility of the University of Saskatchewan, which is supported by the Canada Foundation for Innovation (CFI), the Natural Sciences and Engineering Research Council (NSERC), the National Research Council Canada (NRC), the Canadian Institutes of Health Research (CIHR), the Government of Saskatchewan, and the University of Saskatchewan. Preliminary work was performed at the Linac Coherent Light Source (LCLS), SLAC National Accelerator Laboratory, which is supported by the U.S. Department of Energy, Office of Science, Office of Basic Energy Sciences under Contract No. DE-AC02-76SF00515 (A.H.~Reid, G.L.~Dakovski, G.~Coslovich, Q.~L.~Nguyen, N.~G.~Burdet and M.F.~Lin). J.~J.~Turner also acknowledges support from the U.S. Department of Energy, Office of Science, Basic Energy Sciences, Materials Sciences and Engineering Division, under Contract DE-AC02-76SF00515 through the Early Career Research Program. Q.L.~Nguyen acknowledges support from the Bloch Fellowship in Quantum Science and Engineering by the Stanford-SLAC Quantum Fundamentals, Architectures and Machines Initiative. J.~Geck acknowledges his support by the Deutsche Forschungsgemeinschaft through SFB 1143 (project-id 247310070), the Würzburg-Dresden Cluster of Excellence on Complexity and Topology in Quantum Matter–ct.qmat (EXC 2147, project-id 390858490), and SFB 1415 (project-id 417590517).

\vskip 0.5cm
\vskip 0.5cm
\noindent\textbf{Author contributions}

\noindent M.B and N.K.G. contributed equally to this work. The project was conceived by M.B., A.D. and D.G.H.; single crystal Eu-LSCO samples were grown in the lab of A.R. at the Universit\'e Paris-Saclay and subsequently characterized and prepared by J.G.; the tr-RXS measurements were performed by M.B., N.K.G., H.J. and B.L. with the assistance of A.A.H., A.D. and D.G.H.; the experimental system at PAL-XFEL was operated and maintained by H.J., S.-Y.P., M.K., D.J. and H.C.; preliminary characterization using RXS and tr-RXS was performed by M.B., N.K.G., A.A.H, M.X.N., R.S., A.H.R., G.L.D., G.C., Q.L.N., N.G.B., M.-F.L., J.J.T., A.D. and D.G.H; data analysis was performed by M.B., N.K.G. and H.J. with assistance from B.L., M.N., B.D.R. and Q.L.N.; data interpretation was performed by M.B., N.K.G., H.J., A.A.H., S.S., P.M., B.L., Q.L.N., G.C., J-H.P., J.J.T., A.D. and D.G.H.; the manuscript was written by M.B., N.K.G., A.D. and D.G.H. with contributions from all authors. All authors discussed the underlying physics. 

\vskip 0.5cm
\vskip 0.5cm
\noindent\textbf{Competing interests}

\noindent The authors declare that they have no competing interests.

\end{document}